\documentclass[aps,prd,twocolumn,nofootinbib,showpacs]{revtex4-1}

\usepackage{graphicx,epsfig}

\usepackage[colorlinks=true,linkcolor=blue,citecolor=blue, urlcolor=blue]{hyperref}

\usepackage{ulem}

\usepackage{url}

\usepackage{amsmath,amssymb,bm}

\usepackage{slashed}

\begin{document}

\title{Gauge dependence of the perturbative QCD predictions under the momentum space subtraction scheme}

\author{Jun Zeng}
\email{zengj@cqu.edu.cn}
\author{Xing-Gang Wu}
\email{wuxg@cqu.edu.cn}
\author{Xu-Chang Zheng}
\email{zhengxc@cqu.edu.cn}
\affiliation{Department of Physics, Chongqing University, Chongqing 401331, P.R. China}

\author{Jian-Ming Shen}
\email{cqusjm@cqu.edu.cn}
\affiliation{School of Physics and Electronics, Hunan University, Changsha 410082, P.R. China}

\begin{abstract}

The momentum space subtraction (MOM) scheme is one of the most frequently used renormalization schemes in perturbative QCD (pQCD) theory. In the paper, we make a detailed discussion on the gauge dependence of the pQCD prediction under the MOM scheme. Conventionally, there is renormalization scale ambiguity for the fixed-order pQCD predictions, which assigns an arbitrary range and an arbitrary error for the fixed-order pQCD prediction. The principle of maximum conformality (PMC) adopts the renormalization group equation to determine the magnitude of the coupling constant and hence determines an effective momentum flow of the process, which is independent to the choice of renormalization scale. There is thus no renormalization scale ambiguity in PMC predictions. To concentrate our attention on the MOM gauge dependence, we first apply the PMC to deal with the pQCD series. We adopt the Higgs boson decay width, $\Gamma(H\to gg)$, up to five-loop QCD contributions as an example to show how the gauge dependence behaves before and after applying the PMC. Different interaction vertices have been chosen for defining the MOM-schemes such as the mMOM, the MOMh, the MOMq, the MOMg, and the MOMgg schemes. Under those MOM schemes, we obtain $\Gamma(H \to gg)|^{\rm{mMOM}}_{\rm{PMC}}=332.8{^{+11.6}_{-3.7}}\pm7.3~\rm{KeV}$, $\Gamma(H \to gg)|^{\rm{MOMh}}_{\rm{PMC}}=332.8{^{+27.5}_{-34.6}}\pm7.3~\rm{KeV}$, $\Gamma(H \to gg)|^{\rm{MOMq}}_{\rm{PMC}}=332.9{^{+27.4}_{-34.7}}\pm7.3~\rm{KeV}$, $\Gamma(H \to gg)|^{\rm{MOMg}}_{\rm{PMC}}=332.7{^{+27.5}_{-34.6}}\pm7.3~\rm{KeV}$, $\Gamma(H \to gg)|^{\rm{MOMgg}}_{\rm{PMC}}=337.9{^{+1.2}_{-1.7}}\pm7.7~\rm{KeV}$, where the central values are for the Landau gauge with the gauge parameter $\xi^{\rm MOM}=0$, and the first errors are for $\xi^{\rm MOM}\in[-1,1]$, the second ones are caused by taking $\Delta \alpha_s^{\overline{\rm MS}}(M_Z)=\pm0.0011$. The uncertainty of the Higgs mass $\Delta M_H =0.24~\rm{GeV}$ will cause an extra error $\sim \pm1.7$ (or $\sim\pm1.8$) KeV for all the mentioned MOM schemes. It is found that the Higgs decay width $\Gamma  (H\to gg)$ depends very weakly on the choices of the MOM schemes, being consistent with the renormalization group invariance. It is found that the gauge dependence of $\Gamma(H\to gg)$ under the $\rm{MOMgg}$ scheme is less than $\pm1\%$, which is the smallest gauge dependence among all the mentioned MOM schemes.

\end{abstract}

\maketitle

\section{Introduction}

Quantum chromodynamics (QCD) is believed to be the field theory of hadronic strong interactions. Due to its asymptotic freedom property~\cite{Gross:1973id, Politzer:1973fx}, the QCD strong coupling constant becomes numerically small at short distances, allowing perturbative calculations for the high-energy processes. The QCD theory in a covariant gauge with massless quarks has three fundamental propagators which are for the gluon, the ghost and the quark fields, respeectively, and four fundamental vertices, namely the triple-gluon, the four-gluon, the ghost-gluon and the quark-gluon vertices. In the literature, various renormalization schemes have been adopted to regularize and remove the ultraviolet divergences emerged at higher perturbative orders. Among them, the momentum space subtraction (MOM) scheme~\cite{Celmaster:1979dm, Celmaster:1979km, Celmaster:1979xr, Celmaster:1980ji, Gracey:2013sca, vonSmekal:2009ae} has also been frequently used in addition to the conventional minimum substraction scheme~\cite{tHooft:1973mfk}, which carries considerable information of various quark and gluon interaction vertices at specific momentums and leads to a better convergence for some cases. Initially, the MOM scheme is defined via renormalizing the three-point vertices of the QCD Lagrangian at the completely symmetric point~\cite{Celmaster:1979dm, Celmaster:1979km}, i.e. the squared momentum of each external momentum of the vertex is equal. Lately, the asymmetric point with one of the external momentum vanishes for the three-point vertex has been suggested~\cite{vonSmekal:2009ae, Braaten:1981dv, Chetyrkin:2000dq}, which has the property of avoiding the infrared divergence in massless QCD theory. More explicitly, the minimal MOM (mMOM) scheme~\cite{vonSmekal:2009ae} which subtracts at the asymmetric point where one external momentum vanishes has been suggested as an alternation of the original symmetric MOM scheme. It is an extension of the MOM scheme on the ghost-gluon vertex and allows the strong coupling to be fixed solely through a determination of the gluon and ghost propagators. Further more, there are other four kinds of asymmetric MOM schemes, e.g. the one with vanishing momentum for the incoming ghost in the ghost-gluon vertex, the one with vanishing momentum for the incoming quark in the quark-gluon vertex, and the two schemes in dealing with the case of vanishing momentum for the incoming gluon in the triple-gluon vertex, respectively. Following the same notions as those of Ref.\cite{Chetyrkin:2000dq}, we label the first two MOM schemes as MOMh and MOMq schemes, and the other two schemes as MOMg and MOMgg schemes~\cite{Boucaud:1998xi, Becirevic:1999uc}, respectively. Even though the MOM schemes have been successfully applied in various high-energy processes, in different to the minimum substraction scheme, it has been found that the MOM scheme breaks down the gauge invariance. It is interesting to show whether the gauge dependence exists for all (typical) kinds of MOM schemes, or to find a MOM scheme with minimum gauge dependence.

The strong coupling is the most important component of the pQCD theory, we need to know its exact magnitude at any scale so as to derive an accurate pQCD prediction. The scale running behavior of the strong coupling is controlled by the renormalization group equation (RGE), or the $\beta$-function. The RGE for the MOM scheme can be related to the one under the modified minimal subtraction scheme (e.g. the $\overline{\rm{MS}}$ scheme~\cite{Bardeen:1978yd}) via proper relations. At the present, the explicit expressions of the $\{\beta_i\}$-functions under the $\overline{\rm MS}$ scheme have been known up to five-loop level in Refs.\cite{Caswell:1974gg, Jones:1974mm, Tarasov:1980au, Larin:1993tp, vanRitbergen:1997va, Chetyrkin:2004mf, Czakon:2004bu, Baikov:2016tgj, Herzog:2017ohr, Luthe:2017ttg, Kniehl:2006bg}. Thus the five-loop $\{\beta_i\}$-functions for the MOM schemes (mMOM, MOMh, MOMq, MOMg and MOMgg) can be determined with the help of the known five-loop relations~\cite{vonSmekal:2009ae, Chetyrkin:2000dq, Ruijl:2017eht, Hagiwara:1982ct} to the $\overline{\rm{MS}}$ scheme. Another way of deriving the running behavior of the MOM strong coupling up to five-loop level can be found in Ref.\cite{Chetyrkin:2017bjc}. A key component for solving the $\beta$-function is the QCD asymptotic scale $\Lambda$. The asymptotic scale in $\overline{\rm MS}$-scheme can be fixed by using the PDG world average of the strong coupling constant at the scale of $Z^0$ boson mass, $\alpha^{\overline{\rm MS}}_s(M_Z)=0.1181\pm 0.0011$, which leads to $\Lambda^{n_f=5}_{\overline{\rm MS}}=0.210\pm0.014$ GeV~\cite{Tanabashi:2018oca}. The asymptotic scale for a MOM scheme can be derived by using the Celmaster-Gonsalves relation~\cite{Celmaster:1979km, Celmaster:1979dm, Celmaster:1979xr, Celmaster:1980ji, vonSmekal:2009ae}, e.g.
\begin{eqnarray}
\frac{\Lambda_{\rm{MOM}}}{\Lambda_{\rm{\overline{MS}}}}=\exp \left[\frac{-b_{1}(\xi^{\rm MOM})}{2 \beta_0}\right],
\label{CGLambda}
\end{eqnarray}
where $\xi^{\rm MOM}$ is the gauge parameter, $b_{1}(\xi^{\rm MOM})$ is the next-to-leading order (NLO) coefficient of the perturbative series of $\alpha_{s}^{\rm{MOM}}$ expanded over $\alpha_{s}^{\rm{\overline{MS}}}$, i.e. $\alpha_{s}^{\rm{\overline{MS}}}= \alpha_{s}^{\rm{MOM}}+ b_1(\xi^{\rm MOM}) \alpha_{s}^{2, \rm{MOM}}+ b_2(\xi^{\rm MOM}) \alpha_{s}^{3, \rm{MOM}} +\cdots$. The above relation is correct up to all orders~\cite{Zeng:2015gha}. As an example, for the mMOM scheme, we have
\begin{eqnarray}
&& \frac{\Lambda_{\rm{mMOM}}}{\Lambda_{\rm{\overline{MS}}}} \nonumber\\
&=& \exp \left[\frac{\left(9 \xi^{2, \rm mMOM}+18 \xi^{\rm mMOM} +169\right)C_A-80 T n_f}{264 C_A-96 T n_f}\right],
\end{eqnarray}
where $C_A=3$, $T=1/2$ for SU(3) color group, and $n_f$ is the active flavor number.

The MOM scheme could be a useful alternative to the $\overline{\rm{MS}}$ scheme for studying the behavior and truncation uncertainty of the perturbation series. Many MOM applications have been done in the literature, e.g. two typical MOM applications for the Higgs-boson decays to gluons and the $R$-ratio for the electron-positron annihilation can be found in Refs.\cite{Zeng:2015gha, Zeng:2018jzf, Gracey:2014pba, Celmaster:1980ji, Kataev:2015yha}. Moreover, the processes involving three-gluon or four-gluon vertex provides an important platform for studying the renormalization scale setting problem. For the three-gluon vertex, it has already been pointed out that the typical momentum flow which appears in the three-gluon vertex should be a function of the virtuality of three external gluons~\cite{Binger:2006sj}. As an example, because of the improved convergence, a more accurate and reliable pQCD prediction for Pomeron intercept can be achieved under the MOM scheme other than the $\overline{\rm{MS}}$ scheme~\cite{Brodsky:1998kn, Brodsky:2002ka, Zheng:2013uja, Caporale:2015uva}. The MOM scheme can also be helpful to avoid the small scale problem emerged in $\overline{\rm{MS}}$ scheme~\cite{Deur:2017cvd, Brodsky:1994eh}~\footnote{Because of commensurate scale relations~\cite{Brodsky:1994eh}, one can obtain relations of the scales under various schemes so as to achieve a scheme-independent prediction, and the small scale in one scheme could be amplified in another scheme.}.

The Higgs boson is a crucially important component of the Standard Model (SM), its various decay channels are important components for Higgs phenomenology. Among those decay channels, the decay width of $H\to g g$ have been calculated up to five-loop level under the $\overline{\rm{MS}}$ scheme~\cite{Inami:1982xt, Djouadi:1991tka, Graudenz:1992pv, Dawson:1993qf, Spira:1995rr, Dawson:1991au, Chetyrkin:1997iv, Chetyrkin:1997un, Baikov:2006ch, Herzog:2017dtz}. Using the relations among the strong coupling constants under various renormalization schemes, one can obtain the corresponding five-loop MOM expression for the Higgs boson decay width $\Gamma(H\to gg)$ from the known $\overline{\rm{MS}}$ expression. A way to transform the pQCD predictions from one renormalization scheme to another renormalization scheme has been explained in detail in Ref.\cite{Ma:2017xef}. In the paper, we shall adopt the decay width $\Gamma(H\to gg)$ up to five-loop QCD contributions as an explicit example to show how the gauge dependence of the MOM prediction behaves with increasing known perturbative orders.

Following the standard renormalization group invariance (RGI), a physical observable (corresponding to an infinite order pQCD prediction) should be independent to the choices of renormalization scale and renormalization scheme. For a fixed-order pQCD prediction, conventionally, people uses guessed renormalization scale together with an arbitrary range to estimate its uncertainty, which leads to the mismatch of strong coupling constant with its coefficient at each order and then results as conventional renormalization scheme-and-scale ambiguities. Many scale setting approaches have been suggested to solve the renormalization scale ambiguity. Among them, the principle of maximum conformality (PMC)~\cite{Brodsky:2011ta, Brodsky:2012rj, Brodsky:2011ig, Mojaza:2012mf, Brodsky:2013vpa} has been suggested to eliminate the conventional renormalization scheme-and-scale ambiguities simultaneously. In different to other scale-setting approaches such as the RG-improved effective coupling method~\cite{Grunberg:1980ja, Grunberg:1982fw} and the Principle of Minimum Sensitivity~\cite{Stevenson:1980du, Stevenson:1981vj, Stevenson:1982wn, Stevenson:1982qw, Ma:2014oba} and the sequential BLM~\cite{Mikhailov:2004iq, Kataev:2014jba} or its alternated version Modified seBLM~\cite{Ma:2015dxa}, the purpose of PMC is not to find an optimal renormalization scale but to fix the running behavior of the strong coupling constant with the help of the RGE, whose argument is called as the PMC scale. The PMC scale is physical in the sense that its value reflects the ``correct" typical momentum flow of the process, which is independent to the choice of renormalization scale. After applying the PMC, the convergence of the pQCD series can be greatly improved due to the elimination of divergent renormalon terms. The PMC has a solid theoretical foundation, it satisfies the standard RGI and all the self-consistency conditions of the RGE~\cite{Brodsky:2012ms}. Detailed discussions and many applications of the PMC can be found in the reviews~\cite{Wu:2013ei, Wu:2014iba, Wu:2015rga, Wu:2019mky}. In the paper, we shall first adopt the PMC to eliminate the renormalization scale ambiguity and then discuss the gauge dependence of the MOM predictions on the decay width $\Gamma(H\to gg)$.

The remaining parts of the paper are organized as follows. In Sec.\ref{transform}, we give the basic components and the formulas for transforming the strong coupling constant from various MOM schemes to $\overline{\rm{MS}}$ scheme, which are important to transform the known $\overline{\rm MS}$ pQCD series to MOM one. In Sec.\ref{PMC}, we give a brief review on the PMC single-scale approach, which shall be adopted to do our present PMC analysis. In Sec.\ref{numericalresults}, we discuss the gauge dependence of the decay width $\Gamma(H\to gg)$ under the above mentioned five asymmetric MOM schemes. Sec.\ref{summary} is reserved for a summary. Some detailed formulas are given in the Appendix.

\section{\label{transform} The momentum space subtraction schemes}

The scale dependence of the strong coupling is controlled by the following $\beta$-function,
\begin{eqnarray}
\beta(a(\mu))=\mu^2\frac{\partial a(\mu)}{\partial \mu^2}=-\sum _{i=0}^{\infty } \beta _i a(\mu)^{i+2},
\end{eqnarray}
where $\mu$ is the renormalization scale, $a(\mu) \equiv \alpha _s(\mu)/(4 \pi)$. The $\{\beta_i\}$-functions are scheme dependent, and their expressions up to five-loop level under the $\overline{\rm MS}$-scheme are available in Refs.\cite{Caswell:1974gg, Jones:1974mm, Tarasov:1980au, Larin:1993tp, vanRitbergen:1997va, Chetyrkin:2004mf, Czakon:2004bu, Baikov:2016tgj, Herzog:2017ohr, Luthe:2017ttg, Kniehl:2006bg}. For short, when there is no confusion, we set $a=a(\mu)$ in the following discussions.

For an arbitrary renormalization scheme $R$, the respective renormalization of the gluon, quark and ghost fields are of the form
\begin{eqnarray}
(A^B)^b_\nu &=& \sqrt{Z_3^R} (A^R)^b_\nu , \\
\psi^B &=& \sqrt{Z_2^R} \psi^R, \\
(c^B)^b  &=& \sqrt{\tilde{Z}_3^R} (c^R)^b,
\end{eqnarray}
where $Z_3^R$, $Z_2^R$ and ${\tilde{Z}_3^R}$ are the renormalization constants of the gluon field $A$, the quark field $\psi$, and the ghost field $c$, respectively. The superscripts `$B$' and `$R$' denote the bare and the renormalized fields, respectively. The superscript `$b$' is the color index for the adjoint representation of the gauge group.

By using the usually adopted dimensional regularization~\cite{tHooft:1972tcz} (we work in $D=4-2\epsilon$ dimension), the renormalized strong coupling $a$ and the gauge parameter $\xi$ can be written as follows:
\begin{eqnarray}
a^B   &=& \mu^{2\epsilon} Z_a^R a^R, \label{constantBR}  \\
\xi^B &=& Z_3^R \xi^R, \label{axibar}
\end{eqnarray}
where we have used the fact that the gauge parameter is also renormalized by the gluon field renormalization constant. The bare strong coupling is scale invariant, and the $D$-dimensional $\beta$-function for the renormalized strong coupling can be derived by doing the derivative over both sides of Eq.(\ref{constantBR}):
\begin{eqnarray}
0 &=& \frac{da^{B}}{d \ln \mu^{2}}  \\
  &=& \epsilon Z_{a}^{R} a^R \mu^{2 \epsilon}+\frac{d Z_{a}^R}{d a^R} \frac{d a^R}{d \ln \mu^{2}} a^R\mu^{2 \epsilon}+Z_{a}^R \frac{d a^R}{d \ln \mu^{2}} \mu^{2 \epsilon}.
\end{eqnarray}
Then, we obtain
\begin{eqnarray}
\frac{d a^R}{d \ln \mu^{2}}=-\frac{\epsilon Z_{a}^R a^R}{\frac{d Z_{a}^R}{d a^R} a^R+Z_{a}^R}=-\epsilon a^R+\beta\left(a^R\right).
\end{eqnarray}

The renormalization of the gluon, ghost and quark self-energies can be performed as follows£º
\begin{eqnarray}
1+\Pi_A^R & = Z_3^R (1+\Pi_A^B) \label{z3}, \\
1+\tilde{\Pi}_c^R & = \tilde{Z}_3^R (1 + \tilde{\Pi}_c^B)    \label{tildez3}, \\
1+\Sigma_V^R & = Z_2^R (1 + \Sigma_V^B),
\end{eqnarray}
and the renormalization of the triple-gluon, the ghost-gluon and the quark-gluon vertexes can be performed as follows:
\begin{eqnarray}
\label{TiR}
T_i^R &=& Z_1^R T_i^B, i= 1, 2, \\
\label{tildeGammaR}
{\tilde{\Gamma}}_i^R &=& {\tilde{Z}}_1^R {\tilde{\Gamma}}_i^B, i = h, g, \\
\label{LambdaiR}
\Lambda_i^R &=& \bar{Z}_{1}^R \Lambda_i^B, \Lambda_i^{T,R} = \bar{Z}_1^R \Lambda_i^{T,B}, i = q, g,
\end{eqnarray}
where the vertex renormalization constants are related to the field and coupling renormalization constants via the Ward-Slavnov-Taylor identities (i.e. the generalized Ward-Takahashi identities~\cite{tHooft:1971akt, tHooft:1972qbu, Slavnov:1972fg, Taylor:1971ff}) by
\begin{eqnarray}
\sqrt{Z_3^R Z_a^R}=\frac{Z_1^R}{Z_3^R}=\frac{\tilde{Z}_1^R}{\tilde{Z}_3^R}=\frac{\bar{Z}_1^R}{Z_2^R}.
\label{S-T-identity}
\end{eqnarray}

Under the minimal subtraction scheme ($\rm{MS}$)~\cite{tHooft:1973mfk} in which the ultraviolet divergence (${1}/{\epsilon}$-terms) in pQCD series are directly subtracted, the renormalized parameters $Z^{\rm MS}_{k}$ can be written as
\begin{eqnarray}\label{ZMS}
Z^{\rm MS}_{k}=1+\sum_{n=1}^{\infty} \left(\sum_{m=1}^{n} \frac{b^{\rm MS}_{m,n}}{\epsilon^{m}}\right)a^{{\rm MS}, n},
\end{eqnarray}
where the coefficients $b^{\rm MS}_{m,n}$ are free of $\mu$-dependence~\cite{Buras:1998raa}. The renormalized constant $Z^{\rm MS}_{a}$ is gauge independent, which takes the following form
\begin{eqnarray}
Z^{\rm MS}_{a}=&& 1-\frac{\beta_{0}}{\epsilon} a^{\rm MS}+\left(\frac{\beta_{0}^{2}}{\epsilon^{2}}-\frac{\beta_{1}}{2 \epsilon}\right) a^{2, {\rm MS}}-\left(\frac{\beta_{0}^{3}}{\epsilon^{3}}-\frac{7 \beta_{0} \beta_{1}}{6\epsilon^{2}}\right. \nonumber\\
&& \left.+\frac{\beta_{2}}{3 \epsilon}\right) a^{3, {\rm MS}} +\left(\frac{\beta_{0}^{4}}{\epsilon^{4}}-\frac{23 \beta_{1} \beta_{0}^{2}}{12 \epsilon^{3}}+\frac{20\beta_{2} \beta_{0}+9 \beta_{1}^{2}}{24\epsilon^{2}} \right. \nonumber\\
&& \left.-\frac{\beta_{3}}{4 \epsilon}\right) a^{4, {\rm MS}}-\left(\frac{\beta_{0}^{5}}{\epsilon^{5}} +\frac{172 \beta_{2} \beta_{0}^{2}+157 \beta_{1}^{2} \beta_{0}}{120 \epsilon^{3}} \right.\nonumber\\
&&\left.-\frac{163 \beta_{1} \beta_{0}^{3}}{60 \epsilon^{4}}-\frac{34 \beta_{1} \beta_{2}+39 \beta_{0} \beta_{3}}{60 \epsilon^{2}}-\frac{\beta_{4}}{5 \epsilon}\right)a^{5, {\rm MS}}+\cdots.
\end{eqnarray}
Here the $\{\beta_{i}\}$-functions are for the ${\rm MS}$ scheme, which are the same for all the other dimensional-like renormalization schemes. This is due to the fact that the strong coupling among the dimensional-like schemes can be simply related via a scale shift~\cite{Mojaza:2012mf}, e.g. the $\overline{\rm MS}$ scheme differs from the ${\rm MS}$ scheme by an additional absorbtion of $\ln 4\pi -\gamma_E$, which corresponds to redefining the ${\rm MS}$ scale $\mu_{\rm MS}$ as $\mu^2_{\rm MS}=\mu^2_{\overline{\rm MS}}\exp\left(\ln 4\pi -\gamma_E\right)$. Gross and Wilczek found that the LO $\{\beta_{i}\}$-functions under the dimensional-like renormalization schemes are gauge independent~\cite{Gross:1973ju}, and lately, Caswell and Wilczek gave a proof of such gauge independence up to all orders~\cite{Caswell:1974cj}~\footnote{A demonstration of the gauge independence of the anomalous dimensions which ensure the scale invariance of a physical observable has also been given there.}.

Using Eq.(\ref{axibar}), one obtains the following relations for the strong coupling and gauge parameter between the MOM and $\rm{\overline {MS}}$ schemes:
\begin{eqnarray}
\label{alphasMM@MS}
a^{\rm{MOM}}   &=&\frac{Z_a^{\rm{\overline {MS}}}}{Z_a^{\rm{MOM}}} a^{\rm{\overline {MS}}}, \\
\xi^{\rm{MOM}} &=&\frac{Z_3^{\rm{\overline {MS}}}}{Z_3^{\rm{MOM}}}\xi^{\rm{\overline {MS}}}.
\label{axiMM@MS}
\end{eqnarray}
It has been found that the MOM scheme is gauge dependent. In MOM scheme~\cite{vonSmekal:2009ae, Chetyrkin:2000dq, Ruijl:2017eht}, the gluon, ghost and quark self-energies are absorbed into the field renormalization constants at the subtraction point $q^2 =-\mu^2$:
\begin{eqnarray}
1+\Pi_A^{\rm{MOM}}(-\mu^2)&= Z_3^{\rm{MOM}} \Bigg[1+\Pi_A^B(-\mu^2)\Bigg] = 1\label{pia-mm-muequq}, \\
1+\tilde{\Pi}_c^{\rm{MOM}}(-\mu^2)&= \tilde{Z}_3^{\rm{MOM}}\Bigg[1+\tilde{\Pi}_c^B(-\mu^2)\Bigg]=1\label{tildepic-mm-muequq}, \\
1+\Sigma_V^{\rm{MOM}}(-\mu^2)&= Z_2^{\rm{MOM}} \Bigg[1+\Sigma_V^B(-\mu^2) \Bigg]=1\label{sigmav-mm-muequq}.
\end{eqnarray}

Using Eq.(\ref{axibar}), we obtain the following relationship of the gauge parameters under the $\rm{\overline{MS}}$ scheme and MOM scheme:
\begin{eqnarray}
\label{xiMOM}
\xi^{{\rm{MOM}}}&=&\left(1+\Pi^{{\rm{\overline{MS}}}}_A\right)\xi^{{\rm{\overline{MS}}}}.
\end{eqnarray}

In the following subsections, we make a simple introduction of five asymmetric MOM schemes, giving the relations of the strong couplings under those schemes with the one under the conventional $\overline{\rm MS}$ scheme, and their gauge-dependent basic components, which are done by renormalizing the three-point vertices, such as the ghost-gluon, the gluon-quark and the triple-gluon ones, at the asymmetric point with one of the external momentum of the vertex vanishes, respectively.

\subsection{The propagators}

\begin{figure}[htb]
\centering
\includegraphics[width=0.138\textwidth]{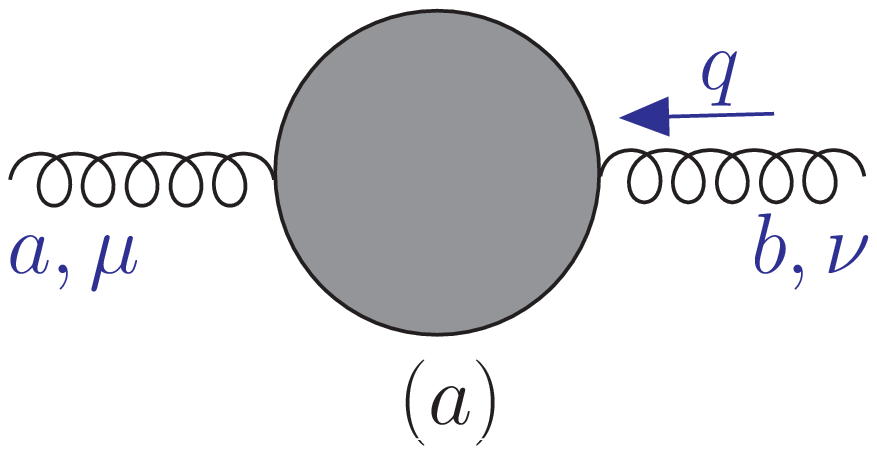}
\includegraphics[width=0.15\textwidth]{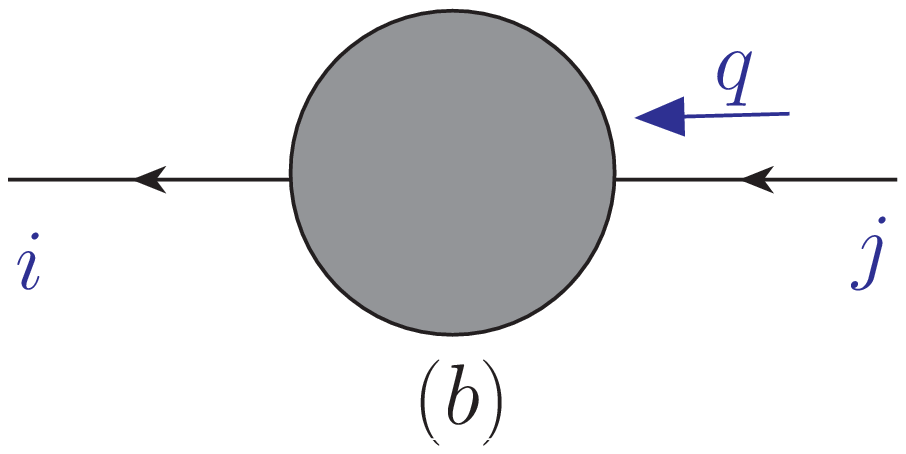}
\includegraphics[width=0.15\textwidth]{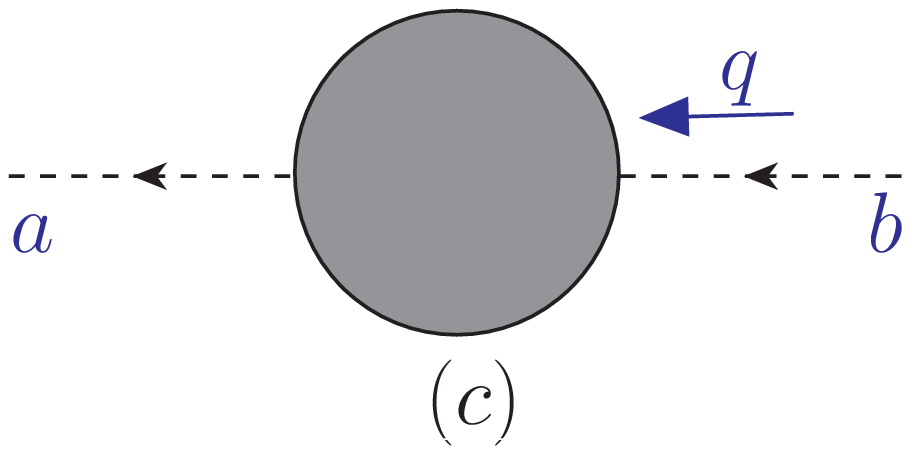}
\caption{The gluon, the quark and the ghost propagators $\Pi_{\mu \nu}^{a b}(q)$, $\Sigma^{i j}(q)$ and ${\tilde{\Pi}}^{a b}(q)$.}
\label{self-energy}
\end{figure}

The gluon, the quark and the ghost propagators, as shown by Fig.\ref{self-energy}, take the following form
\begin{eqnarray}
D^{ab}_{\mu\nu}(q)&=& -\frac{\delta^{ab}}{q^2} \biggl[\Bigl( - g_{\mu\nu} + \frac{q_\mu q_\nu}{q^2}\Bigr)
\frac{1}{1+\Pi_{A}(q^2)}- \xi \,\frac{q_\mu q_\nu}{q^2}\biggr],\\
S^{ij}(q)         &=& -\frac{\delta^{ij} \slashed{q}}{q^2\Bigl(1+\Sigma_V(q^2)\Bigr)},\\
\Delta^{ab}(q)    &=& -\frac{\delta^{ab}}{q^2\Bigl(1+\tilde{\Pi}_{c}(q^2)\Bigr)},
\end{eqnarray}
where $a$ and $b$ are color indices, $i$ and $j$ denote quark flavors. The  gauge parameter $\xi=0$ is the Landau gauge, $\xi=1$ is the Feynman gauge, and etc.
The self-energies $\Pi(q^2)$, $\Sigma_V(q^2)$ and $\tilde{\Pi}(q^2)$ can be extracted from the corresponding one-particle irreducible diagrams by applying proper projection operators~\cite{Chetyrkin:2000dq} (the same holds for the vertex functions discussed below).

\subsection{The ghost-gluon vertex}

\begin{figure}[htb]
\centering
\includegraphics[width=0.4\textwidth]{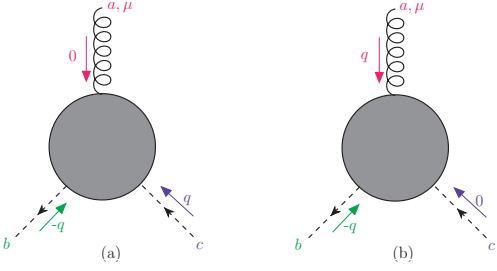}
\caption{ The ghost-gluon vertex: (a) $\tilde{\Gamma}_{\mu}^{a b c}(0; $-$q, q)$ for the case of the incoming gluon has zero momentum and (b) $\tilde{\Gamma}_{\mu}^{a b c}(q; $-$q, 0)$ for the case of one incoming ghost has zero momentum.}
\label{tilde-ghost-gluon-vetex}
\end{figure}

The tree-level ghost-gluon vertex is $-i g_s  f^{abc} q_{\mu}$, where $q_{\mu}$ is the outgoing ghost momentum. There are two possibilities to set one of the external momenta to zero for the ghost-gluon vertex. One is to set the gluon momentum to zero, whose diagram is shown by Fig.\ref{tilde-ghost-gluon-vetex}(a), and the renormalized vertex can be written as
\begin{eqnarray}
\tilde{\Gamma}_{\mu}^{a b c}(0; -q, q)=- i g_s f^{abc} q_{\mu} \tilde{\Gamma}_g(q^2),
\end{eqnarray}
and the other is to set one of the incoming ghost momentum to zero, whose diagram is shown by Fig.\ref{tilde-ghost-gluon-vetex}(b), and the renormalized vertex can be written as
\begin{eqnarray}
\tilde{\Gamma}_{\mu}^{a b c}(q; -q, 0)=- i g_s f^{abc} q_{\mu} \tilde{\Gamma}_h(q^2).
\end{eqnarray}
Here, $\tilde{\Gamma}_g(q^2)$ or $\tilde{\Gamma}_h(q^2)$, is the Lorentz invariant function with vanishing gluon or ghost momentum, respectively. At the tree-level, we have
\begin{eqnarray}
  \tilde{\Gamma}_h(q^2) \bigl|_\text{tree} = \tilde{\Gamma}_g(q^2) \bigl|_\text{tree} = 1.
\end{eqnarray}

The MOMh scheme is defined by renormalizing the ghost-gluon vertex (Fig.\ref{tilde-ghost-gluon-vetex}(b)) with the following condition:
\begin{eqnarray}
\label{tildegammah}
{\tilde{\Gamma}}_{h}^{\rm{MOMh}}(q^2 =-\mu^2)={\tilde{Z}}_1^{\rm{MOMh}}{\tilde{\Gamma}}_{h}^B(q^2 =-\mu^2)=1,
\end{eqnarray}

Using Eqs.(\ref{TiR}, \ref{tildeGammaR}, \ref{LambdaiR}, \ref{S-T-identity}, \ref{axiMM@MS}, \ref{pia-mm-muequq}, \ref{tildepic-mm-muequq}, \ref{sigmav-mm-muequq}), we can connect the strong coupling in the MOMh scheme to the one in the $\rm{\overline{MS}}$ scheme through the following equation:
\begin{eqnarray}
\label{aMOMh}
a^{{\rm{MOMh}}}(\mu)&=&\frac{\left({\tilde{\Gamma}}_{h}^{{\rm{\overline{MS}}}}(-\mu^2)\right)^{2}a^{{\rm{\overline{MS}}}}(\mu)}
{\left(1+\Pi^{{\rm{\overline{MS}}}}_A(-\mu^2) \right)\left(1+\tilde{\Pi}^{{\rm{\overline{MS}}}}_c(-\mu^2)\right)^2}~.
\end{eqnarray}
In addition, motivated by the non-renormalization of the ghost-gluon vertex in the Landau gauge~\cite{Taylor:1971ff}, the vertex renormalization constant for this vertex is chosen as the same as that in $\rm{\overline{MS}}$, i.e.
\begin{eqnarray}
{\tilde{Z}}_1^{\rm{mMOM}} = {\tilde{Z}}_1^{\rm{\overline{MS}}},
\label{tildez1}
\end{eqnarray}
which is equal to $1$ in the Landau gauge. We can then derive the following relation for the coupling constants  in those two schemes,
\begin{eqnarray}
\label{aMOM}
a^{{\rm{mMOM}}}(\mu)&=&\frac{a^{{\rm{\overline{MS}}}}(\mu)}{\left(1+\Pi^{{\rm{\overline{MS}}}}_A(-\mu^2) \right)\left(1+\tilde{\Pi}^{{\rm{\overline{MS}}}}_c(-\mu^2)\right)^2}.
\end{eqnarray}
We put the derivation in Appendix \ref{zggq}. It shows that the MOMh scheme is equivalent to mMOM scheme for the Landau gauge ($\xi^{\rm mMOM}= \xi^{\rm MOMh}=0$).

\subsection{The quark-gluon vertex}

\begin{figure}[htb]
\centering
\includegraphics[width=0.4\textwidth]{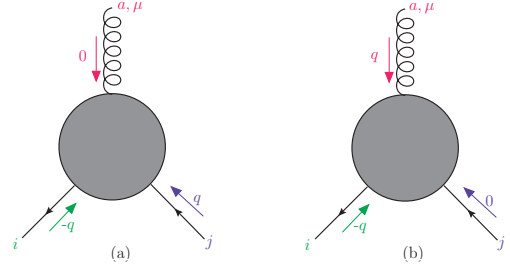}
\caption{The quark-gluon vertex: (a) $\Lambda_{\mu, i j}^{a}(0;  $-$q, q)$ with zero incoming gluon momentum and (b) $\Lambda_{\mu, i j}^{a}(q; $-$q, 0)$ with the vanishing quark momentum.}
\label{tilde-qurak-gluon-vetex}
\end{figure}

There are two non-trivial cases with vanishing incoming external momentum for the quark-gluon vertex, the case of a vanishing incoming gluon momentum as shown by Fig.\ref{tilde-qurak-gluon-vetex}(a) and the case of a vanishing quark momentum as shown by Fig.\ref{tilde-qurak-gluon-vetex}(b). It is clear that nullifying the incoming quark momentum is equal to the result of nullifying the outgoing quark momentum, therefore, the vertex Fig.\ref{tilde-qurak-gluon-vetex}(a) can be written as
\begin{eqnarray}
&& \Lambda^a_{\mu,ij}(0; -q, q) \nonumber\\
&=& g T^a_{ij} \Bigl[\gamma_\mu \Lambda_g(q^2) + \gamma^\nu \biggl( g_{\mu\nu} - \frac{q_\mu q_\nu}{q^2} \biggr)\Lambda_g^T(q^2)\Bigr], \label{lambdag}
\end{eqnarray}
and the vertex Fig.\ref{tilde-qurak-gluon-vetex}(b) can be written as
\begin{eqnarray}
&& \Lambda^a_{\mu,ij}(q; -q, 0) \nonumber\\
&=& g T^a_{ij} \Bigl[\gamma_\mu \Lambda_q(q^2) + \gamma^\nu \biggl( g_{\mu\nu} - \frac{q_\mu q_\nu}{q^2} \biggr)\Lambda_q^T(q^2)\Bigr]. \label{lambdaq}
\end{eqnarray}
The subscript `$g$' in Eq.(\ref{lambdag}) and `$q$' in Eq.(\ref{lambdaq}) indicate the functions with vanishing gluon momentum and incoming quark momentum, respectively. $T_{ij}^{a}$ is the ${\rm SU(3)}$ color group generator for the quark. At the tree-level, we have
\begin{eqnarray}
  \Lambda_g(q^2) \bigl|_\text{tree}   = \Lambda_q(q^2) \bigl|_\text{tree} &\,=\, 1 , \\
  \Lambda_g^T(q^2) \bigl|_\text{tree} = \Lambda_q^T(q^2) \bigl|_\text{tree} &\,=\, 0 .
\label{lambdagq}
\end{eqnarray}

The MOMq scheme is defined by renormalizing the quark-gluon vertex with vanishing incoming quark momentum, e.g.,
\begin{eqnarray}
\Lambda_{q}^{\rm{MOMq}}(q^2 =-\mu^2)={\overline{Z}}_{1}^{\rm{MOMq}}(\mu^2)\Lambda_{q}^B(q^2 =-\mu^2)=1,
\end{eqnarray}
Therefore, the relation of the coupling constants in the MOMq scheme and the $\rm{\overline{MS}}$ scheme is
\begin{eqnarray}
\label{aMOMq}
a^{{\rm{MOMq}}}(\mu)&=&\frac{\left(\Lambda_{q}^{{\rm{\overline{MS}}}}(-\mu^2)\right)^{2} a^{{\rm{\overline{MS}}}}(\mu)}{\left(1+\Sigma_V^{{\rm{\overline{MS}}}} (-\mu^2)\right)^{2}\left(1+\Pi^{{\rm{\overline{MS}}}}_{A}(-\mu^2)\right)}.
\end{eqnarray}

\subsection{The triple-gluon vertex}

\begin{figure}[htb]
\centering
\includegraphics[width=0.2\textwidth]{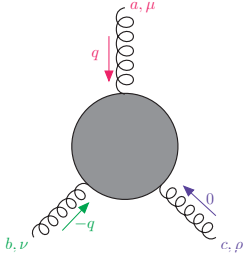}
\caption{The triple-gluon vertex with one zero momentum, $\Gamma_{\mu \nu \rho}^{a b c}(q, -q, 0)$.}
\label{tilde-triple-gluon-vetex}
\end{figure}

The triple-gluon vertex is symmetric under the exchange of any two of the gluons. As shown in Fig.\ref{tilde-triple-gluon-vetex}, one can set the momentum of the right-hand gluon to zero without loss of generality. Under this condition, the triple-gluon vertex generally takes the following form
\begin{eqnarray}
&& \Gamma_{\mu \nu \rho}^{a b c}(q, -q, 0) \nonumber\\
&=& - i g_s f^{abc} \biggl[(2 g_{\mu\nu} q_\rho - g_{\rho\nu} q_\mu- g_{\mu\rho} q_\nu ) T_1(q^2)\nonumber\\  &&
-\Bigl( g_{\mu\nu} - \frac{q_\mu q_\nu}{q^2} \Bigr)\, q_\rho T_2(q^2)+q_\mu q_\nu q_\rho T_3(q^2)\biggr],
\label{triple-gluon-formulae}
\end{eqnarray}
where $f^{abc}$ is the structure constant of the ${\rm SU(3)}$ color group. $T_{1}(q^2)$ corresponds to tree-level vertex, i.e. $T_1(q^2) \bigl|_\text{tree}=1$. $T_{2}(q^2)$ is always absent at tree-level but arises from radiative corrections. $T_{3}(q^2)$ vanishes due to the Ward-Slavnov-Taylor identity for the triple gluon vertex~\cite{Chetyrkin:2000dq, Ruijl:2017eht}. The MOMg scheme is defined by renormalizing the above triple-gluon vertex with vanishing incoming gluon momentum, e.g.,
\begin{eqnarray}
T_{1}^{\rm{MOMg}}(q^2=-\mu^2)=Z_{1}^{\rm{MOMg}}(\mu^2) T_{1}^B(q^2=-\mu^2)=1.
\end{eqnarray}
Therefore, the relation of the coupling constants in the MOMg scheme and the $\rm{\overline{MS}}$ scheme is
\begin{eqnarray}
\label{aMOMg}
a^{{\rm{MOMg}}}(\mu)&=&\frac{\left(T_{1}^{{\rm{\overline{MS}}}}(-\mu^2)\right)^{2}a^{{\rm{\overline{MS}}}}(\mu)}
{\left(1+\Pi^{{\rm{\overline{MS}}}}_{A}(-\mu^2)\right)^3}~.
\end{eqnarray}

Another MOM scheme, which is also based on the triple-gluon vertex is MOMgg scheme, is defined by the following renormalization condition~\cite{Boucaud:1998xi, Becirevic:1999uc}:
\begin{eqnarray}
T_{1}^{\rm{MOMgg}}(q^2=-\mu^2)-\frac{1}{2}T_{2}^{\rm{MOMgg}}(q^2=-\mu^2)=1.
\end{eqnarray}
This gives the following coupling relations between MOMgg scheme and the $\rm{\overline{MS}}$ scheme:
\begin{eqnarray}
\label{aMOMgg}
a^{{\rm{MOMgg}}}(\mu)=\frac{\left(T_{1}^{{\rm{\overline{MS}}}}(-\mu^2)-\frac{1}{2}T_{2}^{{\rm{\overline{MS}}}}(-\mu^2)\right)^{2}
a^{{\rm{\overline{MS}}}}(\mu)}{\left(1+\Pi^{{\rm{\overline{MS}}}}_{A}(-\mu^2)\right)^3}~.
\end{eqnarray}

\subsection{The MOM scheme $\beta$-function}

At the present, the gluon self-energies $\Pi^{{\rm{\overline{MS}}}}_A(-\mu^2)$, the ghost self-energies $\tilde{\Pi}^{{\rm{\overline{MS}}}}_c(-\mu^2)$, the quark self-energies $\Sigma_V^{{\rm{\overline{MS}}}}(-\mu^2)$, the quark-gluon vertex with vanishing incoming quark momentum $\Lambda_{q}^{{\rm{\overline{MS}}}}(-\mu^2)$, the ghost-gluon vertex with vanishing incoming ghost moment, and the two functions ${\tilde{\Gamma}}_{h}^{{\rm{\overline{MS}}}}(-\mu^2)$ and $T_1^{{\rm{\overline{MS}}}}(-\mu^2)$ defined in Eq.(\ref{aMOMg}) and the function $T_2^{{\rm{\overline{MS}}}}(-\mu^2)$ defined in Eq.(\ref{aMOMgg}) have been calculated up to four-loop QCD corrections under the $\overline{\rm MS}$ scheme, c.f. Refs.\cite{Chetyrkin:2000dq, Ruijl:2017eht}. Using the formulas given in those two references and using the relations (\ref{aMOMh}, \ref{aMOM}, \ref{aMOMq}, \ref{aMOMg}, \ref{aMOMgg}) and the equation  (\ref{xiMOM}), we can obtain the expressions for the strong couplings and the gauge parameters under various MOM schemes. For convenience, we put those relations in the Appendix \ref{MSandmMOM}.

Those relations are helpful to transform the conventional $\overline{\rm MS}$ series to the one under a specific MOM scheme. They are also important to get the MOM scheme $\beta$-function and hence determine the correct $\alpha_s$-running behavior in MOM scheme. The MOM $\beta$-function is explicitly gauge dependent, which can be written as
\begin{eqnarray}
\beta^{\rm{MOM}}=\beta^{\rm{\overline{MS}}}\frac{\partial a^{\rm{MOM}}}{\partial a^{\rm{\overline{MS}}}}+ \frac{\partial \xi^{\rm{\overline{MS}}}}{\partial \ln {\mu^2} }\frac{\partial a^{\rm{MOM}}}{\partial \xi^{\rm{\overline{MS}}}}.
\label{mom-beta1}
\end{eqnarray}
The anomalous dimension of gauge parameter $\gamma_{\xi}^{\rm{\overline{MS}}} =\frac{1}{\xi^{\rm{\overline{MS}}}} \frac{\partial \xi^{\rm{\overline{MS}}}}{\partial \ln {\mu^2} }$is equal to $\left(-\gamma_{A}^{\rm{\overline{MS}}}\right)$, where $\gamma_{A}^{\rm{\overline{MS}}}$ is the gluon field anomalous dimension. Therefore, the MOM-scheme $\beta$-function takes the form:
\begin{eqnarray}
\beta^{\rm{MOM}}=\Bigg(\beta^{\rm{\overline{MS}}}\frac{\partial a^{\rm{MOM}}}{\partial a^{\rm{\overline{MS}}}}-\xi^{\rm{\overline{MS}}}\gamma_{A}^{\rm{\overline{MS}}} \frac{\partial a^{\rm{MOM}}}{\partial \xi^{\rm{\overline{MS}}}}\Bigg)\Bigg |_{\xi^{\rm{\overline{MS}}} \rightarrow \xi^{\rm{MOM}}}^{a^{\rm{\overline{MS}}} \rightarrow a^{\rm{MOM}}},
\label{mom-beta2}
\end{eqnarray}

It has been stated that the property of the gauge invariance of the renormalization schemes is a sufficient but not a necessary property for the factorization of the QCD $\beta$-function~\cite{Garkusha:2018mua}. Thus a reliable $\alpha_s$-behavior can be determined and hence a reliable pQCD prediction for various MOM schemes by applying a proper scale-setting approach to deal with the $\{\beta_i\}$-terms of the process.

\section{General PMC analysis over the perturbative series}
\label{PMC}

Conventionally, a pQCD approximant, $\delta(Q)$, of a physical observable takes the form
\begin{eqnarray}
\delta(Q)=a^{p}(\mu)\sum _{i=1}^{\infty}C_{i}(\mu) a^{i-1}(\mu),
\label{conv-formulae}
\end{eqnarray}
where $Q$ represents the scale at which the observable is measured, the index $p$ indicates the $\alpha_s$-order of the leading-order (LO) prediction. Here the perturbative coefficients $C_i$ are usually in $n_f$ power series, where $n_f$ is the number of light flavors involved in the process. Using the degeneracy relations among different orders~\cite{Mojaza:2012mf, Brodsky:2013vpa, Bi:2015wea}, the pQCD series can be written as the following $\beta_i$ series:
\begin{widetext}
\begin{eqnarray}
\delta(Q) &=& r_{1,0}a(\mu)^p+\Bigg[ r_{2,0} + p \beta_0 r_{2,1} \Bigg]a(\mu)^{p+1} + \Bigg[r_{3,0} + p \beta_1 r_{2,1} + (p+1){\beta _0}r_{3,1} + \frac{p(p+1)}{2} \beta_0^2 r_{3,2} \Bigg]a(\mu)^{p+2} \nonumber\\
&& + \Bigg[ r_{4,0} + p{\beta_2}{r_{2,1}} + (p+1){\beta_1}{r_{3,1}} + \frac{p(3+2p)}{2}{\beta_1}{\beta_0}{r_{3,2}} + (p+2){\beta_0}{r_{4,1}} + \frac{(p+1)(p+2)}{2}\beta_0^2{r_{4,2}} \nonumber\\
&& + \frac{p(p+1)(p+2)}{3!}\beta_0^3{r_{4,3}} \Bigg]a(\mu)^{p+3}+\bigg[r_{5,0}+(p+3) r_{5,1} \beta _0+\frac{(p+2)(p+3)}{2} r_{5,2} \beta _0^2+(p+2) r_{4,1} \beta _1\nonumber \\
&&+\frac{p(p+1)(p+2)(p+3)}{24} r_{5,4} \beta _0^4+\frac{(p+1)(p+2)(p+3)}{6} r_{5,3} \beta _0^3+\frac{(p+1)(2p+5)}{2} r_{4,2} \beta _1 \beta _0\nonumber \\
&&+\frac{p(3p^2+12p+11)}{6} r_{4,3} \beta _1 \beta _0^2+(p+1) r_{3,1} \beta _2+\frac{p(p+2)}{2} r_{3,2}(2\beta _2 \beta _0+\beta _1^2)+p r_{2,1} \beta _3\bigg]a(\mu)^{p+4}+ \cdots,
\label{PMCrij}
\end{eqnarray}
\end{widetext}
where $r_{i,0}~(i=1,2,3\ldots)$ are conformal coefficients which are generally free from renormalization scale dependence, and $r_{i,j}$ ($1\leq j<i$) are non-conformal coefficients, namely to $r_{i,j} = \sum_{k=0}^{j} \left(\begin{array}{l}j\\k\end{array}\right) \ln^{k}(\mu^2/Q^2) \overline{r}_{i-k,j-k}$, where $\overline{r}_{m,n}=r_{m,n}|_{\mu=Q}$. As a subtle point, any $n_f$ -terms that are irrelevant to determine the $\alpha_s$-running behavior should be kept as a conformal coefficient and cannot be transfomred into $\{\beta_i\}$-terms~\cite{Wu:2013ei}.

For the standard PMC multi-scale approach described in Refs.\cite{Brodsky:2011ta, Mojaza:2012mf}, one needs to absorb the same type of $\{\beta_i\}$-terms at various orders into the strong coupling constant via an order-by-order manner. Different types of $\{\beta_i\}$-terms as determined from the RGE lead to different running behaviors of the strong coupling constant, and hence, determine the distinct PMC scales at each order. Because the precision of the PMC scale for high-order terms decreases at higher orders due to the less known $\{\beta_i\}$-terms in its higher-order terms. Due to the unknown perturbative terms, the PMC prediction has residual scale dependence~\cite{Zheng:2013uja}, which is however quite different from the arbitrary conventional renormalization scale dependence. The PMC scale, reflecting the correct momentum flow of the process, is independent to the choice of renormalization scale, and its resultant residual scale dependence is generally small due to both the exponential suppression and the $\alpha_s$ suppression~\cite{Wu:2019mky}. As an alteration, the PMC single-scale approach has been suggested to suppress the residual scale dependence~\cite{Shen:2017pdu}. It effectively replaces the individual PMC scales derived under the multi-scale approach by a single scale in the sense of a mean value theorem. The PMC single scale can be regarded as the overall effective momentum flow of the process; it shows stability and convergence with increasing order in pQCD via the pQCD approximates. The prediction of the PMC single-scale approach is scheme-independent up to any fixed order~\cite{Wu:2018cmb}, thus its value satisfies the standard RGI. The examples collected in Ref.\cite{Wu:2018cmb} show that the residual scale dependence emerged in PMC multi-scale approach can indeed be greatly suppressed. In the present paper, we adopt the PMC single-scale approach to do our discussions.

Using the standard procedures for the PMC single-scale approach, we can eliminate all the non-conformal $\{\beta_i\}$-terms and rewrite Eq.(\ref{conv-formulae}) as the following conformal series:
\begin{eqnarray}
\delta(Q) = \sum\limits_{n \ge 1} {r_{n,0}}{a(\overline Q)^{n+p-1}},
\label{deltaQ}
\end{eqnarray}
where the PMC scale $\overline Q$ is fixed by requiring all the non-conformal $\{\beta_i\}$-terms vanish. The perturbative series of $\ln{\overline Q^2}/{Q^2}$ over $a(Q)$ up to next-to-next-to-next-to-leading log ($\rm{N^3LL}$) accuracy takes the following form:
\begin{eqnarray}
\ln\frac{\overline Q^2}{Q^2} = \lambda_{0} + \lambda_{1}a(Q) + \lambda_{2} a^2(Q)+\lambda_{3} a^3(Q).  \label{eq:PMCsscaleaQ}
\end{eqnarray}
For convenience, we put the perturbative coefficients $\lambda_{i} (i=0,1,2,3)$ in the Appendix \ref{lambda}. One may observe that both the resultant PMC conformal series (\ref{deltaQ}) and the scale $\overline{Q}$ are free of renormalization scale ($\mu$), and thus, the conventional renormalization scale dependence has been eliminated. There is residual dependence for $\delta(Q)$ due to the unknown terms (e.g. the unknown N$^4$LL-terms and higher) in the perturbative series (\ref{eq:PMCsscaleaQ}).

\section{Gauge dependence of the total decay width $\Gamma(H\to gg)$}
\label{numericalresults}

In the present section, we adopt the total decay width $\Gamma(H\to gg)$ under various MOM schemes as an explicit example to show how the gauge dependence behaves with increasing known perturbative orders before and after applying the PMC.

Up to $\alpha_s^6$-order level, the decay width of $H\to gg$ takes the following form
\begin{eqnarray}
\Gamma (H\to gg)=\frac{M_H^3 G_F}{36 \sqrt{2} \pi }\sum _{i=0}^{4}  C_{i}(\mu) a^{i+2}(\mu),
\label{cij}
\end{eqnarray}
where $a(\mu)=\alpha_s(\mu)/(4\pi)$, $\mu$ is the renormalization scale, $M_H$ is the Higgs boson mass, and $G_F=1.16638\times10^{-5}{\rm GeV}^{-2}$ is the Fermi coupling constant. The coefficients $C_{i\in[0,4]}(M_H)$ under the $\rm{\overline {MS}}$ scheme can be read from Refs.\cite{Inami:1982xt, Djouadi:1991tka, Graudenz:1992pv, Dawson:1993qf, Spira:1995rr, Dawson:1991au, Chetyrkin:1997iv, Chetyrkin:1997un, Baikov:2006ch, Herzog:2017dtz}. Those coefficients are usually given in $n_f$-power series, and before applying the PMC, the perturbative series (\ref{cij}) should be transformed into the $\{\beta_i\}$-series of Eq.(\ref{PMCrij}) with $p=2$. For convenience, we put the required coefficients $r_{i,j}(M_H)$ in Appendix~\ref{MShgg}. And then by using the formulas given in Appendix~\ref{MSandmMOM}, which give the perturbative transformations of the strong couplings and the gauge parameters among the MOM scheme and the $\overline{\rm MS}$ scheme, one can conveniently transform the perturbative series of $\Gamma  (H\to gg)$ from the $\rm{\overline {MS}}$ scheme into the MOM scheme. The coefficients at any renormalization scale can be obtained from $C_{i\in[0,4]}(M_H)$ by using the RGE. Finally, by applying the PMC single-scale setting approach, the pQCD series Eq.(\ref{cij}) can be rewritten as the following conformal series
\begin{eqnarray}
\Gamma  (H\to \ gg) = \frac{M_H^3 G_F}{36 \sqrt{2} \pi } \sum _{i=1}^{5}  r_{i,0} a^{i+1}(\overline Q),
\end{eqnarray}
and the solution of $\ln{{\overline Q}^2}/{ M_H^2}$ can be written as a power series in $a(M_H)$ at the $\rm{N^3LL}$ accuracy.

To do the numerical calculation, we adopt: the top-quark pole mass $m_t = 173.3~\rm{GeV}$~\cite{CMS:2012awa, ATLAS:2012coa}, the Higgs mass $M_H = 125.09 \pm 0.21 \pm 0.11$~$\rm{GeV}$ and the MOM QCD asymptotic scale is fixed by $\alpha^{\rm \overline{MS}}_s(M_Z)=0.1181$ together with Celmaster-Gonsalves relation (\ref{CGLambda})~\cite{Aad:2015zhl}.

\subsection{The gauge dependence of the effective scale $\bar{Q}$ and the effective coupling $\alpha_s(\bar{Q})$ for the five asymmetric MOM schemes}

\begin{figure}[htb]
\centering
\includegraphics[width=0.235\textwidth]{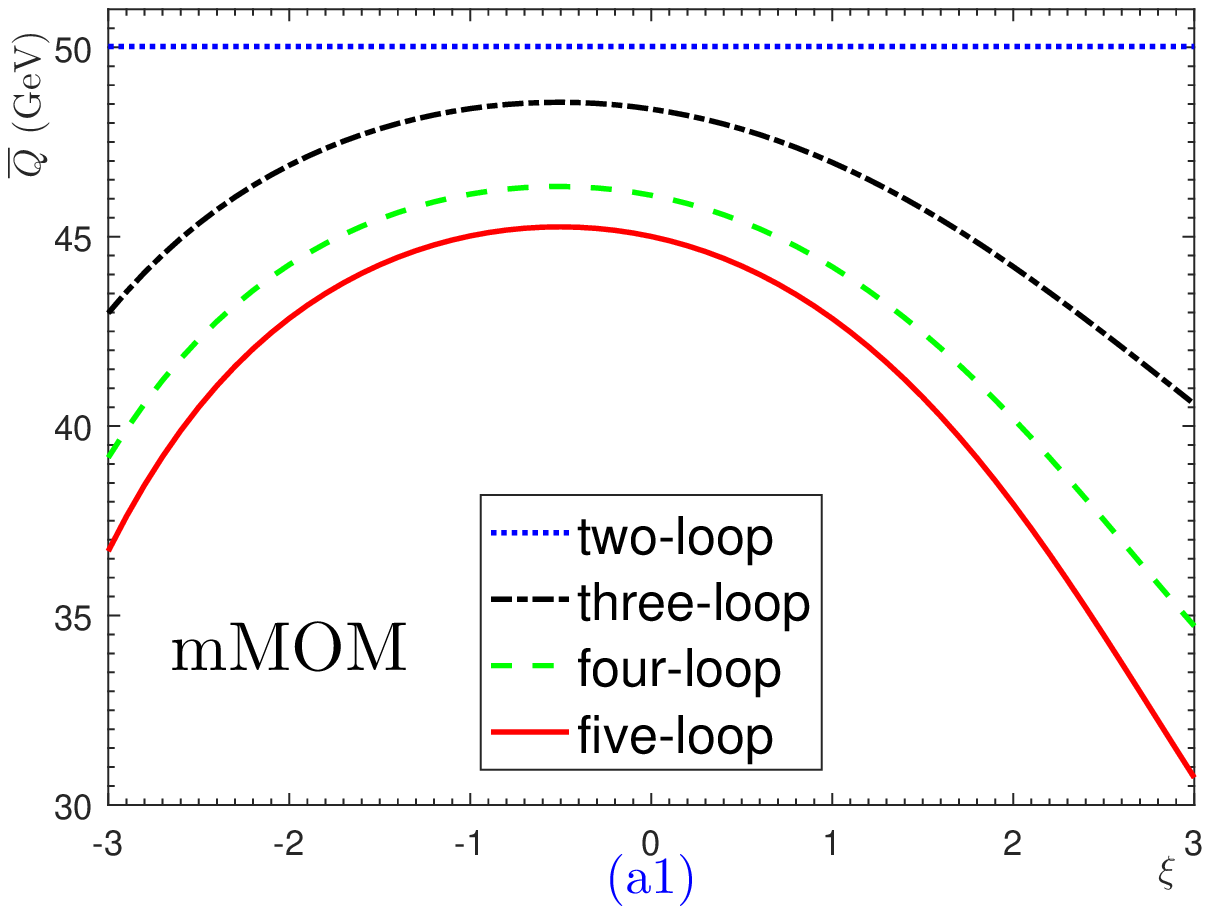}
\includegraphics[width=0.235\textwidth]{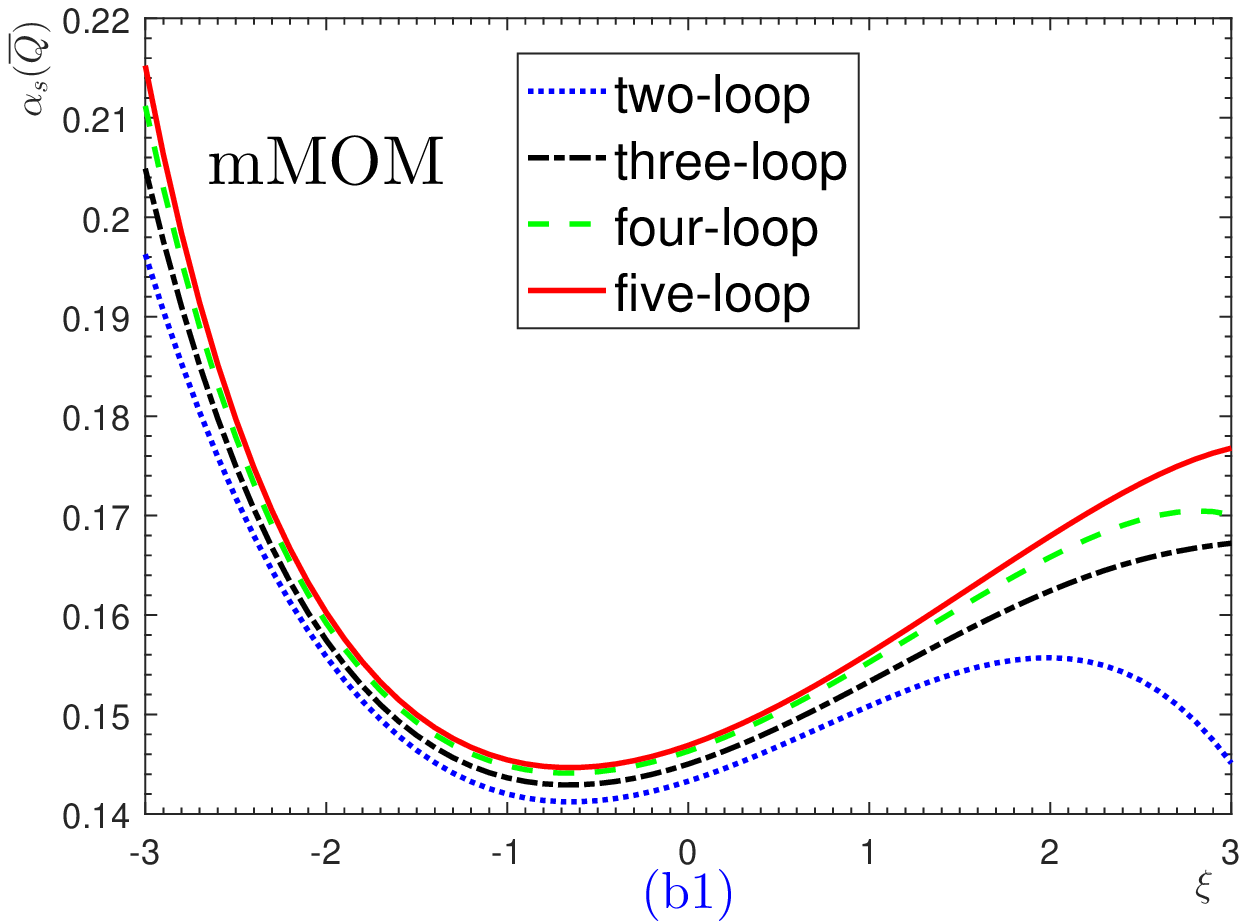}
\includegraphics[width=0.235\textwidth]{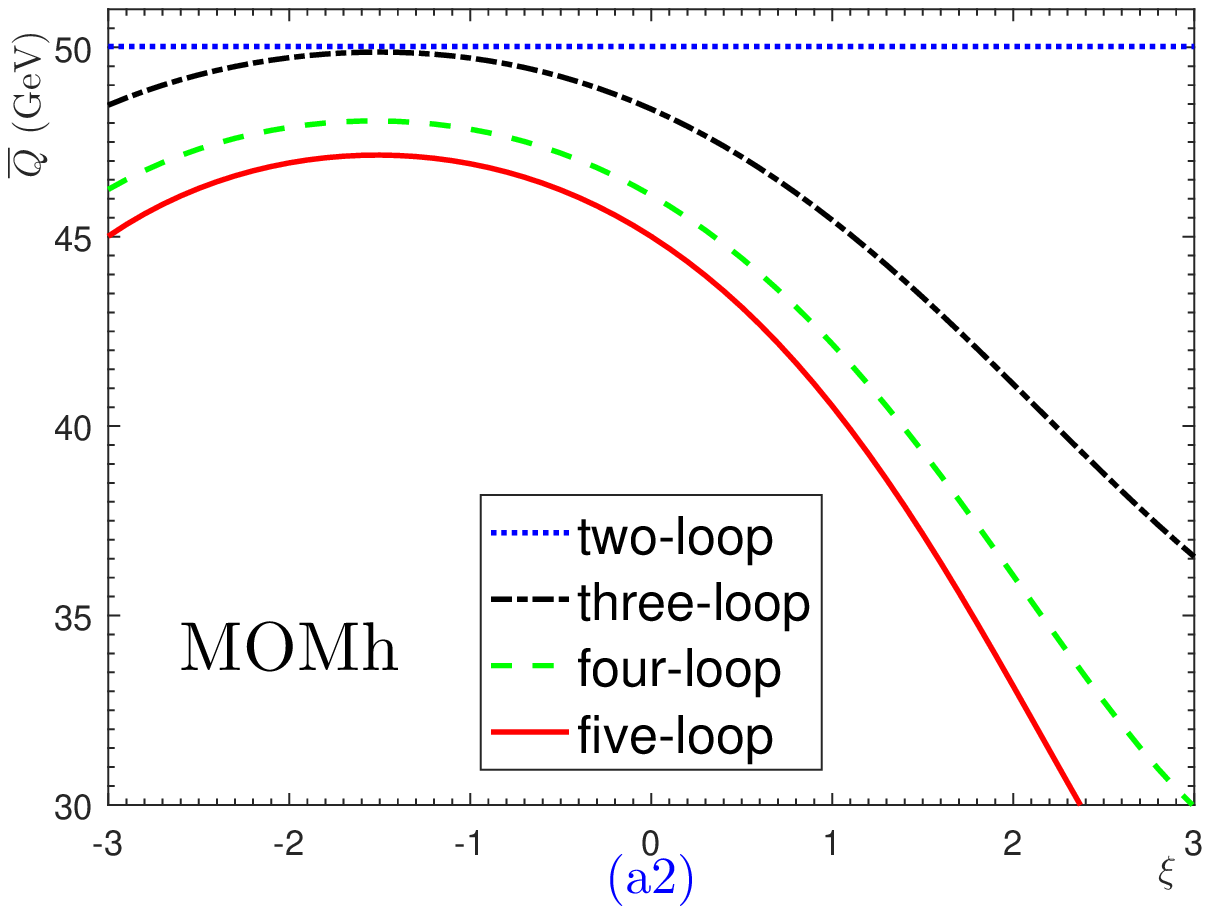}
\includegraphics[width=0.235\textwidth]{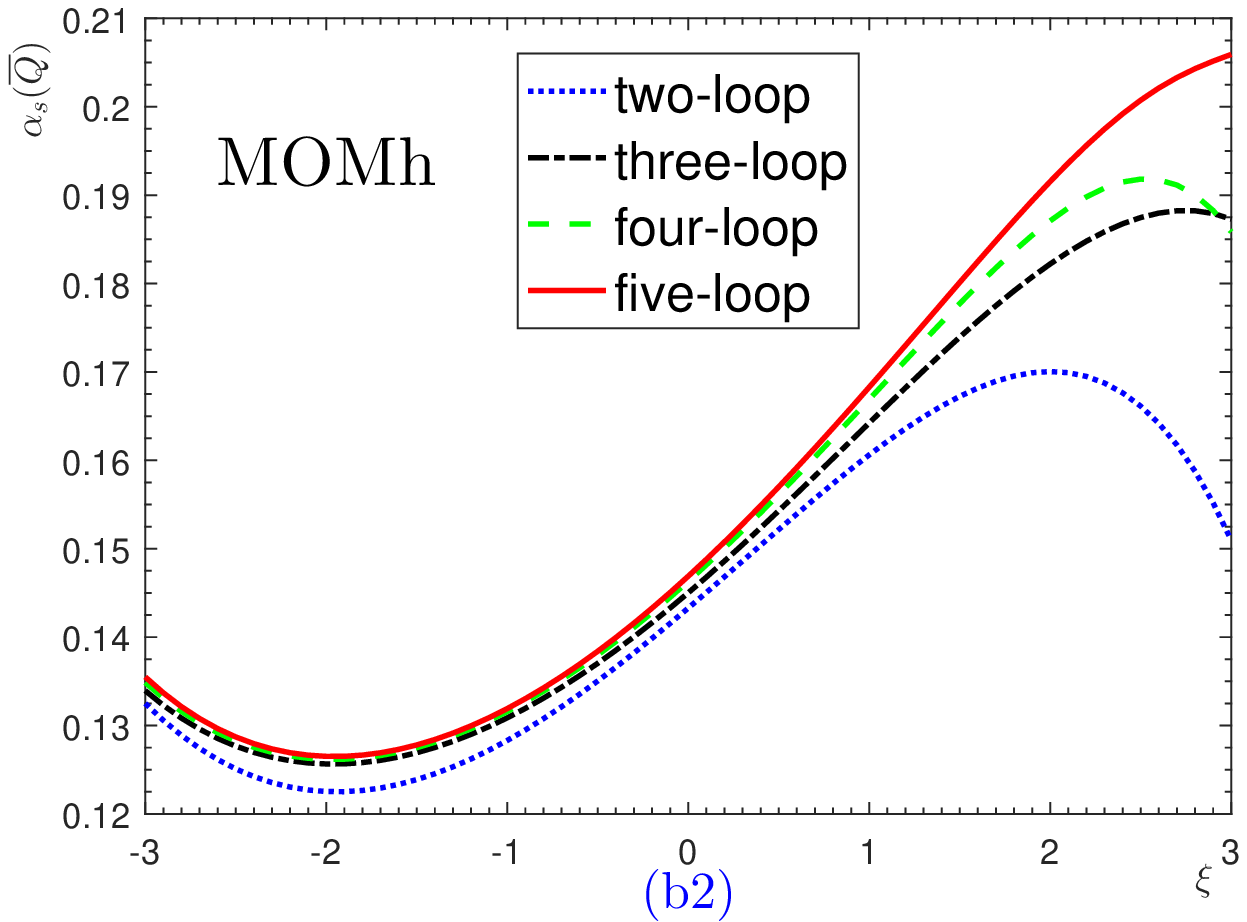}
\includegraphics[width=0.235\textwidth]{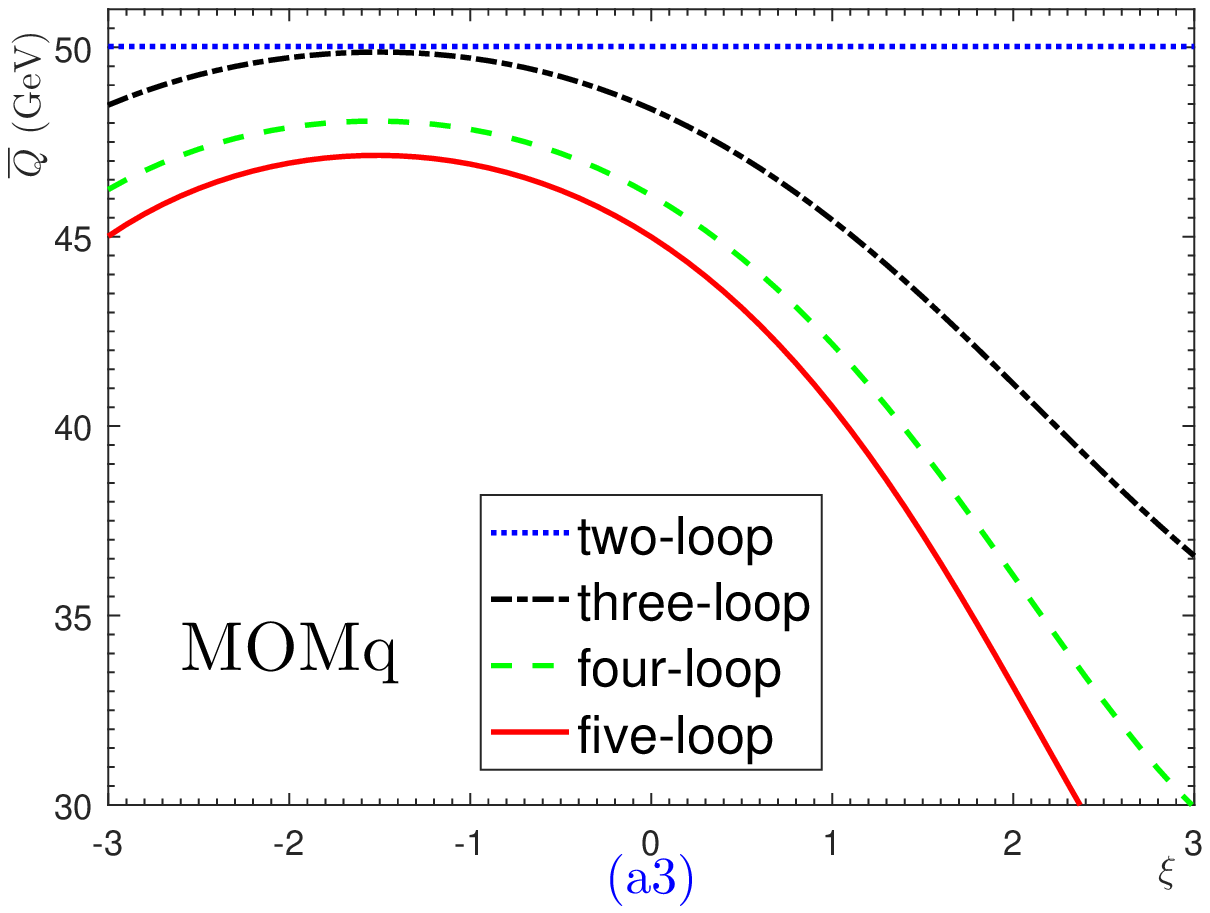}
\includegraphics[width=0.235\textwidth]{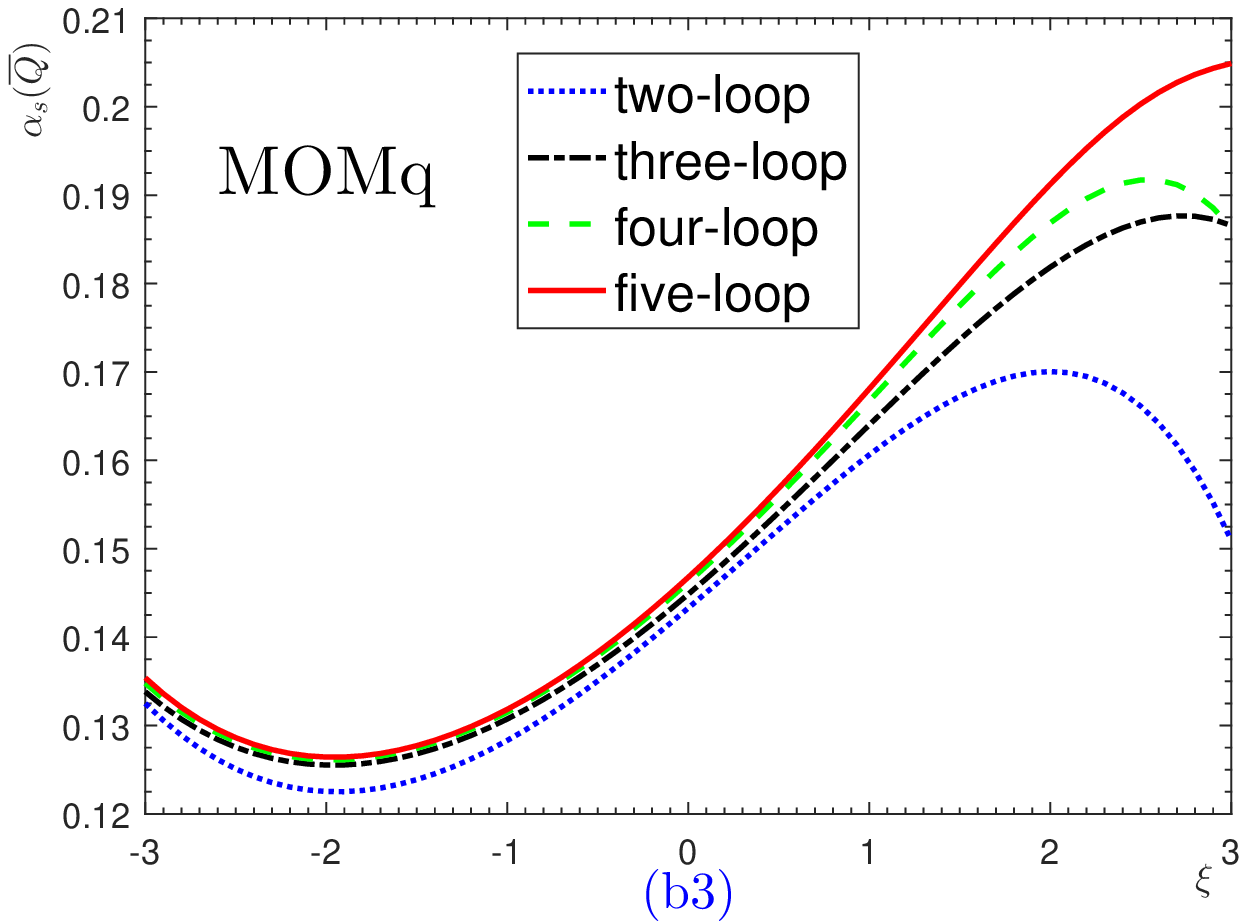}
\includegraphics[width=0.235\textwidth]{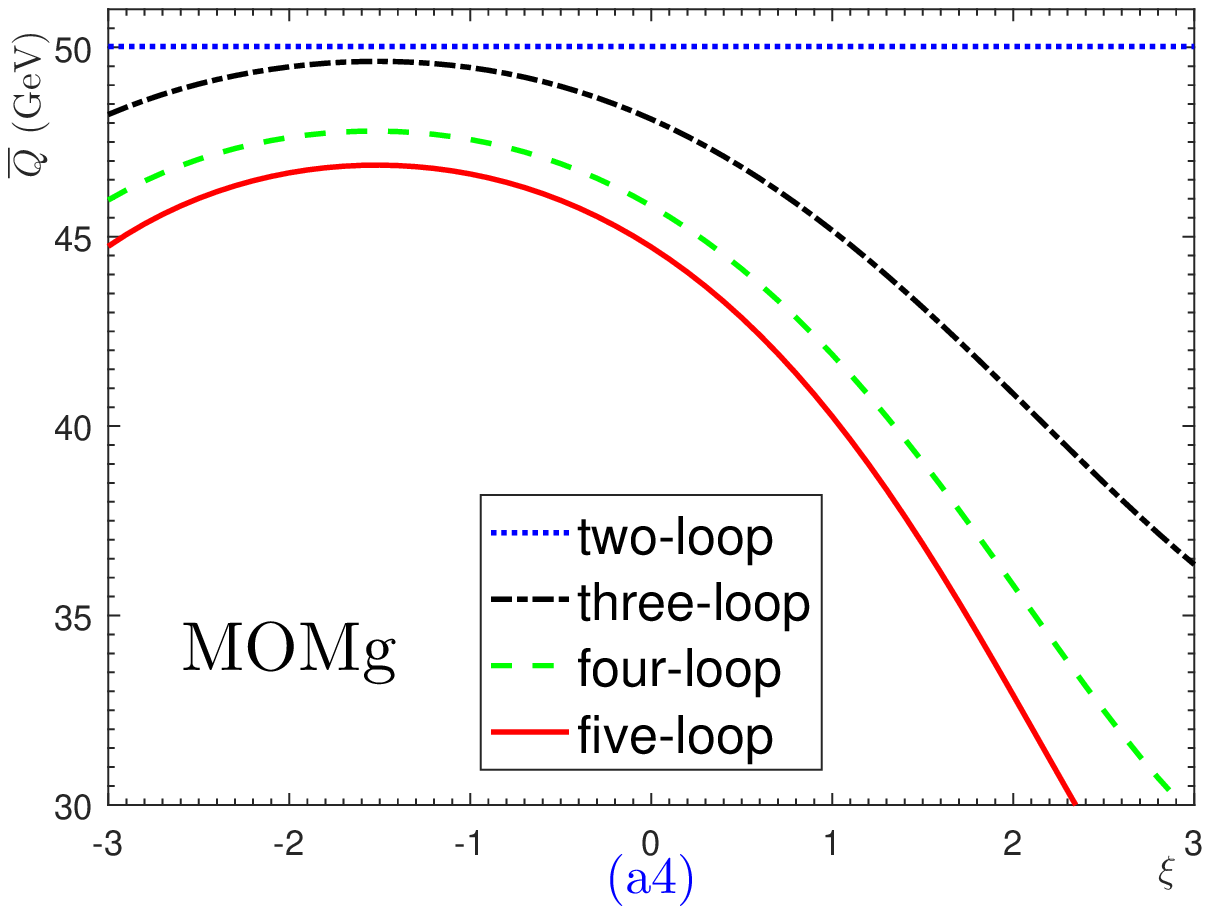}
\includegraphics[width=0.235\textwidth]{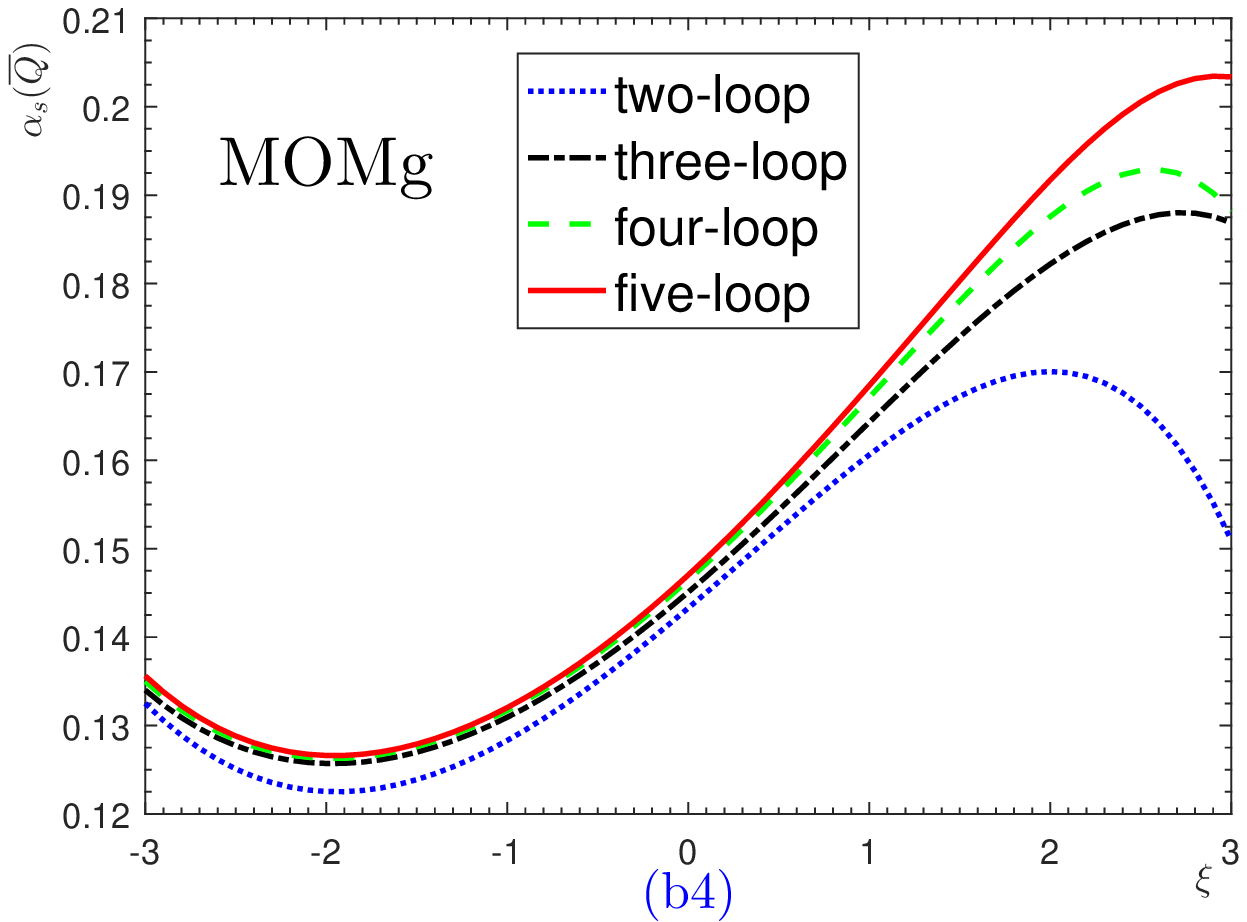}
\includegraphics[width=0.235\textwidth]{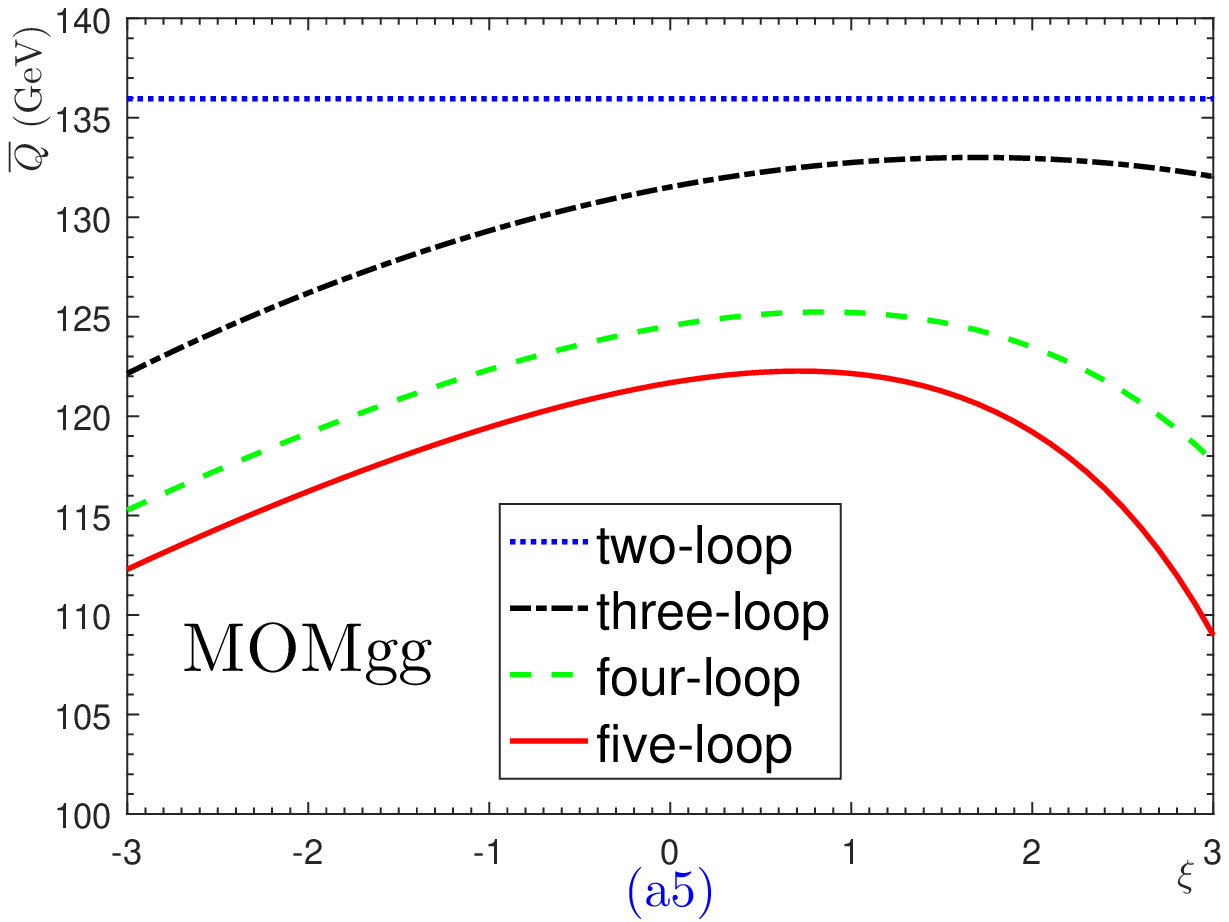}
\includegraphics[width=0.235\textwidth]{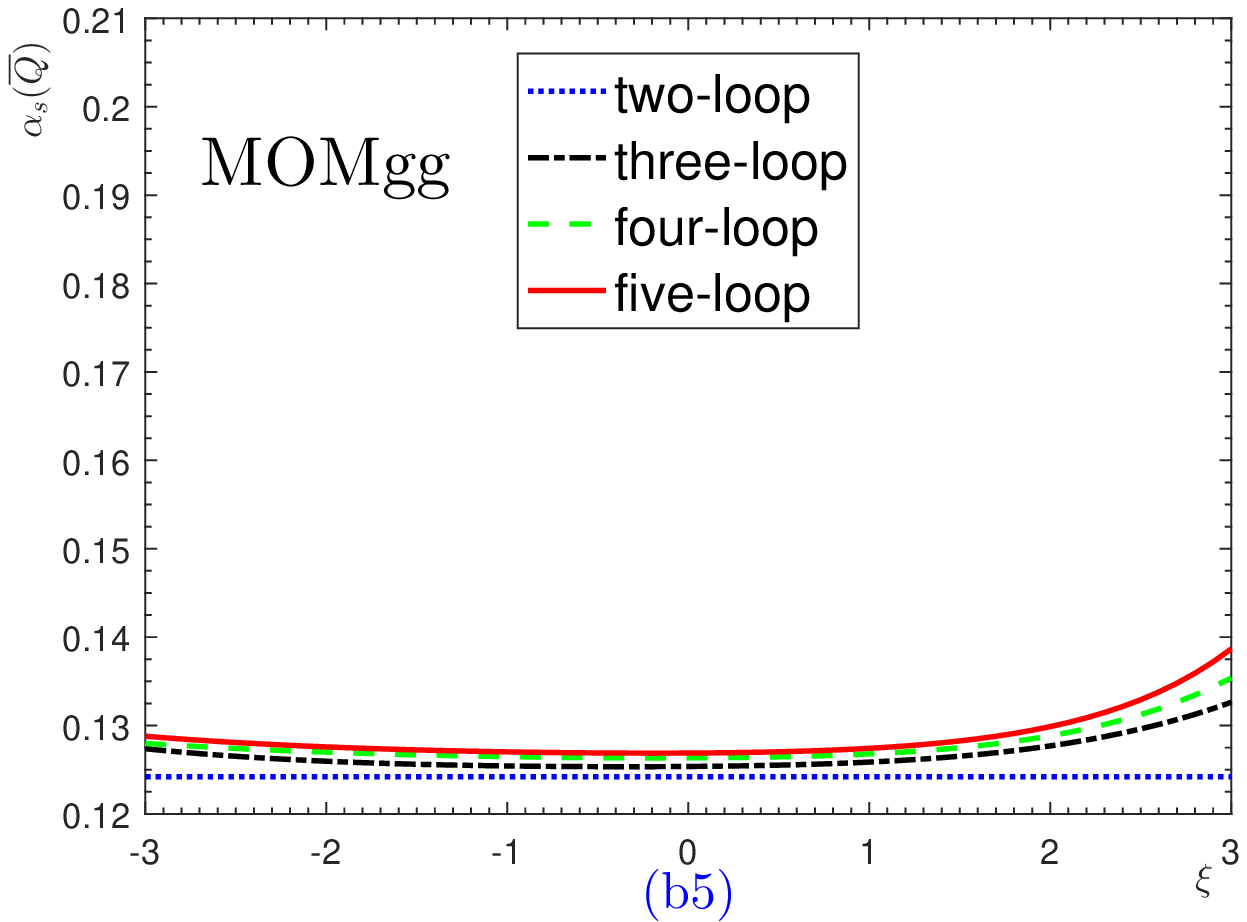}
\caption{The PMC effective scales $\bar{Q}$ (Left column) and their corresponding coupling constants $\alpha_s(\bar{Q})$ (Right column) versus the gauge parameter ($\xi$). Five asymmetric MOM schemes, e.g. mMOM, MOMh, MOMq, MOMg, MOMgg, are adopted. The dotted, the dash-dot, the dashed and the solid lines are results up to two-loop, three-loop, four-loop and five loop QCD corrections, respectively. }
\label{single-scale-coupling}
\end{figure}

The effective scales $\bar{Q}$ (Left column) and their corresponding coupling constants $\alpha_s(\bar{Q})$ (Right column) versus the gauge parameter are shown in Fig.\ref{single-scale-coupling}. At the two-loop level, the determined scale $\bar{Q}$ is gauge independent; However at the three-loop level or even higher, Fig.\ref{single-scale-coupling}(a1-a5) show that $\overline{Q}$ is gauge dependent under the five mentioned MOM schemes. This is due to the fact that all the $r_{i,j}$ terms of MOM schemes are gauge dependent except for those of $i-j=1$ terms. For a specific gauge, we observe that the difference between the two nearby values of $\overline{Q}$ becomes smaller when more loop terms have been included, indicating the precision of $\bar{Q}$ is improved by knowing more loop terms, which agrees with the perturbative nature of $\bar{Q}$. At the present, $\bar{Q}$ can be fixed up to N$^3$LL-accuracy, which is of high accuracy, e.g. the N$^3$LL-term shall only shift the N$^2$LL-accurate $\bar{Q}$ by $\sim +1$ GeV for mMOM, MOMh, MOMq and MOMg, and $\sim +3$ GeV for MOMgg, respectively. Moreover, if setting $\xi^{\rm MOM}\in[-3, 3]$, the effective scale $\bar{Q}$ is $\sim [31, 45]$ GeV for mMOM scheme, $\sim[25, 47]$ GeV for MOMh, MOMq and MOMg schemes, $\sim [109, 122]$ GeV for MOMgg scheme; if setting $\xi^{\rm MOM}\in[-1, 1]$, the effective scale $\bar{Q}$ is $\sim [43, 45]$ GeV for mMOM scheme, $\sim[40, 47]$ GeV for MOMh, MOMq and MOMg scheme, and $\sim [119, 122]$ GeV for MOMgg scheme. This indicates that the perturbative behavior is better for a smaller magnitude of the gauge parameter $|\xi^{\rm MOM}|$ for all those MOM schemes. Fig.\ref{single-scale-coupling}(b1-b5) show that the effect coupling ($\alpha_s({\overline{Q}})$) is also gauge dependent for all the five MOM schemes, the only exception is the MOMgg scheme whose gauge dependence is small and is even zero at the two-loop level. Numerically, the scale $\bar{Q}$ and the effect coupling $\alpha_s(\bar{Q})$ are almost the same for the three MOM schemes, e.g. the MOMh, the MOMq and the MOMg schemes; and if the magnitude of the gauge parameter $|\xi^{\rm MOM}|$ for those three schemes are less than $1$, the differences between the two nearby values of $\alpha_s({\overline{Q}})$ at different orders are almost unchanged for a fixed gauge parameter, indicating the effective couplings for those three schemes quickly achieves its accurate value at lower orders.

\begin{figure}[htb]
\centering
\includegraphics[width=0.235\textwidth]{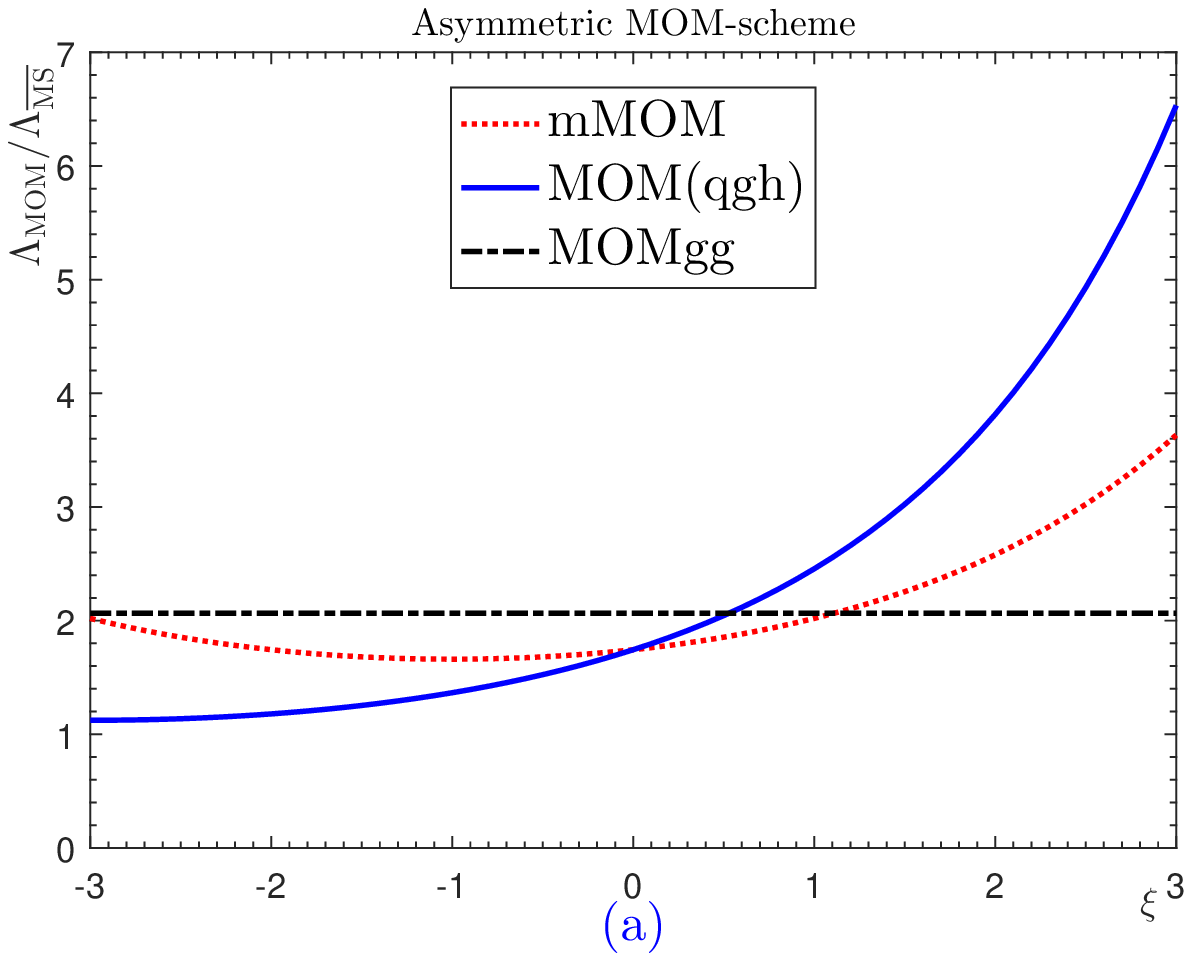}
\includegraphics[width=0.235\textwidth]{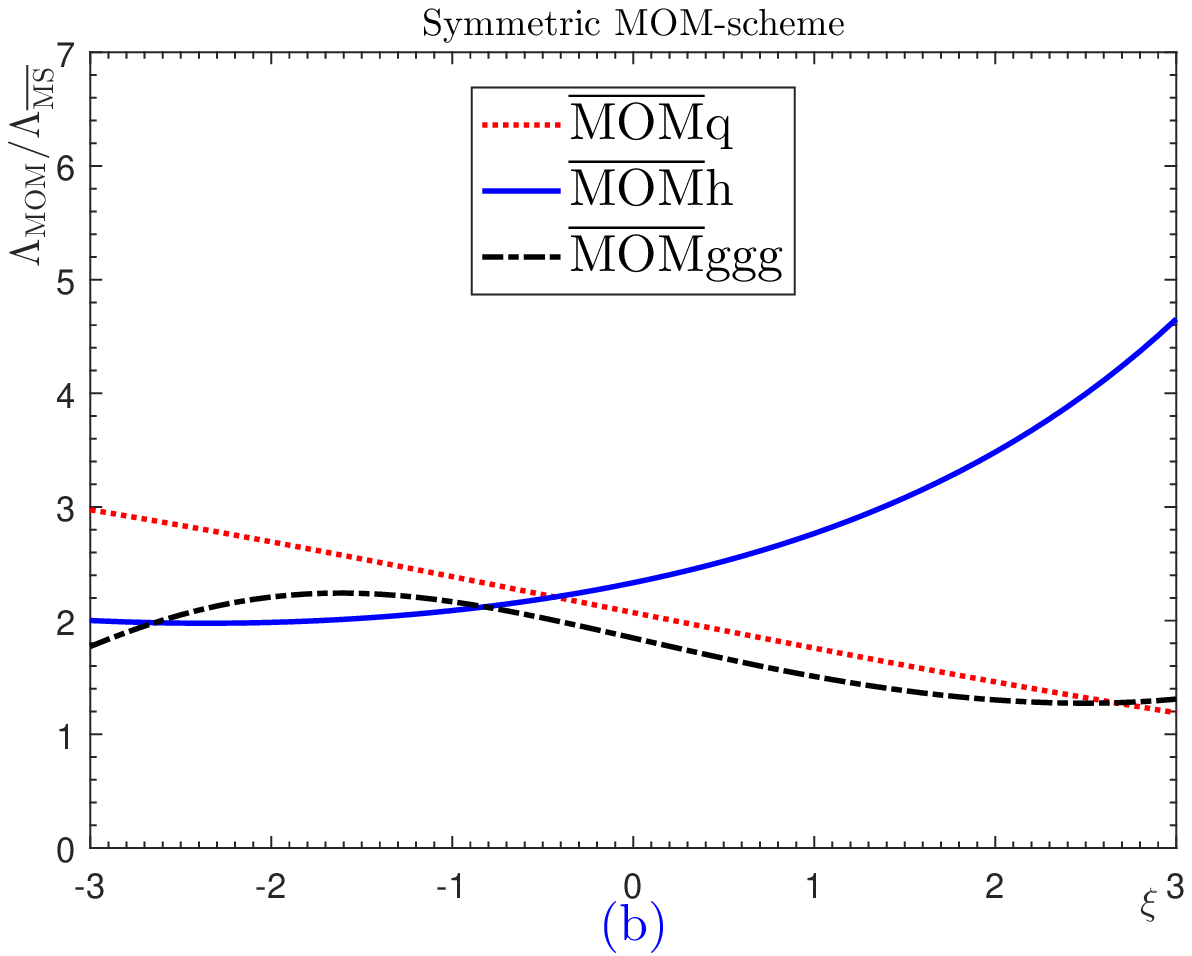}
\caption{Gauge dependence of the ratio $\Lambda_{\rm{MOM}}/\Lambda_{\overline{\rm MS}}$ by varying the gauge parameter $\xi=\xi^{\rm MOM} \in [-3,3]$. $n_f=5$. (a) The ratios for five asymmetric MOM-schemes, where the asymptotic scales of MOMg, MOMq, and MOMh schemes are the same and are shown by solid line; (b) The ratios for three symmetric MOM-schemes. }
\label{RatioLambda}
\end{figure}

It is interesting to find that under the MOMgg scheme, the effective coupling $\alpha_s(\bar{Q})$ also can achieve its accurate value at lower orders, whose value is almost gauge independent for $|\xi^{\rm MOMgg}|\leq 1$, indicating the gauge dependence of $\bar{Q}$ is well compensated by the gauge dependence of $\Lambda^{\rm MOMgg}_{\rm QCD}$. More explicitly, the asymptotic scale for various MOM schemes can be derived from $\Lambda_{\overline{\rm MS}}$ by using the Celmaster-Gonsalves relation~\cite{Celmaster:1979km, Celmaster:1979dm, Celmaster:1979xr, Celmaster:1980ji, vonSmekal:2009ae}. We present the ratios of ${\Lambda_{\rm{MOM}}}/{\Lambda_{\rm{\overline{MS}}}}$ for $n_f=5$ in Fig.~\ref{RatioLambda}, where the ratios for the following mentioned three symmetric MOM schemes are also presented. It is found that the asymptotic scales of MOMg, MOMq, and MOMh schemes are the same, together with the close values of $\bar{Q}$, one can explain the close behaviors of $\alpha_s(\bar{Q})$ and then close $\Gamma(H\to gg)$ under those schemes. Almost all of the ratios show explicit gauge dependence; the only exception is the ratio of MOMgg scheme, whose value is free of $\xi^{\rm MOMgg}$ and is fixed to be $e^{50/69}\approx 2.06$ for $n_f=5$.

\subsection{The gauge dependence of $\Gamma  (H\to gg)$ for the five asymmetric MOM schemes}

\begin{figure}[htb]
\centering
\includegraphics[width=0.235\textwidth]{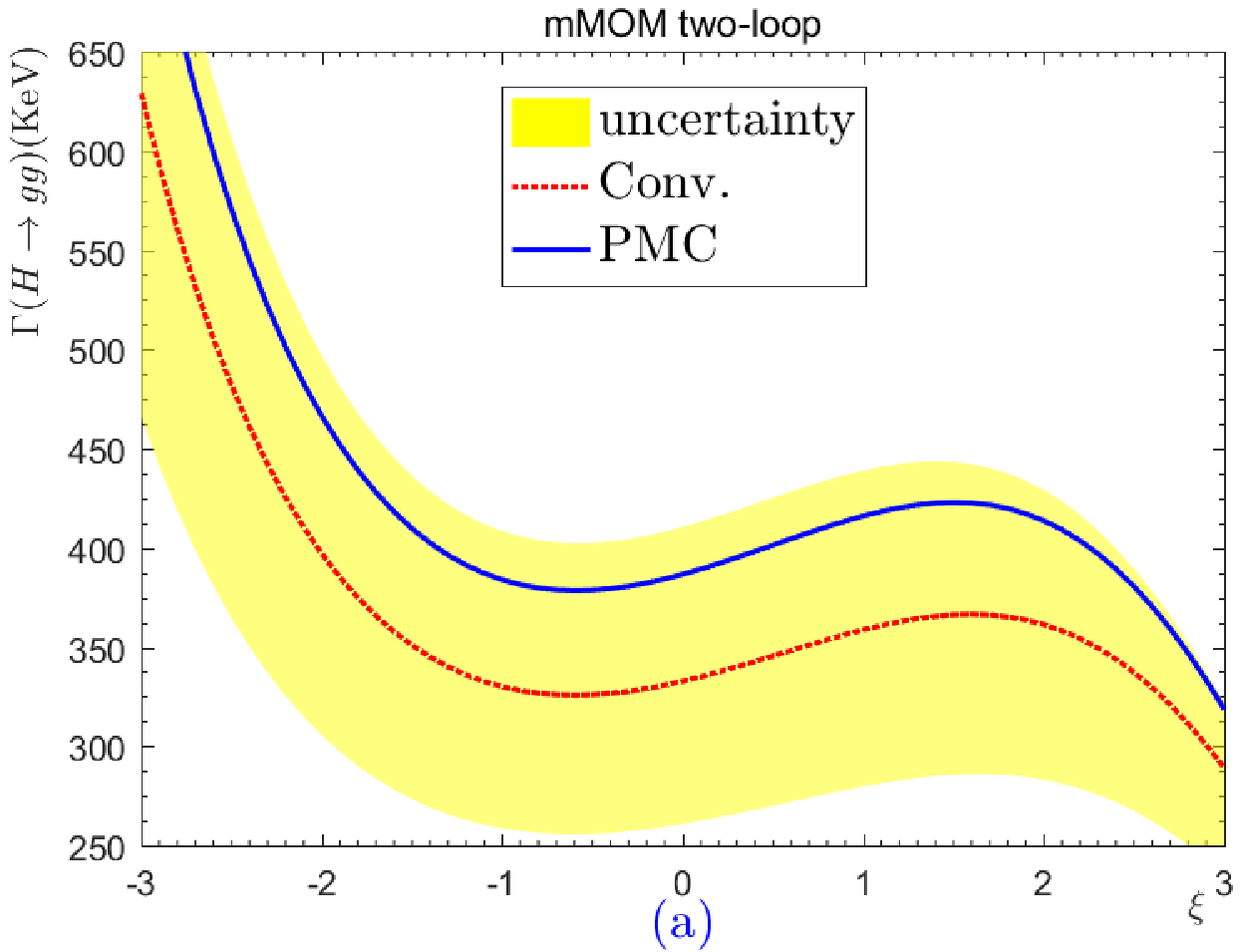}
\includegraphics[width=0.235\textwidth]{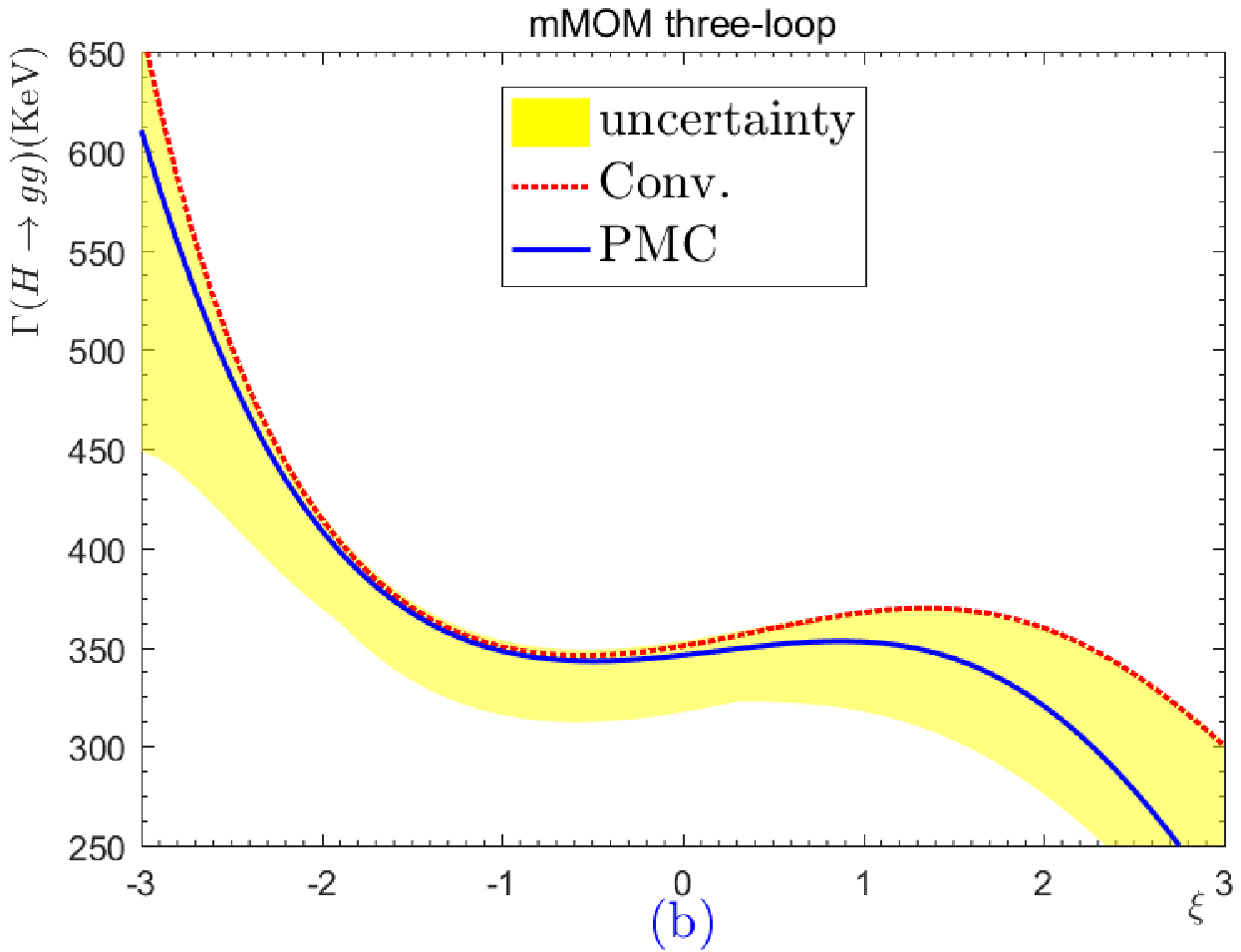}
\includegraphics[width=0.235\textwidth]{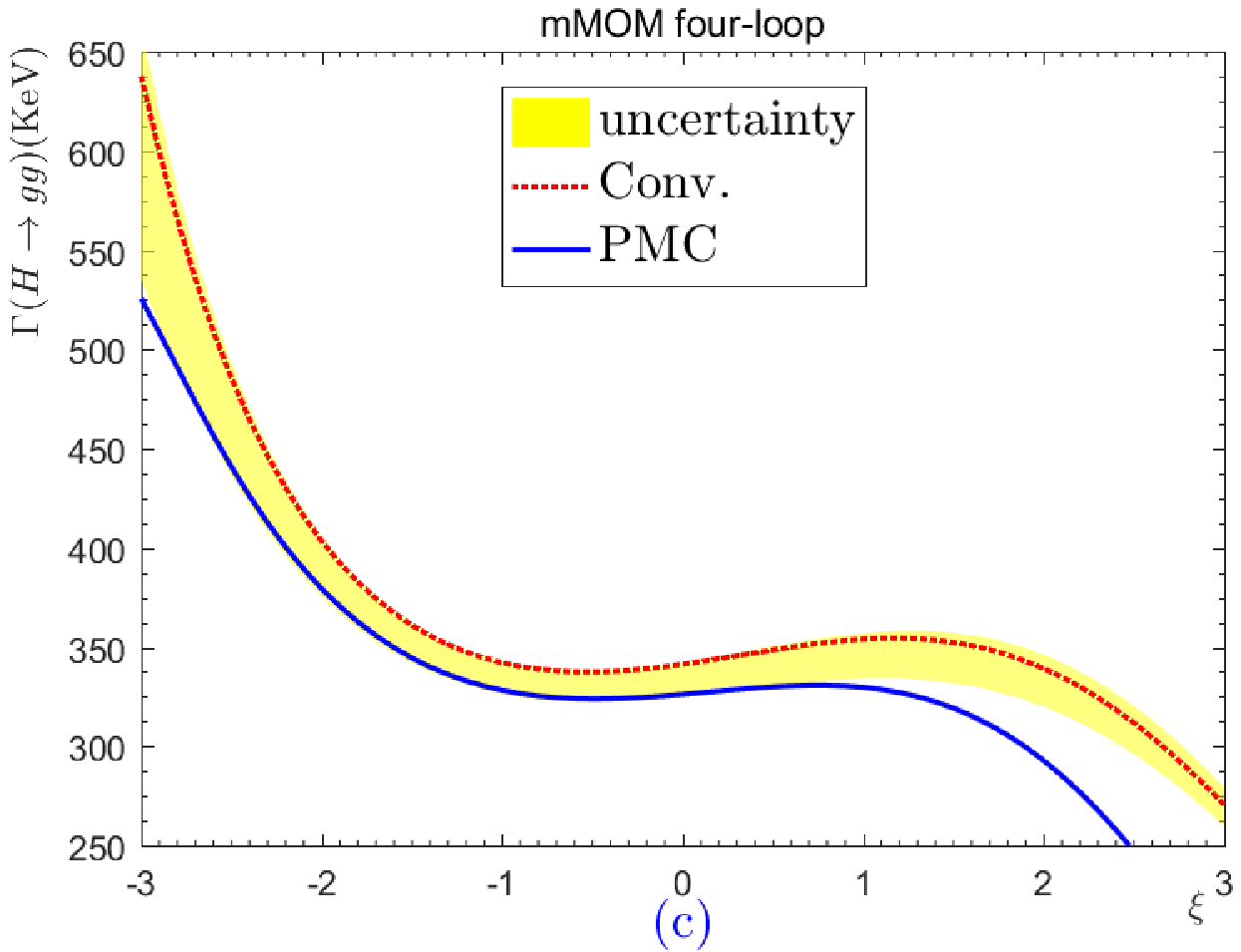}
\includegraphics[width=0.235\textwidth]{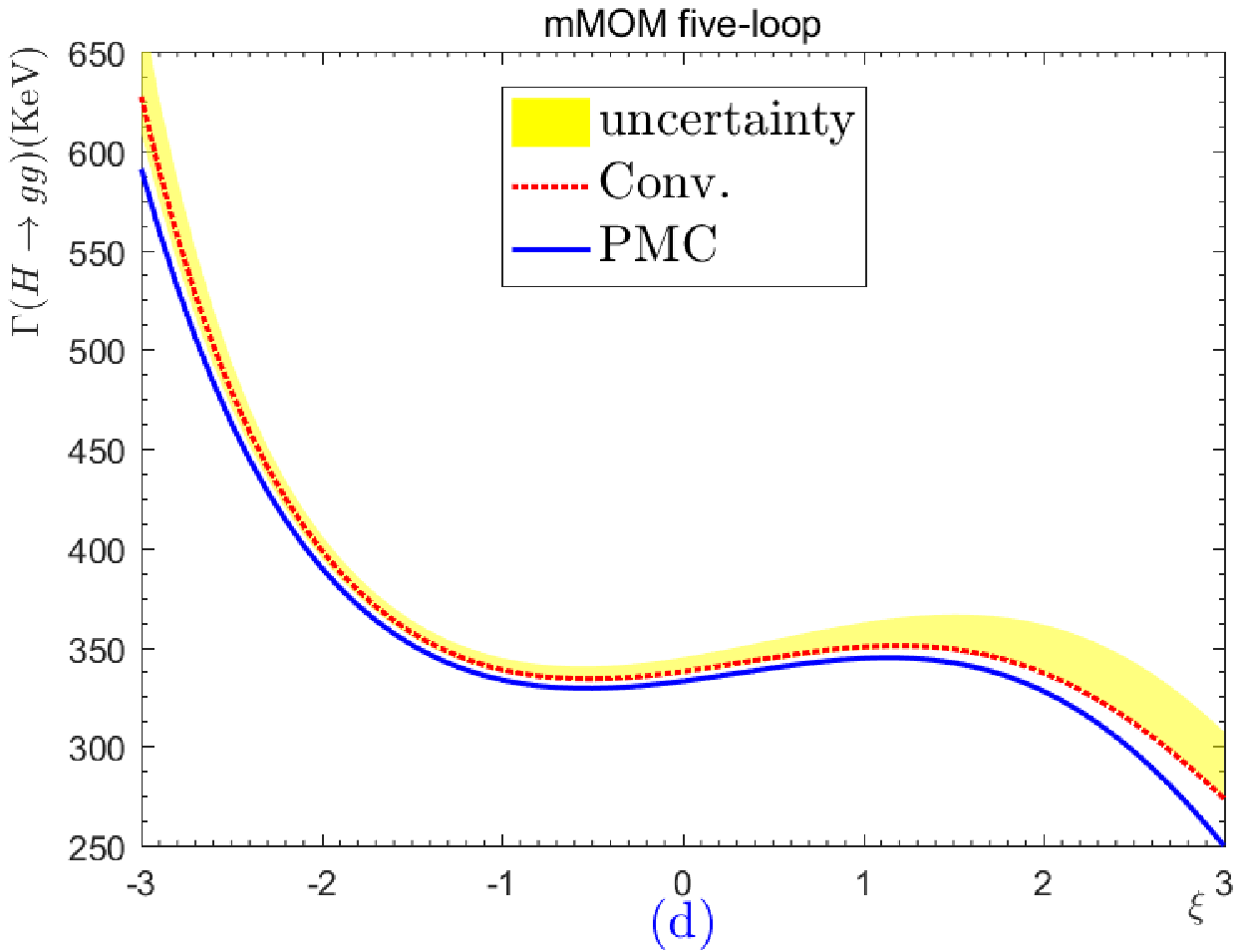}
\caption{Total decay width $\Gamma  (H\to gg)$ versus gauge parameter $\xi=\xi^{\rm mMOM}$ up to two-loop, three-loop, four-loop, and five-loop levels, respectively, under the mMOM scheme. The dotted line is for conventional scale setting approach with $\mu = M_H$ and the shaded band shows its renormalization scale uncertainty by varying $\mu \in [M_H/4,4M_H]$. The solid line is the prediction for the PMC, which is independent to the choice of renormalization scale.}
\label{two-fiveloop}
\end{figure}

\begin{figure}[htb]
\centering
\includegraphics[width=0.235\textwidth]{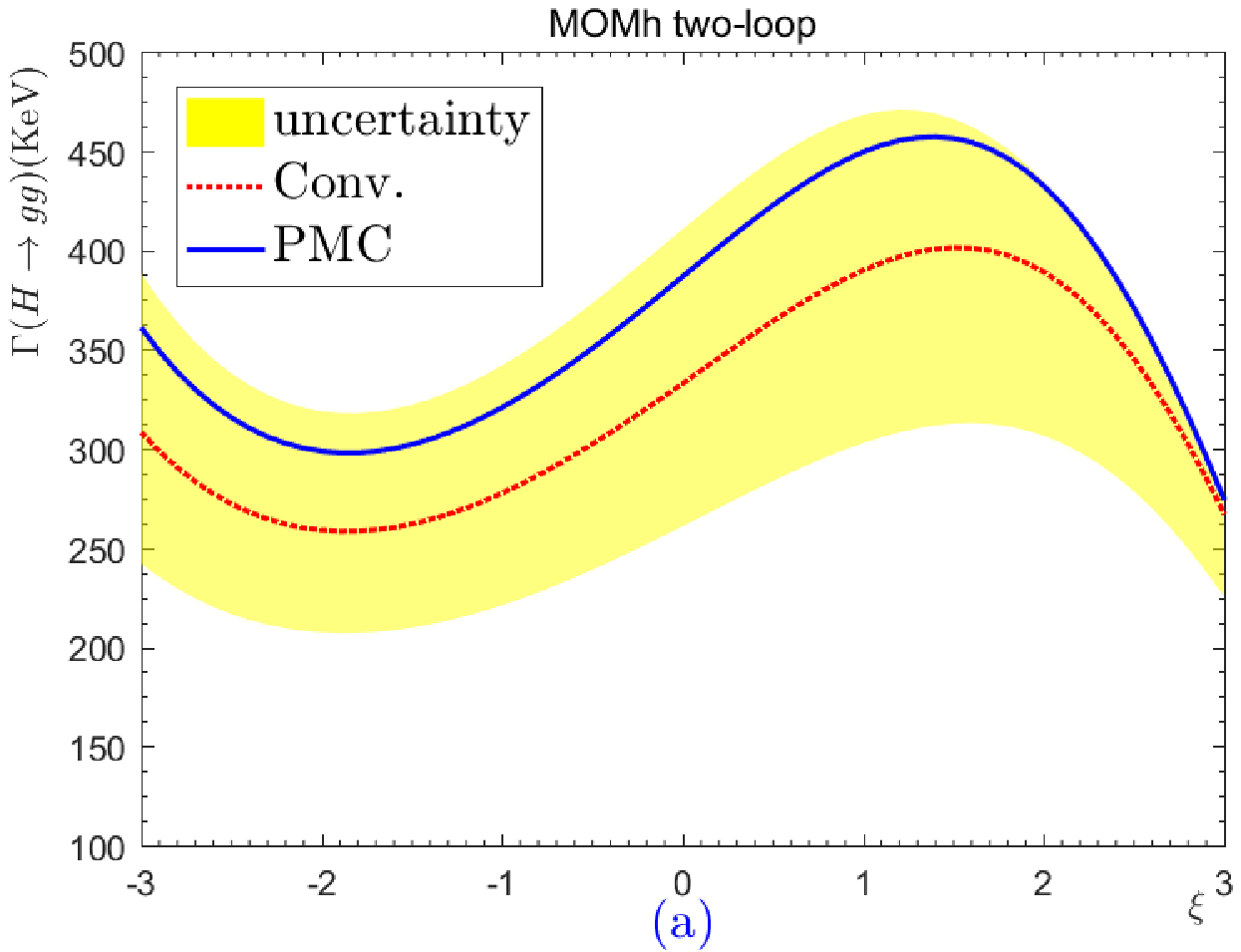}
\includegraphics[width=0.235\textwidth]{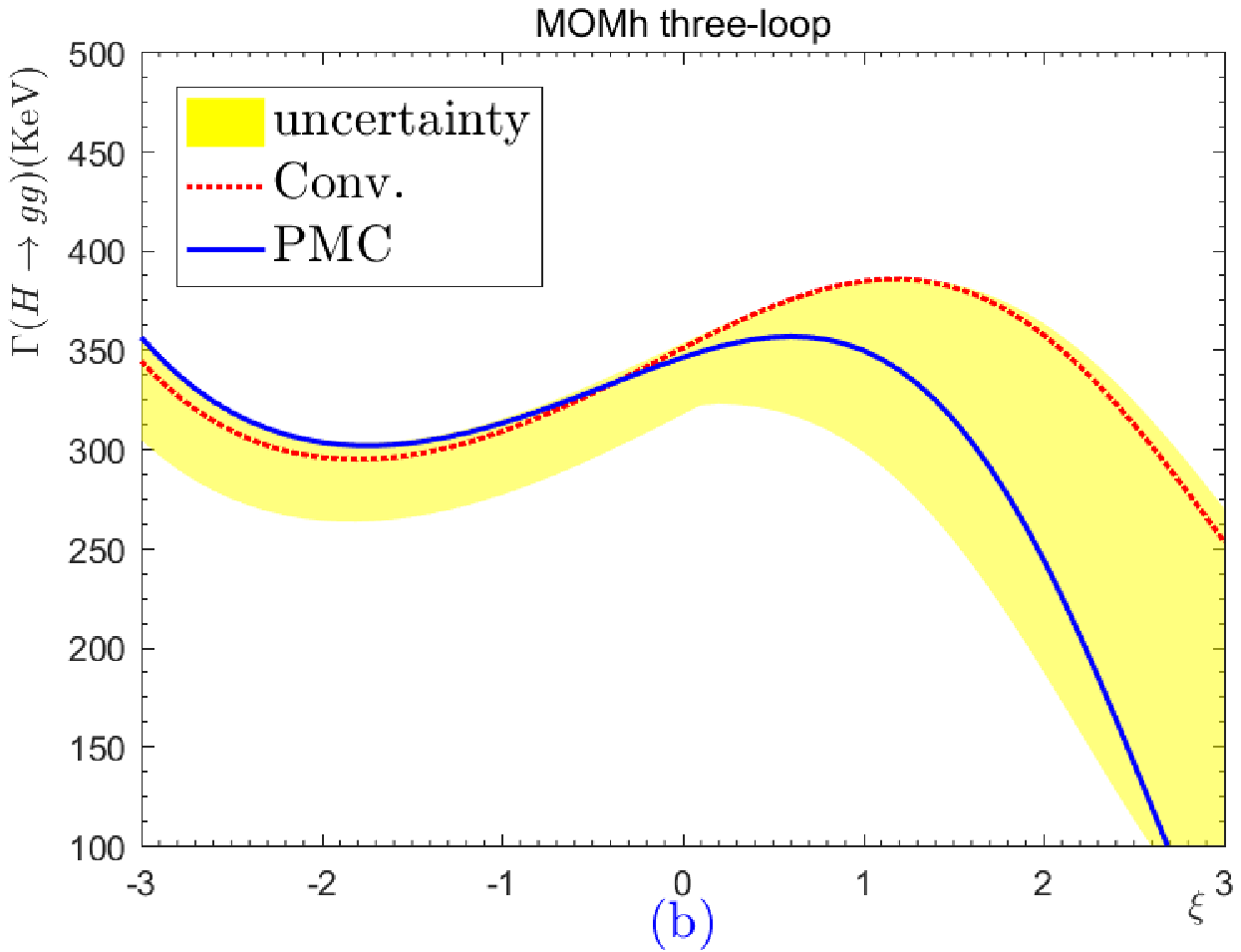}
\includegraphics[width=0.235\textwidth]{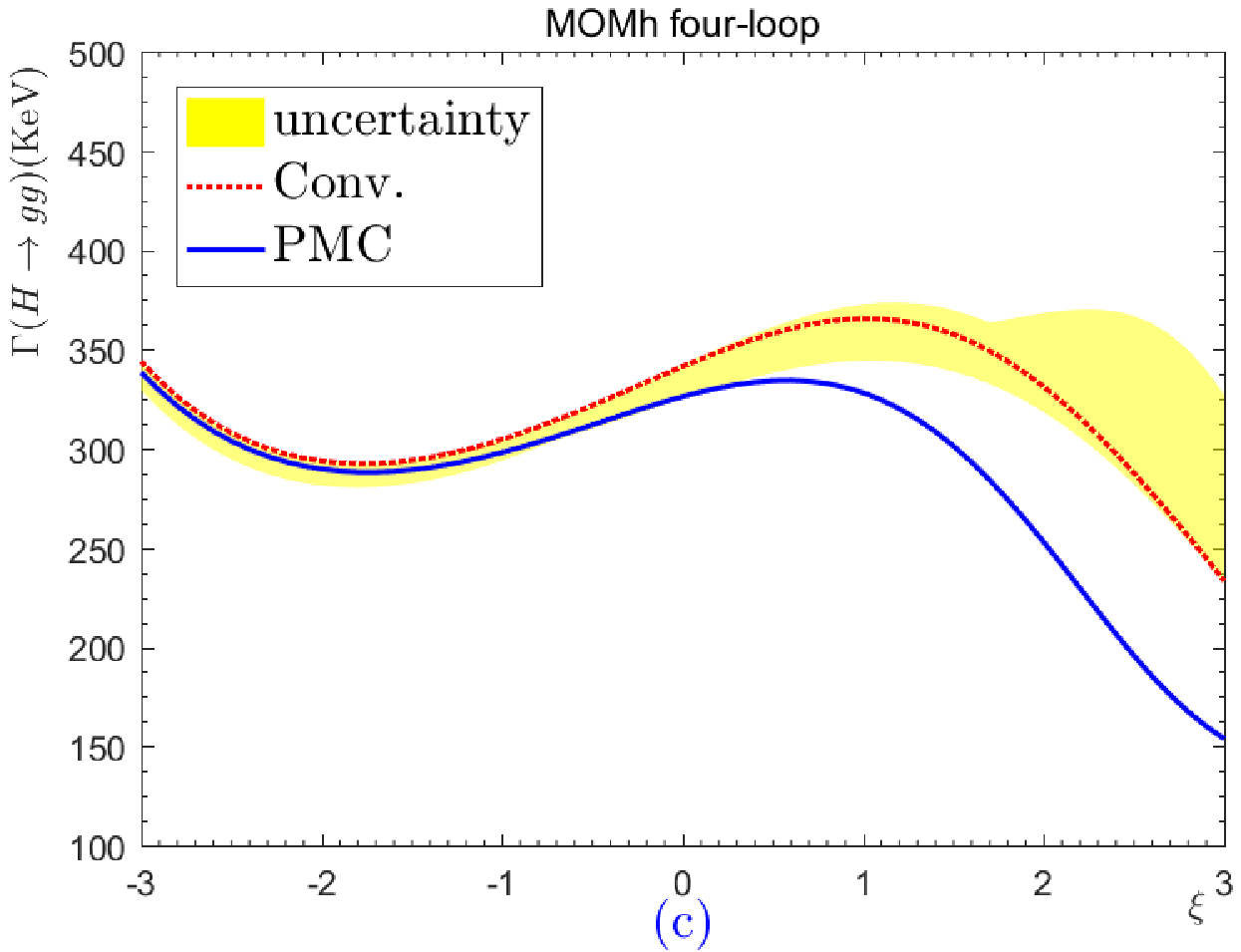}
\includegraphics[width=0.235\textwidth]{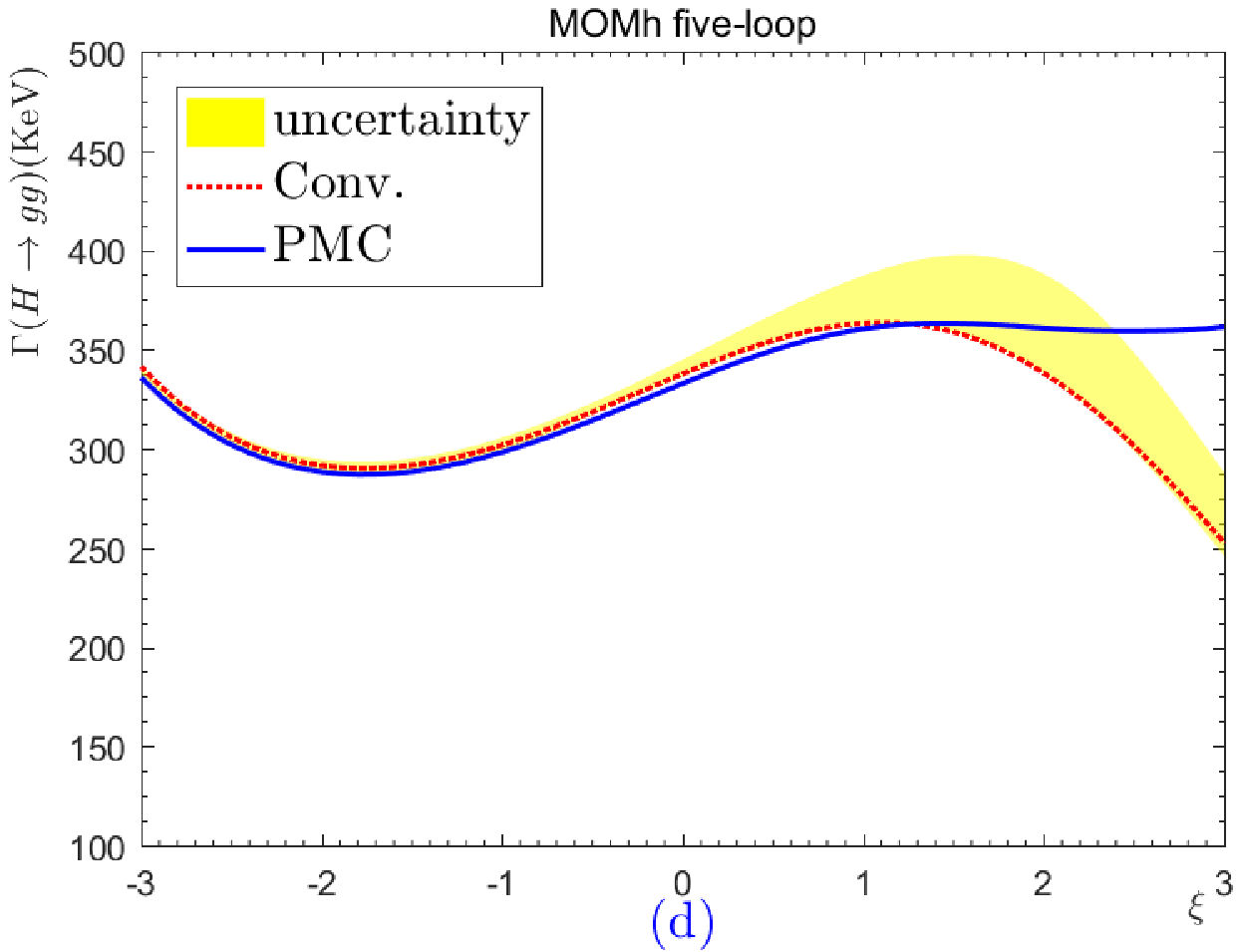}
\caption{Total decay width $\Gamma  (H\to gg)$ versus gauge parameter $\xi=\xi^{\rm MOMh}$ up to two-loop, three-loop, four-loop, and five-loop levels, respectively, under the MOMh scheme. The dotted line is for conventional scale setting approach with $\mu = M_H$ and the shaded band shows its renormalization scale uncertainty by varying $\mu \in [M_H/4,4M_H]$. The solid line is the prediction for PMC scale-setting, which is independent to the choice of renormalization scale.}
\label{two-fiveloopMOMh}
\end{figure}

\begin{figure}[htb]
\centering
\includegraphics[width=0.235\textwidth]{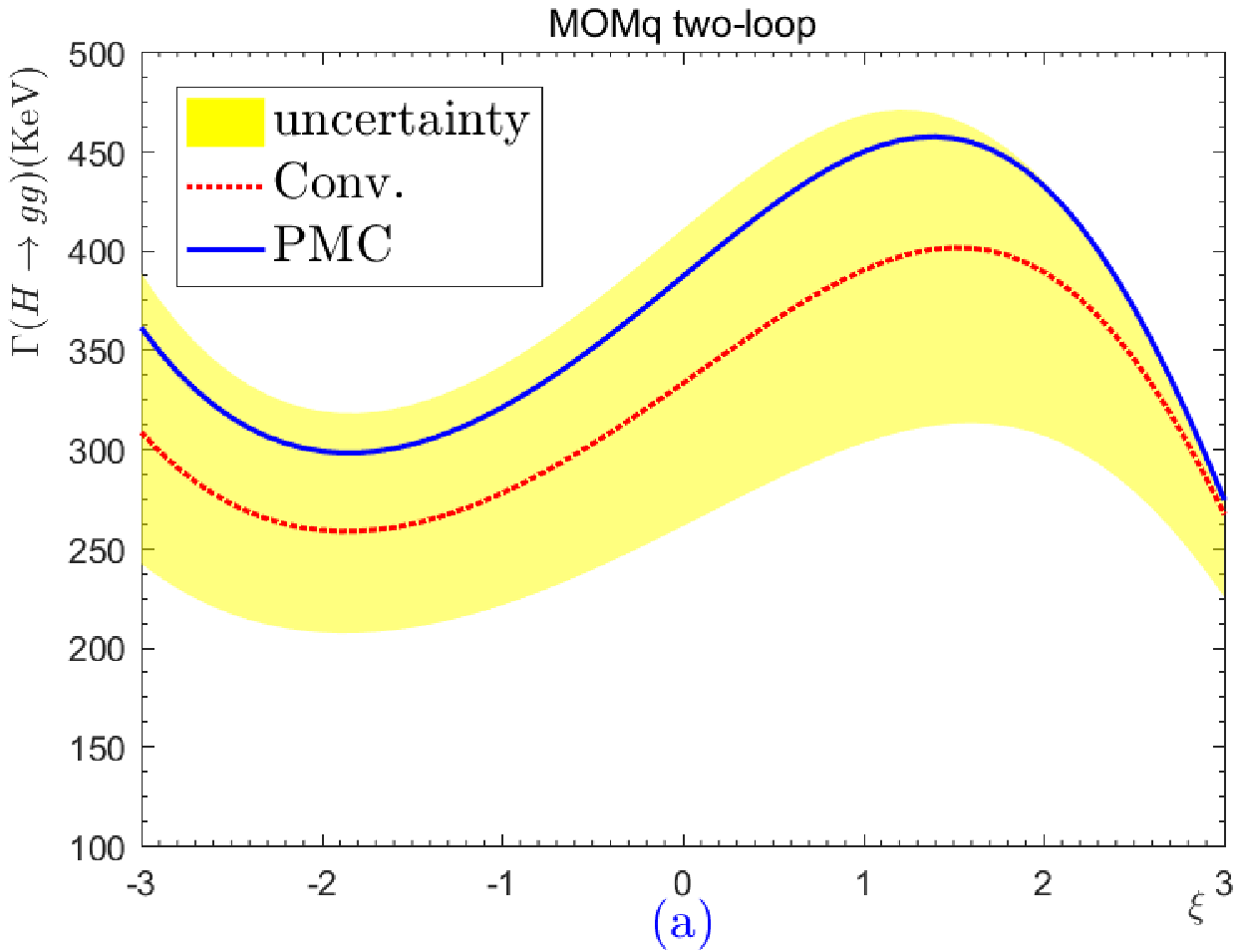}
\includegraphics[width=0.235\textwidth]{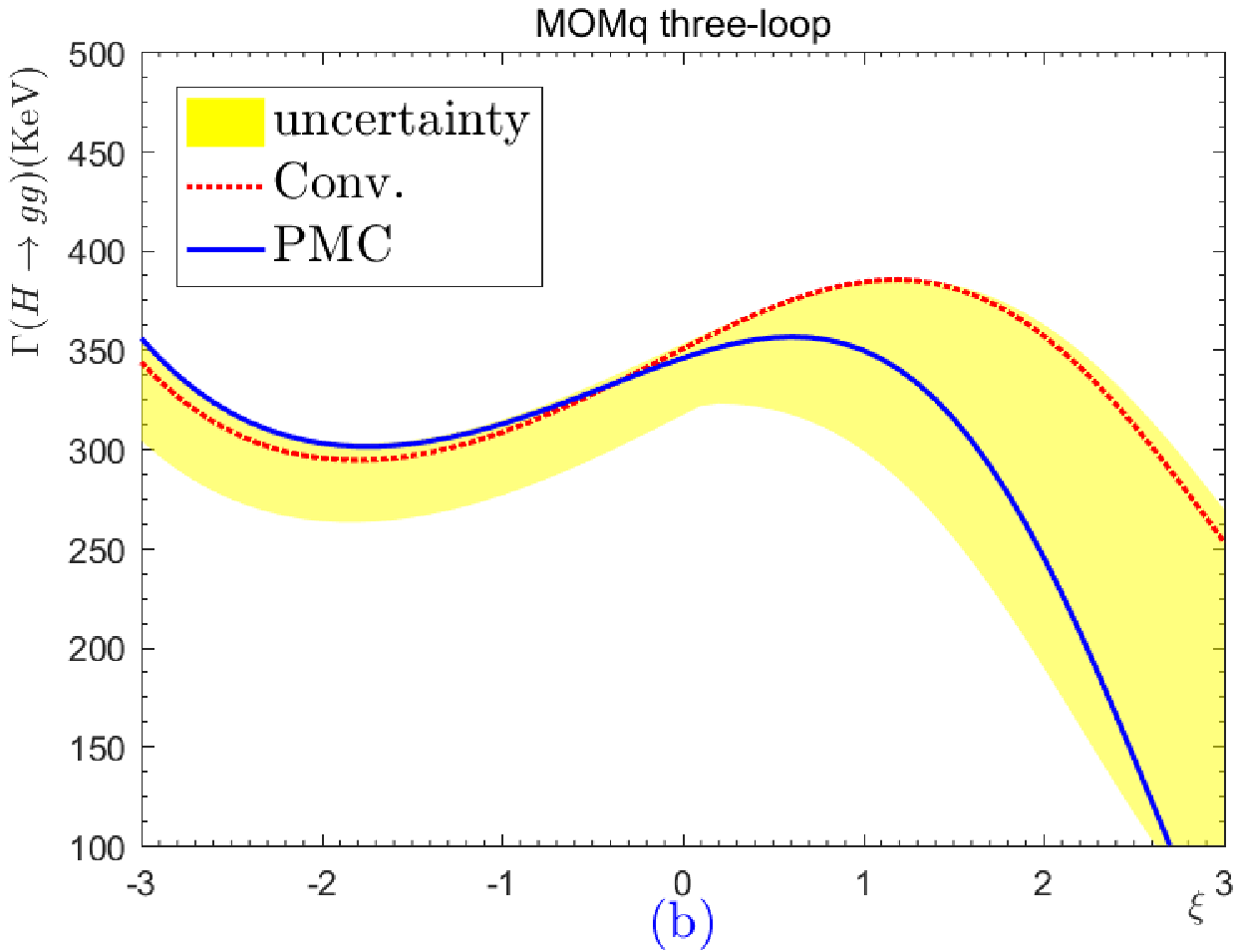}
\includegraphics[width=0.235\textwidth]{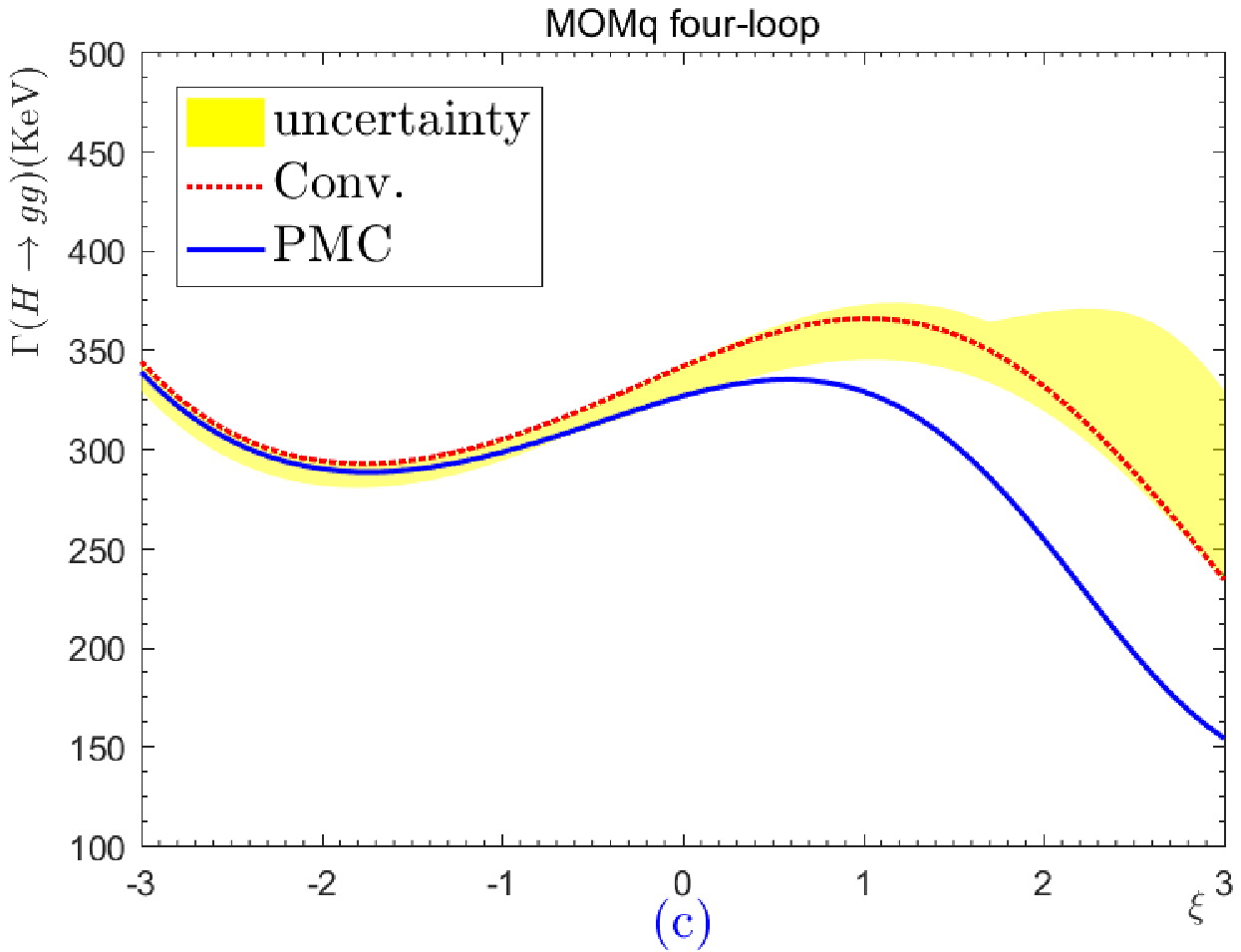}
\includegraphics[width=0.235\textwidth]{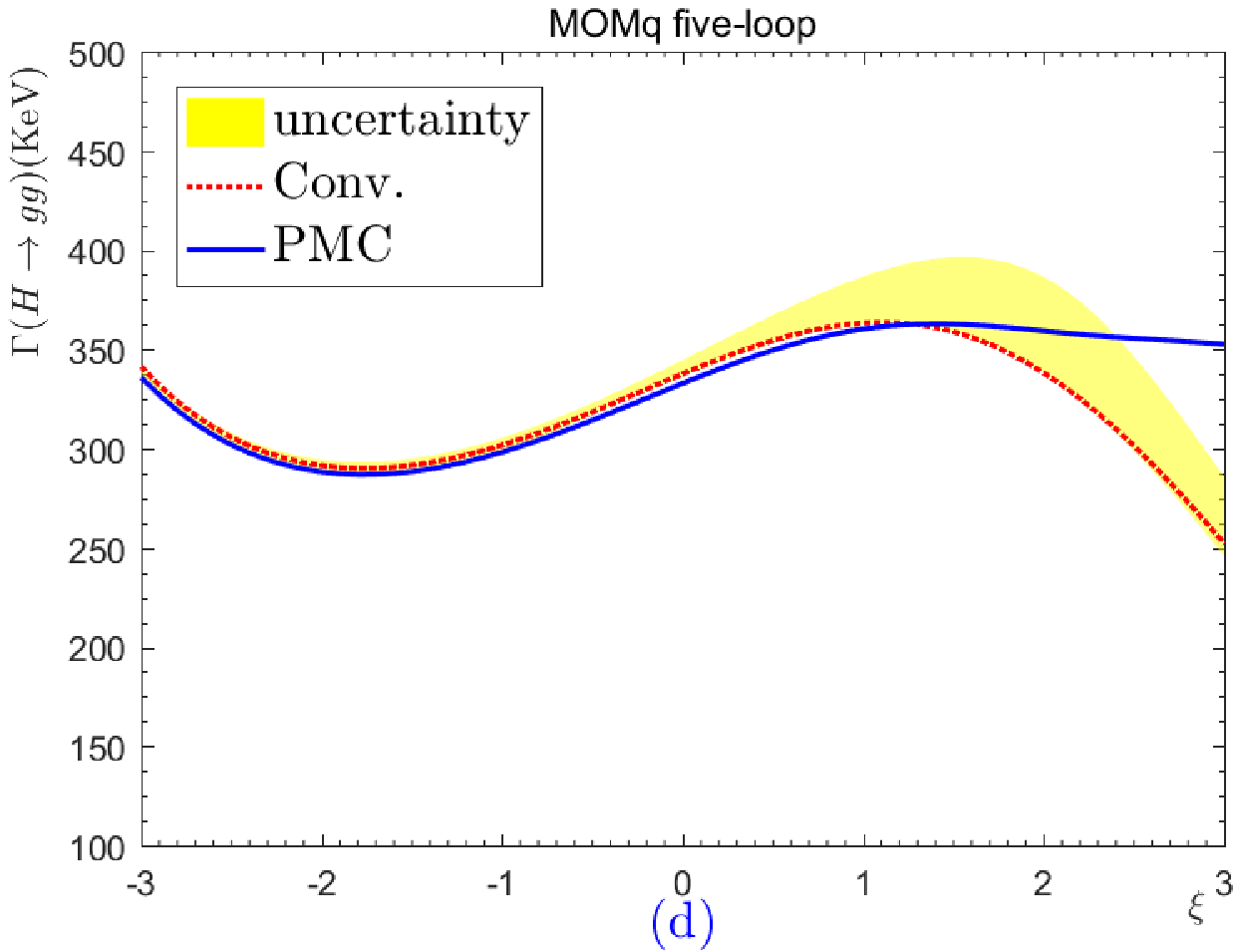}
\caption{Total decay width $\Gamma  (H\to gg)$ versus gauge parameter $\xi=\xi^{\rm MOMq}$ up to two-loop, three-loop, four-loop, and five-loop levels, respectively, under the MOMq scheme. The dotted line is for conventional scale setting approach with $\mu = M_H$ and the shaded band shows its renormalization scale uncertainty by varying $\mu \in [M_H/4,4M_H]$. The solid line is the prediction for the PMC, which is independent to the choice of renormalization scale.}
\label{two-fiveloopMOMq}
\end{figure}

\begin{figure}[htb]
\centering
\includegraphics[width=0.235\textwidth]{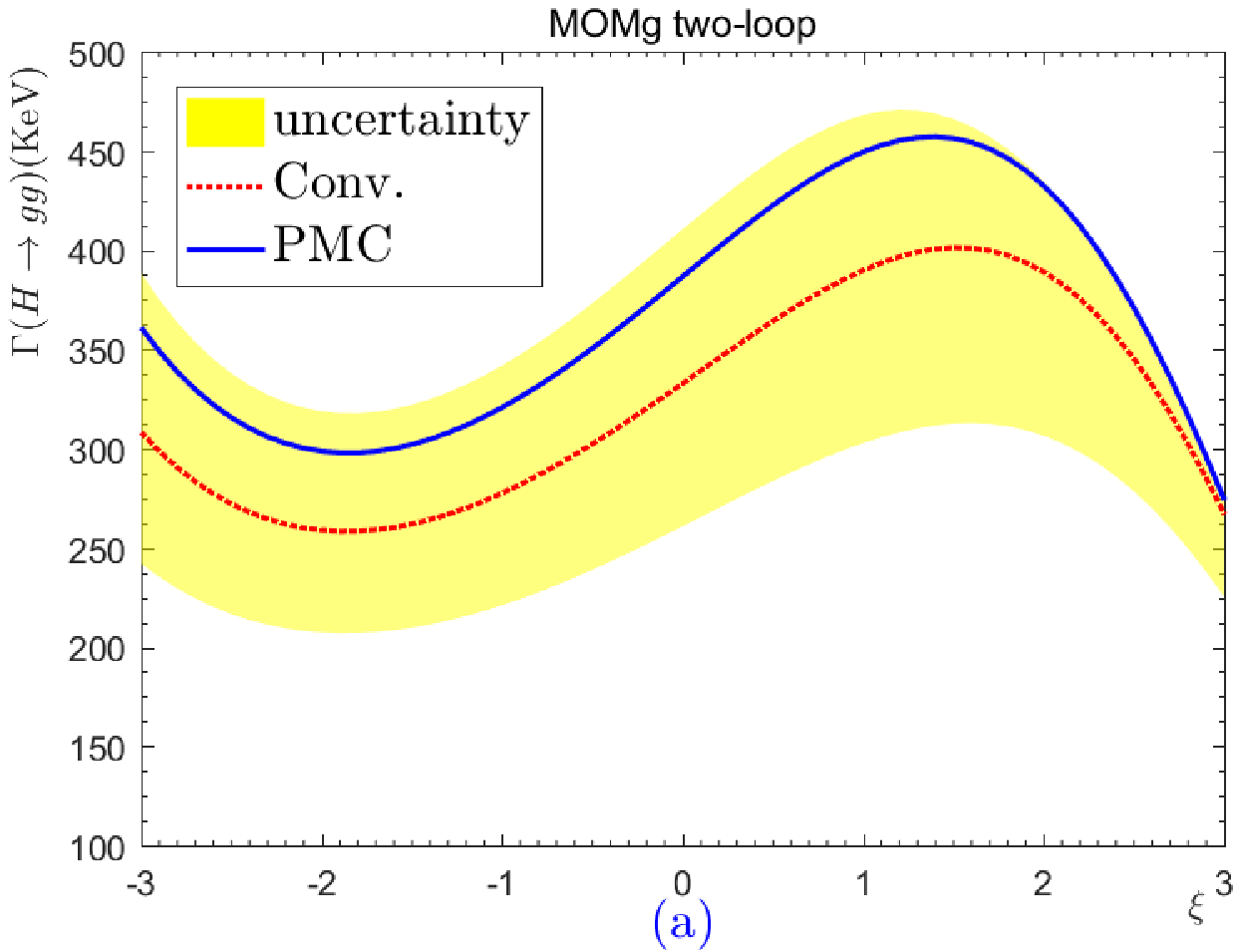}
\includegraphics[width=0.235\textwidth]{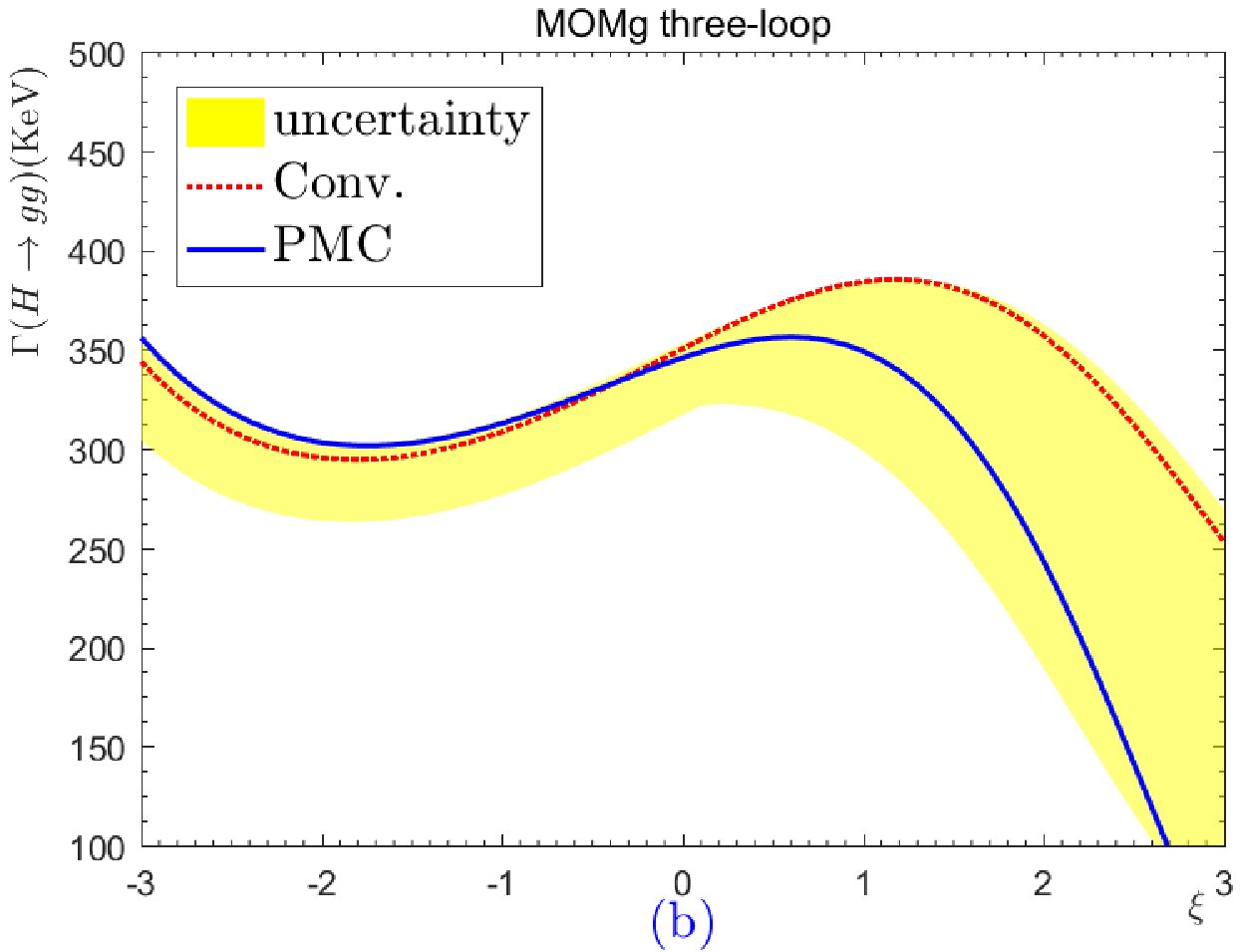}
\includegraphics[width=0.235\textwidth]{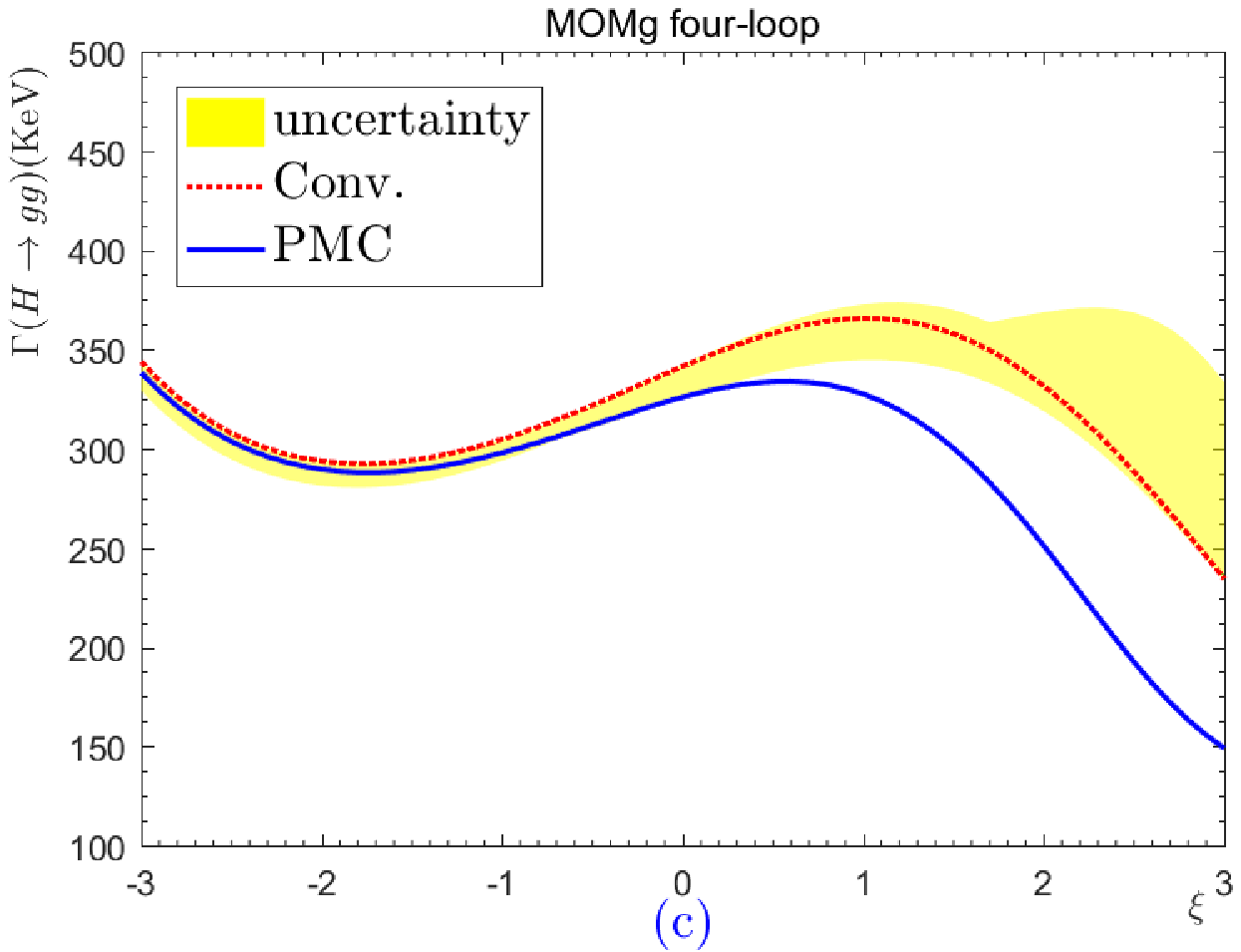}
\includegraphics[width=0.235\textwidth]{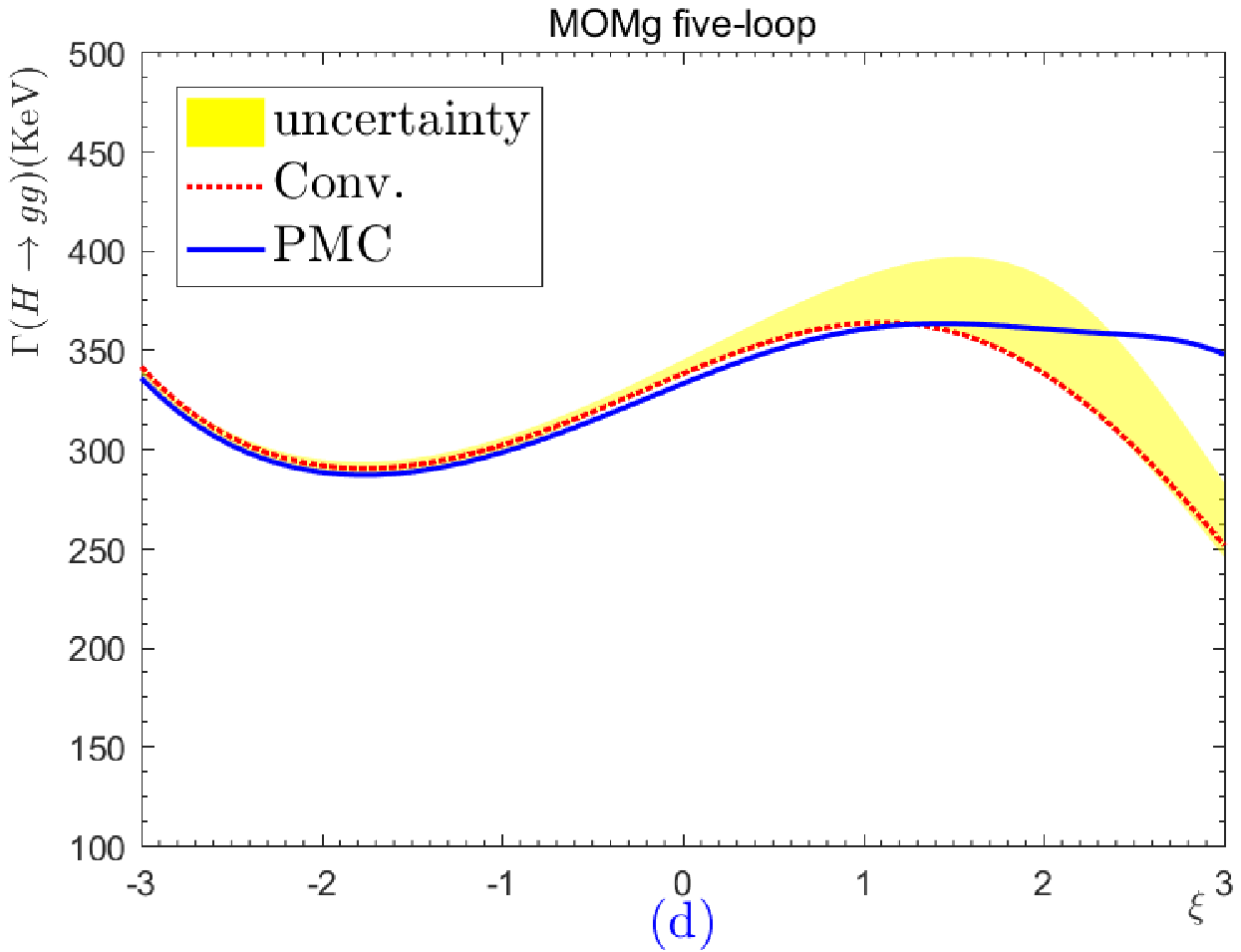}
\caption{Total decay width $\Gamma  (H\to gg)$ versus gauge parameter $\xi=\xi^{\rm MOMg}$ up to two-loop, three-loop, four-loop, and five-loop levels, respectively, under the MOMg scheme. The dotted line is for conventional scale setting approach with $\mu = M_H$ and the shaded band shows its renormalization scale uncertainty by varying $\mu \in [M_H/4,4M_H]$. The solid line is the prediction for the PMC, which is independent to the choice of renormalization scale.}
\label{two-fiveloopMOMg}
\end{figure}

\begin{figure}[htb]
\centering
\includegraphics[width=0.235\textwidth]{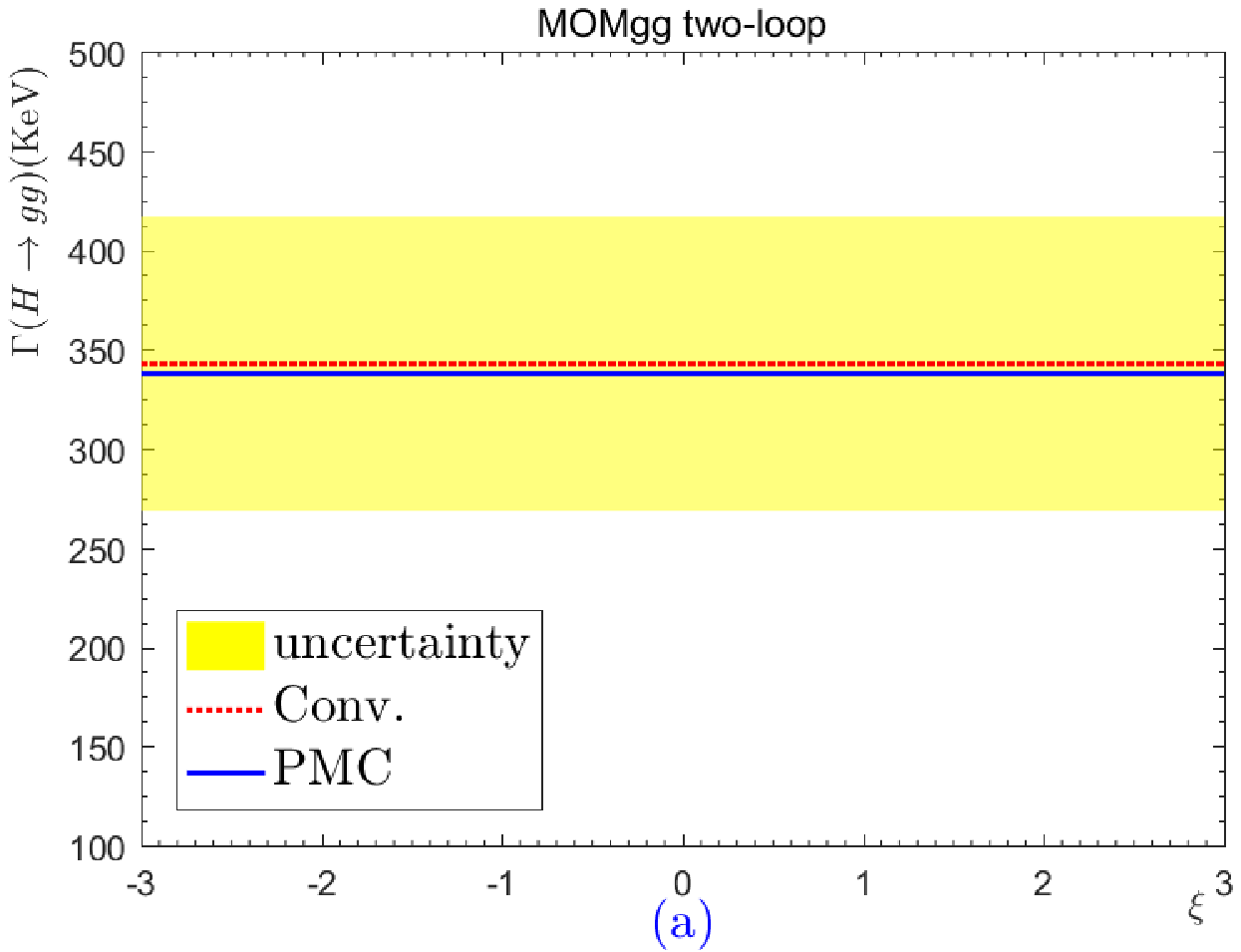}
\includegraphics[width=0.235\textwidth]{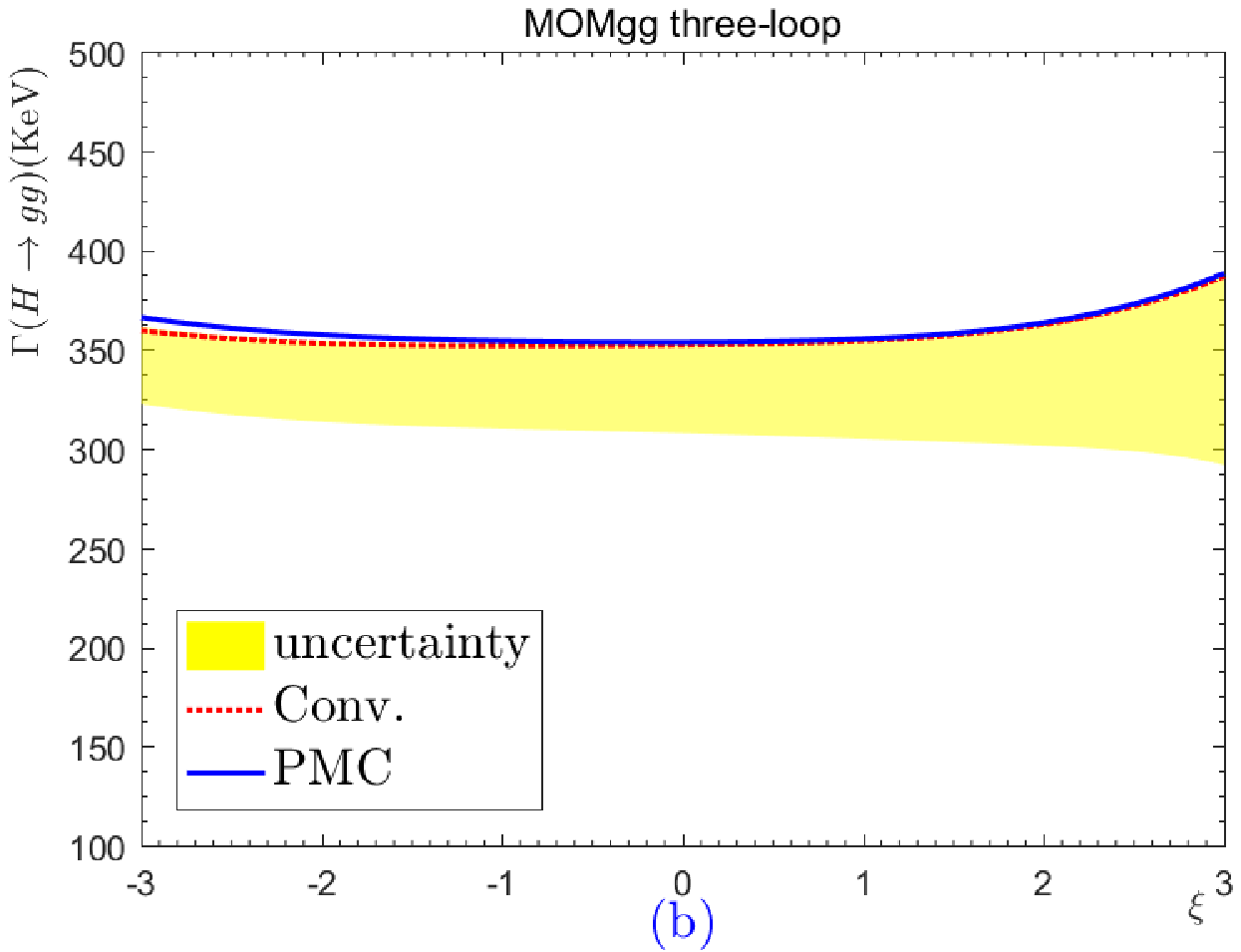}
\includegraphics[width=0.235\textwidth]{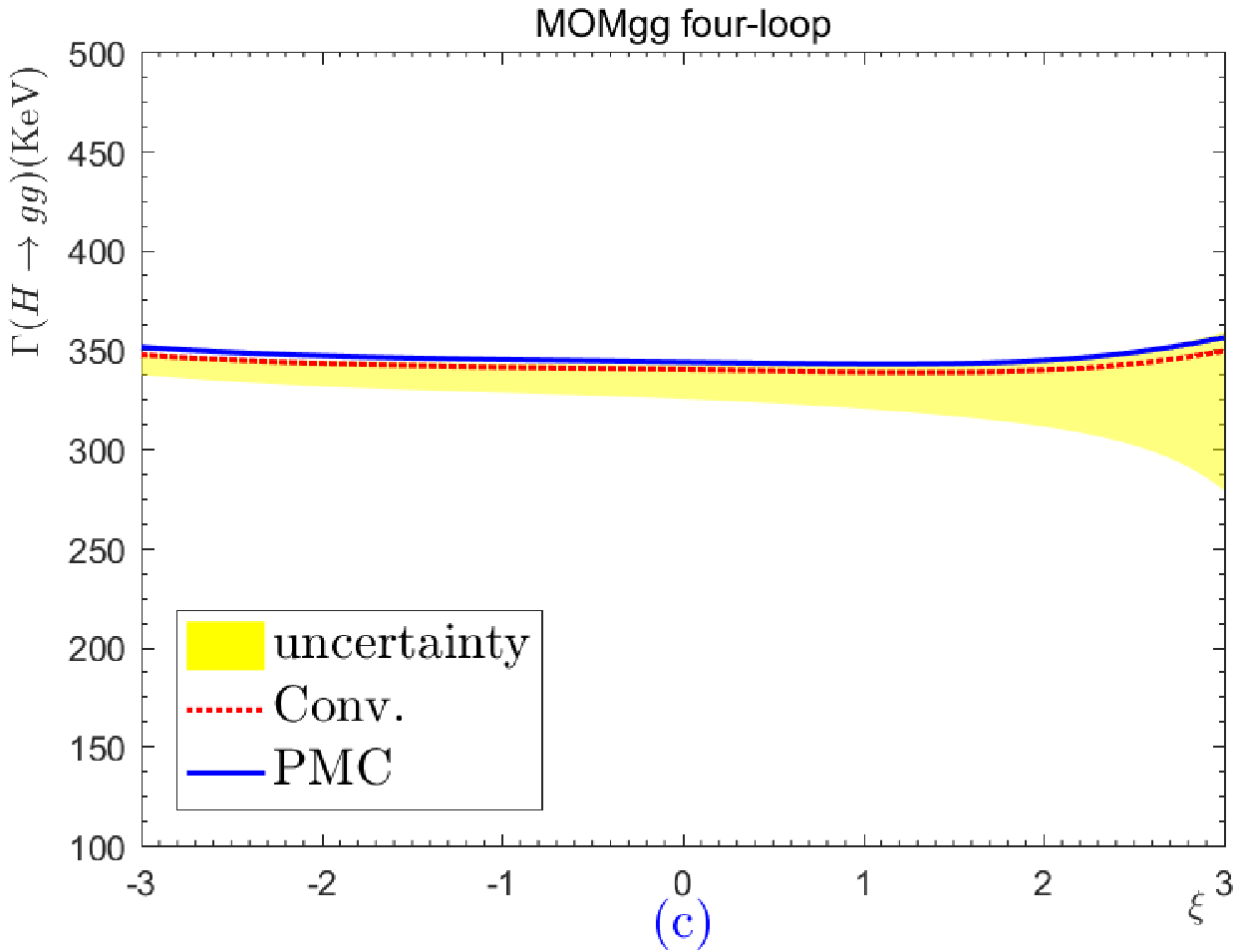}
\includegraphics[width=0.235\textwidth]{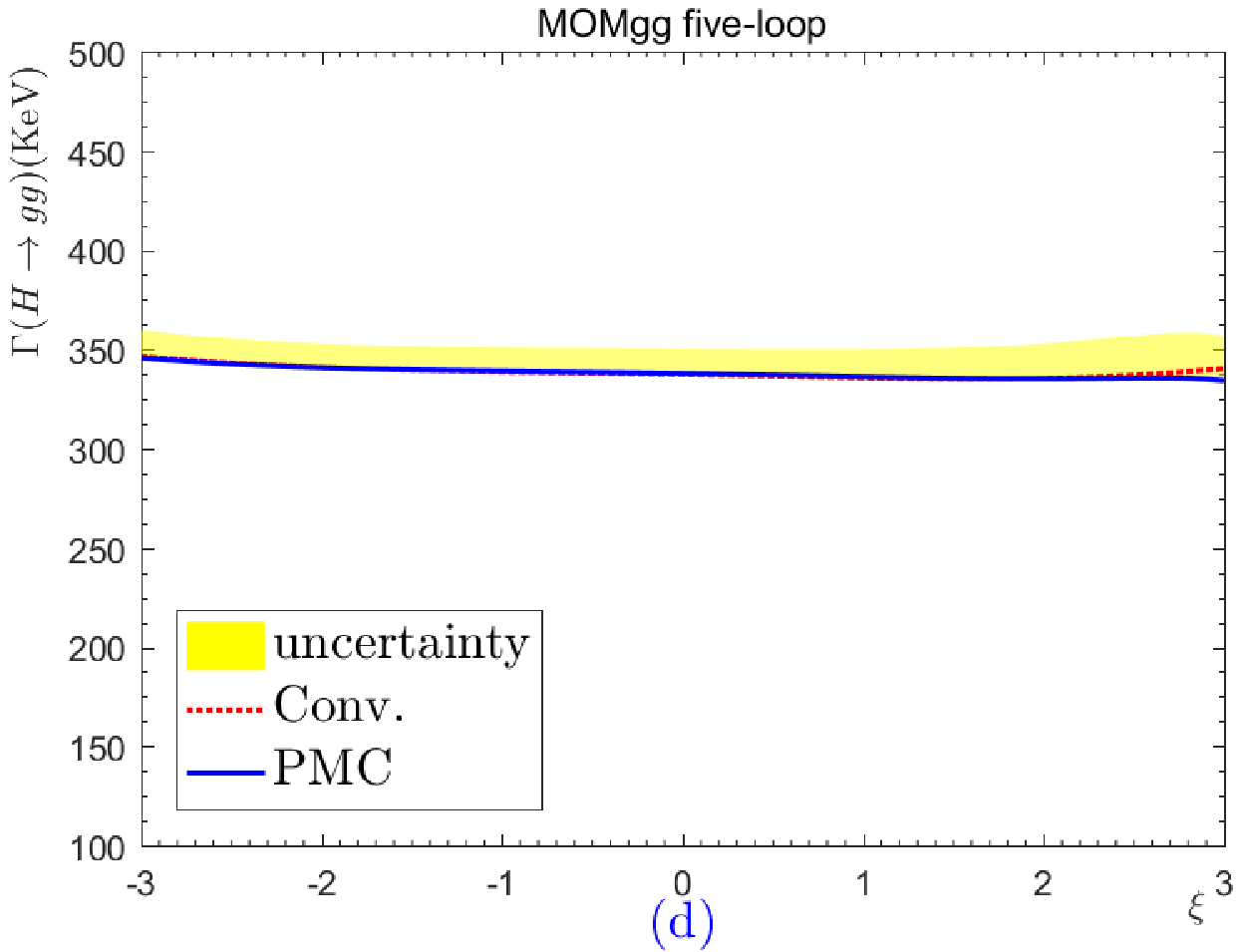}
\caption{Total decay width $\Gamma  (H\to gg)$ versus gauge parameter $\xi=\xi^{\rm MOMgg}$ up to two-loop, three-loop, four-loop, and five-loop levels, respectively, under the MOMg scheme. The dotted line is for conventional scale setting approach with $\mu = M_H$ and the shaded band shows its renormalization scale uncertainty by varying $\mu \in [M_H/4,4M_H]$. The solid line is the prediction for the PMC, which is independent to the choice of renormalization scale.}
\label{two-fiveloopMOMgg}
\end{figure}

We present the total decay width $\Gamma (H \to gg)$ up to two-loop, three-loop, four-loop and five-loop level before and after applying the PMC in Figs.(\ref{two-fiveloop}, \ref{two-fiveloopMOMh}, \ref{two-fiveloopMOMq}, \ref{two-fiveloopMOMg}, \ref{two-fiveloopMOMgg}). Agreeing with conventional wisdom, the conventional renormalization scale dependence, estimated by varying $\mu \in [M_H/4,4M_H]$, becomes smaller when more loop terms have been included, e.g. the shaded bands becomes narrower when more loop terms have been included. At the same time, one may observe that the PMC predictions under various MOM schemes are scale independent at any fixed order, but becomes more accurate when more loop terms have been included. Thus the conventional renormalization scale uncertainty is eliminated by applying the PMC, which is consistent with previous PMC examples done in the literature. As mentioned in the Introduction, the scale independence of the PMC prediction is reasonable, since the determined scale $\bar{Q}$ reflects the overall typical momentum flow of the process which should be independent to the induced parameters.

\begin{table}[htb]
\begin{center}
\begin{tabular}{  c c c  c  c c c c c c }
\hline
& ~$ \xi^{\rm MOM}$~          & ~$ -3 $~        & ~$ -2 $~    & ~$ -1 $~        & ~$0$~      & ~$1$~    & ~$2$~ & ~$3$    \\
\hline
& $\Gamma|^{\rm{mMOM}}$  &627.3  & 398.7  &338.5  &337.7 &350.1 &336.9  &273.6 \\
\hline
& $\Gamma|^{\rm{MOMh}}$  & 341.3  &291.5  &301.8  &337.7 &363.2 &338.1  &252.5 \\
\hline
& $\Gamma|^{\rm{MOMq}}$  & 341.3  &291.4
 &301.8  &337.7 &363.3 &338.1  &252.1 \\
\hline
& $\Gamma|^{\rm{MOMg}}$  & 341.1  &291.4  &301.8  &337.7 &363.2 &337.9  &251.2 \\
\hline
& $\Gamma|^{\rm{MOMgg}}$  & 346.4  &340.9  &338.8  &337.4 &335.7 &335.3  &340.3 \\
\hline
\end{tabular}
\caption{Total decay width (in unit: KeV) of $H \to gg$ under conventional scale-setting approach up to five-loop level under the mMOM, MOMh, MOMq, MOMg, and MOMgg schemes. Typical gauge of conventional scale setting. All the other input parameters are set to be their central values.}
\label{tableconv}
\end{center}
\end{table}

\begin{table}[htb]
\begin{center}
\begin{tabular}{  c c c  c  c c c c c c }
\hline
& ~$ \xi^{\rm MOM}$~          & ~$ -3 $~        & ~$ -2 $~    & ~$ -1 $~        & ~$0$~      & ~$1$~    & ~$2$~ & ~$3$    \\
\hline
& $\Gamma|^{\rm{mMOM}}_{\rm PMC}$  & 590.9  &389.8  &333.4  &332.8 &344.3 &327.7  &249.8 \\
\hline
& $\Gamma|^{\rm{MOMh}}$  & 335.8  &288.2  &298.2  &332.8 &360.3 &360.7  &361.4 \\
\hline
& $\Gamma|^{\rm{MOMq}}$  & 335.8  &288.2
 &298.2  &332.9 &360.2 &359.2  &352.6 \\
\hline
& $\Gamma|^{\rm{MOMg}}$  & 335.5  &288.0  &298.1  &332.7 &360.1 &360.2  &347.5 \\
\hline
& $\Gamma|^{\rm{MOMgg}}$  & 345.7  &340.8  &339.1  &337.9 &336.2 &335.2  &334.2 \\
\hline
\end{tabular}
\caption{Total decay width (in unit: KeV) of $H \to gg$ under PMC scale-setting approach up to five-loop level under the mMOM, MOMh, MOMq, MOMg, and MOMgg schemes. Typical gauge of conventional scale setting. All the other input parameters are set to be their central values.}
\label{tablepmc}
\end{center}
\end{table}

Figs.(\ref{two-fiveloop}, \ref{two-fiveloopMOMh}, \ref{two-fiveloopMOMq}, \ref{two-fiveloopMOMg}, \ref{two-fiveloopMOMgg}) show that the gauge dependence cannot be eliminated by including more and more higher-order terms for both the conventional and the PMC scale-setting approaches. More explicitly, the total decay width of $H \to gg$ up to five-loop level under the mMOM, MOMh, MOMq, MOMg, and MOMgg schemes are presented in Tables \ref{tableconv} and \ref{tablepmc}. The present prediction of $\Gamma(H \to gg)$ by the ``LHC Higgs Cross Section Working Group'' is about $335.4~\rm{KeV}$ for $M_H=125.09$ GeV~\footnote{This value can be extracted from the Webpage: https://twiki .cern.ch/twiki/bin/view/LHCPhysics/CERNYellowReportPageBR}. In different to the dimensional like schemes which are gauge independent, the gauge dependence is the intrinsic nature of MOM schemes and one cannot expect to eliminate it by including higher order terms. As discussed in Introduction, the MOM schemes have some advantages in dealing with the perturbative series; And as will be shown below, the scheme independence of variance MOM predictions can be greatly suppressed at higher orders. It is thus interesting to find in which region of $\xi^{\rm MOM}$, the MOM prediction is more reliable. Being consistent with the suggestion of Ref.\cite{Braaten:1981dv}, Figs.(\ref{two-fiveloop}, \ref{two-fiveloopMOMh}, \ref{two-fiveloopMOMq}, \ref{two-fiveloopMOMg}, \ref{two-fiveloopMOMgg}) show that the gauge dependence is relatively weaker within the region of $\xi^{\rm MOM} \in [-1,1]$. As an exception, Fig.\ref{two-fiveloopMOMgg} indicates that at enough higher orders, the MOMgg prediction could be unchanged for a wide range of $|\xi^{\rm MOMgg}|$.Thus among all the MOM schemes, we prefer the MOMgg scheme.  More explicitly, the coefficient functions $C_{i\geq2}$ in Eq.(\ref{conv-formulae}) and $r_{i\geq2,0}$ in Eq.(\ref{PMCrij}) are series in powers of $\xi^{\rm MOM}$, thus a small magnitude such as $|\xi^{\rm MOM}|\leq1$ could lead to a better convergence and a more steady prediction over the change of $\xi^{\rm MOM}$; this is why the Landau gauge with $\xi^{\rm MOM}=0$ is usually adopted in the literature~\cite{Celmaster:1979km}. We also agree that it is better to choose a smaller value of $|\xi^{\rm MOM}|$ for various MOM schemes. And in the following discussion, we shall adopt $\xi^{\rm MOM} \in [-1,1]$ to do our discussion.

To show how the scheme-and-scale dependence varies with the increasing orders more explicitly, under conventional scale-setting approach, we obtain
\begin{eqnarray*}
\Gamma(H \to gg)|^{\rm{mMOM}, 2l}_{\rm{Conv.}}&=&332.9{^{+25.9}_{-7.2}}\pm7.7\pm1.7{^{+77.7}_{-72.0}}~\rm{KeV}, \\
\Gamma(H \to gg)|^{\rm{mMOM}, 3l}_{\rm{Conv.}}&=&350.6{^{+16.9}_{-5.0}}\pm8.4\pm1.8{^{+2.6}_{-33.4}}~\rm{KeV}, \\
\Gamma(H \to gg)|^{\rm{mMOM}, 4l}_{\rm{Conv.}}&=&341.4{^{+12.7}_{-3.9}}\pm7.9\pm1.8{^{+1.1}_{-16.3}}~\rm{KeV}, \\
\Gamma(H \to gg)|^{\rm{mMOM}, 5l}_{\rm{Conv.}}&=&337.7{^{+12.5}_{-3.8}}\pm7.7\pm1.7{^{+7.2}_{-1.4}}~\rm{KeV}.
\end{eqnarray*}
\begin{eqnarray*}
\Gamma(H \to gg)|^{\rm{MOMh}, 2l}_{\rm{Conv.}}&=&332.9{^{+57.0}_{-55.3}}\pm7.7\pm1.7{^{+77.7}_{-72.0}}~\rm{KeV}, \\
\Gamma(H \to gg)|^{\rm{MOMh}, 3l}_{\rm{Conv.}}&=&350.6{^{+33.8}_{-42.0}}\pm8.4\pm1.8{^{+2.6}_{-33.4}}~\rm{KeV}, \\
\Gamma(H \to gg)|^{\rm{MOMh}, 4l}_{\rm{Conv.}}&=&341.4{^{+24.1}_{-36.8}}\pm7.9\pm1.8{^{+1.1}_{-16.3}}~\rm{KeV}, \\
\Gamma(H \to gg)|^{\rm{MOMh}, 5l}_{\rm{Conv.}}&=&337.7{^{+25.6}_{-35.9}}\pm7.7\pm1.7{^{+7.2}_{-1.4}}~\rm{KeV}.
\end{eqnarray*}
\begin{eqnarray*}
\Gamma(H \to gg)|^{\rm{MOMq}, 2l}_{\rm{Conv.}}&=&332.9{^{+57.0}_{-55.3}}\pm7.7\pm1.7{^{+77.7}_{-72.0}}~\rm{KeV}, \\
\Gamma(H \to gg)|^{\rm{MOMq}, 3l}_{\rm{Conv.}}&=&350.2{^{+33.8}_{-42.0}}\pm8.4\pm1.8{^{+2.6}_{-33.3}}~\rm{KeV}, \\
\Gamma(H \to gg)|^{\rm{MOMq}, 4l}_{\rm{Conv.}}&=&341.4{^{+24.2}_{-36.8}}\pm7.9\pm1.8{^{+1.0}_{-16.0}}~\rm{KeV}, \\
\Gamma(H \to gg)|^{\rm{MOMq}, 5l}_{\rm{Conv.}}&=&337.7{^{+25.7}_{-35.9}}\pm7.7\pm1.7{^{+7.1}_{-1.4}}~\rm{KeV}.
\end{eqnarray*}
\begin{eqnarray*}
\Gamma(H \to gg)|^{\rm{MOMg}, 2l}_{\rm{Conv.}}&=&332.9{^{+57.0}_{-55.3}}\pm7.7\pm1.7{^{+77.7}_{-72.0}}~\rm{KeV}, \\
\Gamma(H \to gg)|^{\rm{MOMg}, 3l}_{\rm{Conv.}}&=&350.4{^{+33.8}_{-42.0}}\pm8.4\pm1.8{^{+2.6}_{-33.4}}~\rm{KeV}, \\
\Gamma(H \to gg)|^{\rm{MOMg}, 4l}_{\rm{Conv.}}&=&341.4{^{+24.2}_{-36.8}}\pm7.9\pm1.8{^{+1.0}_{-16.2}}~\rm{KeV}, \\
\Gamma(H \to gg)|^{\rm{MOMg}, 5l}_{\rm{Conv.}}&=&337.7{^{+25.6}_{-35.9}}\pm7.7\pm1.7{^{+7.1}_{-1.4}}~\rm{KeV}.
\end{eqnarray*}
\begin{eqnarray*}
\Gamma(H \to gg)|^{\rm{MOMgg}, 2l}_{\rm{Conv.}}&=&342.7^{+0.0}_{-0.0}\pm8.1\pm1.8{^{+74.3}_{-74.0}}~\rm{KeV}, \\
\Gamma(H \to gg)|^{\rm{MOMgg}, 3l}_{\rm{Conv.}}&=&352.2{^{+2.2}_{-0.5}}\pm8.5\pm1.8{^{+0.9}_{-44.4}}~\rm{KeV}, \\
\Gamma(H \to gg)|^{\rm{MOMgg}, 4l}_{\rm{Conv.}}&=&339.8{^{+1.3}_{-1.3}}\pm7.8\pm1.8{^{+2.3}_{-14.8}}~\rm{KeV}, \\
\Gamma(H \to gg)|^{\rm{MOMgg}, 5l}_{\rm{Conv.}}&=&337.4{^{+1.5}_{-1.7}}\pm7.7\pm1.7{^{+12.9}_{-0.7}}~\rm{KeV}.
\end{eqnarray*}
The central values are for all input parameters to be their central values. Here,  $nl$ ($n=(2,...,5)$) stands for the result up to $n$-loop QCD corrections, and `Conv.' is the short notation for `Conventional scale-setting approach'. The first error shows the gauge dependence by varying $\xi^{\rm MOM}$ within the region of $[-1,1]$ (the central value is for the Landau gauge, $\xi^{\rm MOM}=0$); the second error is caused by the $\Delta\alpha_s(M_Z)=\pm0.0011$; the third error is caused by the Higgs mass uncertainty $\Delta M_H =\pm 0.24~\rm{GeV}$; the last error is caused by varying the renormalization scale $\mu \in [M_H/4,4M_H]$.

Under the convention scale-setting approach, the uncertainty caused by $\Delta\alpha_s(M_Z)$ is about $4\%$ for all orders. The uncertainty caused by the Higgs mass $\Delta M_H =0.24~\rm{GeV}$ is about $1\%$ for all order. The uncertainties caused by the choices of the gauge parameter and the renormalization scale are somewhat larger. By taking $\xi^{\rm MOM}\in[-1,1]$, the total decay width $\Gamma(H \to gg)|^{\rm mMOM}_{\rm Conv.}$ shall be changed by about $10\%$, $6\%$, $5\%$ and $5\%$ for $n=2,3,4,5$, respectively; the total decay widthes $\Gamma(H \to gg)|^{\rm MOMh, MOMq, MOMg}_{\rm Conv.}$ behave closely, which shall be changed by about $34\%$, $22\%$, $18\%$ and $18\%$ for $n=2,3,4,5$, respectively; and the total decay width $\Gamma(H \to gg)|^{\rm MOMgg}_{\rm Conv.}$ is free of gauge dependence at the two level, which shall be changed by about $1\%$ for $n=3,4,5$. Thus the gauge dependence of the MOMgg scheme is the smallest. And by taking $\mu\in [M_H/4,4M_H]$, the total decay width for mMOM, MOMh, MOMq and MOMg schemes behave closely, which shall be changed by about $45\%$, $10\%$, $5\%$ and $3\%$ for $n=2,3,4,5$, respectively; and the total decay width $\Gamma(H \to gg)|^{\rm MOMgg}_{\rm Conv.}$ shall be changed by about $43\%$, $13\%$, $5\%$ and $4\%$ for $n=2,3,4,5$, respectively. Thus by including more loop terms, the renormalization scale error does become smaller and smaller.

After applying the PMC, the renormalization scale dependence is exactly removed, and we have
\begin{eqnarray*}
\Gamma(H \to gg)|^{\rm{mMOM}, 2l}_{\rm{PMC}}&=&386.7{^{+29.4}_{-8.3}}\pm10.0\pm1.9~\rm{KeV}, \\
\Gamma(H \to gg)|^{\rm{mMOM}, 3l}_{\rm{PMC}}&=&345.9{^{+7.2}_{-3.0}}\pm8.1\pm1.8~\rm{KeV}, \\
\Gamma(H \to gg)|^{\rm{mMOM}, 4l}_{\rm{PMC}}&=&326.2{^{+4.6}_{-2.4}}\pm6.9\pm1.7~\rm{KeV}, \\
\Gamma(H \to gg)|^{\rm{mMOM}, 5l}_{\rm{PMC}}&=&332.8{^{+11.6}_{-3.7}}\pm7.3\pm1.7~\rm{KeV}.
\end{eqnarray*}
\begin{eqnarray*}
\Gamma(H \to gg)|^{\rm{MOMh}, 2l}_{\rm{PMC}}&=&386.7{^{+62.8}_{-66.0}}\pm10.0\pm1.9~\rm{KeV}, \\
\Gamma(H \to gg)|^{\rm{MOMh}, 3l}_{\rm{PMC}}&=&345.9{^{+10.7}_{-33.1}}\pm8.1\pm1.8~\rm{KeV}, \\
\Gamma(H \to gg)|^{\rm{MOMh}, 4l}_{\rm{PMC}}&=&326.2{^{+8.2}_{-28.2}}\pm6.9\pm1.7~\rm{KeV}, \\
\Gamma(H \to gg)|^{\rm{MOMh}, 5l}_{\rm{PMC}}&=&332.8{^{+27.5}_{-34.6}}\pm7.3\pm1.7~\rm{KeV}.
\end{eqnarray*}
\begin{eqnarray*}
\Gamma(H \to gg)|^{\rm{MOMq}, 2l}_{\rm{PMC}}&=&386.7{^{+62.8}_{-66.0}}\pm10.0\pm1.9~\rm{KeV}, \\
\Gamma(H \to gg)|^{\rm{MOMq}, 3l}_{\rm{PMC}}&=&345.6{^{+10.8}_{-33.2}}\pm8.1\pm1.8~\rm{KeV}, \\
\Gamma(H \to gg)|^{\rm{MOMq}, 4l}_{\rm{PMC}}&=&326.5{^{+8.4}_{-28.3}}\pm6.9\pm1.7~\rm{KeV}, \\
\Gamma(H \to gg)|^{\rm{MOMq}, 5l}_{\rm{PMC}}&=&332.9{^{+27.4}_{-34.7}}\pm7.3\pm1.7~\rm{KeV}.
\end{eqnarray*}
\begin{eqnarray*}
\Gamma(H \to gg)|^{\rm{MOMg}, 2l}_{\rm{PMC}}&=&386.7{^{+62.8}_{-66.0}}\pm10.0\pm1.9~\rm{KeV}, \\
\Gamma(H \to gg)|^{\rm{MOMg}, 3l}_{\rm{PMC}}&=&345.8{^{+10.5}_{-33.0}}\pm8.1\pm1.8~\rm{KeV}, \\
\Gamma(H \to gg)|^{\rm{MOMg}, 4l}_{\rm{PMC}}&=&326.0{^{+8.0}_{-28.1}}\pm6.8\pm1.7~\rm{KeV}, \\
\Gamma(H \to gg)|^{\rm{MOMg}, 5l}_{\rm{PMC}}&=&332.7{^{+27.5}_{-34.6}}\pm7.3\pm1.7~\rm{KeV}.
\end{eqnarray*}
\begin{eqnarray*}
\Gamma(H \to gg)|^{\rm{MOMgg}, 2l}_{\rm{PMC}}&=&337.8^{+0.0}_{-0.0}\pm7.9\pm1.7~\rm{KeV}, \\
\Gamma(H \to gg)|^{\rm{MOMgg}, 3l}_{\rm{PMC}}&=&353.6{^{+1.7}_{-0.1}}\pm8.6\pm1.8~\rm{KeV}, \\
\Gamma(H \to gg)|^{\rm{MOMgg}, 4l}_{\rm{PMC}}&=&343.6{^{+1.3}_{-1.1}}\pm8.1\pm1.8~\rm{KeV}, \\
\Gamma(H \to gg)|^{\rm{MOMgg}, 5l}_{\rm{PMC}}&=&337.9{^{+1.2}_{-1.7}}\pm7.7\pm1.8~\rm{KeV}.
\end{eqnarray*}
Similar to the results under conventional scale-setting approach, the uncertainties caused by $\Delta\alpha_s(M_Z)$ and $\Delta M_H$ are also small, which are around $4\%$ and $1\%$, respectively. However, the uncertainties caused by different choices of the gauge parameter are still sizable for various MOM schemes. For example, by taking $\xi^{\rm MOM}\in[-1,1]$, the total decay width $\Gamma(H \to gg)|^{\rm mMOM}_{\rm PMC}$ shall be changed by about $10\%$, $3\%$, $2\%$ and $5\%$ for $n=2,3,4,5$, respectively; the total decay widthes $\Gamma(H \to gg)|^{\rm MOMh, MOMq, MOMg}_{\rm PMC}$ behave closely, which shall be changed by about $33\%$, $12\%$, $11\%$ and $19\%$ for $n=2,3,4,5$, respectively; and the total decay width $\Gamma(H \to gg)|^{\rm MOMgg}_{\rm PMC}$ is almost independent to the choice of gauge parameter, which shall be changed by less than $1\%$ for $n=3,4,5$.

Moreover, by using the guessed scale, the convergence of the conventional pQCD series shall generally change greatly under different choices of the renormalization scale, due to the mismatching of the perturbative coefficient with the $\alpha_s$-value at the same order. For example, there is quite large scale uncertainty for each term of the pQCD series of $\Gamma(H\to gg)$~\cite{Zeng:2015gha, Zeng:2018jzf}; thus, even if by choosing a proper scale, a better convergence can be achieved~\footnote{In the literature, the renormalization scale is usually chosen as the one so as to eliminate large logs with the purpose of improving the pQCD convergence; And some scale-setting approaches have been invented to find an optimal scale with the purpose of improve the pQCD convergence but not to solve the renormalization scale ambiguity.}, one cannot decide whether such a choice leads to the correct pQCD prediction. On the other hand, after applying the PMC, the scale-independent coupling $\alpha_s(\bar{Q})$ can be determined, and together with the scale-invariant conformal coefficients, one can thus obtain the intrinsic perturbative nature of the pQCD series. By defining a $K$ factor, $K =\Gamma(H \to gg)/\Gamma(H \to gg)|_{\rm{Born}}$, one can obtain the relative importance of the high-order terms to the leading-order terms. More explicitly, under the Landau gauge with $\xi^{\rm MOM}=0$, we obtain
\begin{eqnarray}
K^{\rm{mMOM}}  &\simeq& 1.07=1+0.31-0.17-0.08+0.01,\\
K^{\rm{MOMh}}   &\simeq& 1.07=1+0.31-0.17-0.08+0.01,\\
K^{\rm{MOMq}}   &\simeq& 1.07=1+0.31-0.17-0.08+0.01,\\
K^{\rm{MOMg}}   &\simeq& 1.06=1+0.31-0.17-0.09+0.01,\\
K^{\rm{MOMgg}} &\simeq& 1.45=1+0.53+0.04-0.08-0.04.
\end{eqnarray}
Those results show satisfactorily convergent behavior for $\Gamma(H\to gg)$, especially for the mMOM, MOMh, MOMq and MOMg schemes. Similar to the condition of total decay width, the $K$ factor is also gauge dependent. By varying $\xi^{\rm MOM}\in[-1,1]$, we obtain $K^{\rm{mMOM}}=1.07{^{+0.03}_{-0.10}}$, $K^{\rm{MOMh}}=1.07{^{+0.12}_{-0.19}}$, $K^{\rm{MOMq}}=1.07{^{+0.12}_{-0.19}}$, $K^{\rm{MOMg}}=1.06{^{+0.12}_{-0.19}}$ and $K^{\rm{MOMgg}}=1.45{^{+0.00}_{-0.02}}$.

As a final remark, one usually wants to know the magnitude of the ``unknown" high-order pQCD corrections. The conventional error estimate obtained by varying the scale over a certain range is usually treated as such an estimation, which is however unreliable, since it only partly estimates the non-conformal contribution but not the conformal one. In contrast, after applying the PMC, the correct momentum flow of the process and hence the correct $\alpha_s$-value is fixed by the RGE and cannot be varied; otherwise, one will explicitly break the RGI, leading to an unreliable prediction. As a conservative estimation of the magnitude of the unknown perturbative contributions for the PMC series, it is helpful to use the magnitude of the last known term as the contribution of the unknown perturbative term~\cite{Wu:2014iba}; As for the present case, we adopt $\pm |r_{5,0}a^6_s(\bar{Q})|$ as the estimation of the unknown ${\cal O}(\alpha_s^6)$ contribution, which is $\pm 3.1$ KeV for the mMOM, MOMh, MOMq and MOMg schemes, and $\pm 9.3$ KeV for MOMgg scheme.

\subsection{A simple discussion on the symmetric MOM schemes}

In addition to the asymmetric MOM schemes, several symmetric MOM schemes have also been suggested in the literature. In the original symmetric MOM scheme~\cite{Celmaster:1979km}, the triple-gluon vertex function $\Gamma_{\mu \nu \rho}^{a b c}(k, p, l)$ is defined to be the value at the symmetric point $k^2=p^2=l^2 = -\mu^2$, i.e. the Feynman diagram of the vertex is the same as Fig.\ref{tilde-triple-gluon-vetex} but should replace the three external momentum ($q, -q, 0$) there by the present ($k, p, l$). Similarly, the ghost-gluon vertex function $\tilde{\Gamma}_{\mu}^{a b c}(k, p, l)$ and the quark-gluon vertex function $\Lambda^a_{\mu,ij}(k, p, l)$ can also be defined at the symmetric point $k^2=p^2=l^2 = -\mu^2$~\cite{Chetyrkin:2000fd, Gracey:2011vw, Gracey:2014pba}. For simplicity, we label those symmetric MOM schemes which are defined at the triple-gluon vertex, the ghost-gluon vertex and the quark-gluon vertex as the $\rm{\overline{MOM}ggg}$, the $\rm{\overline{MOM}h}$ and the $\rm{\overline{MOM}q}$ schemes, respectively. Using the relations (\ref{alphasMM@MS}, \ref{axiMM@MS}), together with the known three-loop vertex functions under the $\overline{\rm MS}$ scheme collected in Ref.\cite{Gracey:2014pba}, one can obtain the expressions for the strong couplings and the gauge parameters of those symmetric MOM schemes up to three-loop level~\cite{Gracey:2011vw}~\footnote{The pQCD correction for the symmetric MOM scheme is much harder to be calculated than the case of asymmetric MOM scheme due to its complexity, and at present, those vertexes have only been calculated up to three-loop level.}. By further using those expressions, we can obtain pQCD series for $\Gamma(H \to gg)$ under the $\rm{\overline{MOM}ggg}$, the $\rm{\overline{MOM}h}$ and the $\rm{\overline{MOM}q}$ schemes up to three-loop level.

After applying the conventional and PMC scale-setting approaches, we obtain
\begin{eqnarray*}
\Gamma(H \to gg)|^{\rm{\overline{MOM}ggg}, 2l}_{\rm{Conv.}}&=&336.2{^{+36.2}_{-26.9}}\pm7.8\pm1.7{^{+76.7}_{-72.7}}~\rm{KeV}, \\
\Gamma(H \to gg)|^{\rm{\overline{MOM}ggg}, 3l}_{\rm{Conv.}}&=&349.5{^{+24.6}_{-14.5}}\pm8.3\pm1.8{^{+1.9}_{-32.1}}~\rm{KeV},  \\
\Gamma(H \to gg)|^{\rm{\overline{MOM}h}, 2l}_{\rm{Conv.}}&=&350.0{^{+29.3}_{-26.1}}\pm8.4\pm1.8{^{+70.5}_{-75.4}}~\rm{KeV},  \\
\Gamma(H \to gg)|^{\rm{\overline{MOM}h}, 3l}_{\rm{Conv.}}&=&351.6{^{+20.2}_{-20.1}}\pm8.4\pm1.8{^{+0.2}_{-55.9}}~\rm{KeV},  \\
\Gamma(H \to gg)|^{\rm{\overline{MOM}q}, 2l}_{\rm{Conv.}}&=&342.9{^{+42.8}_{-24.0}}\pm8.1\pm1.8{^{+74.2}_{-74.1}}~\rm{KeV}, \\
\Gamma(H \to gg)|^{\rm{\overline{MOM}q}, 3l}_{\rm{Conv.}}&=&342.5{^{+32.2}_{-14.4}}\pm8.0\pm1.8{^{+0.9}_{-35.0}}~\rm{KeV}
\end{eqnarray*}
and
\begin{eqnarray*}
\Gamma(H \to gg)|^{\rm{\overline{MOM}ggg}, 2l}_{\rm{PMC}}&=&291.5{^{+30.1}_{-22.6}}\pm6.1\pm1.5~\rm{KeV}, \\
\Gamma(H \to gg)|^{\rm{\overline{MOM}ggg}, 3l}_{\rm{PMC}}&=&323.6{^{+26.0}_{-14.2}}\pm7.1\pm1.7~\rm{KeV}, \\
\Gamma(H \to gg)|^{\rm{\overline{MOM}h}, 2l}_{\rm{PMC}}&=&402.2{^{+30.9}_{-29.8}}\pm10.8\pm2.0~\rm{KeV}, \\
\Gamma(H \to gg)|^{\rm{\overline{MOM}h}, 3l}_{\rm{PMC}}&=&332.5{^{+6.9}_{-11.1}}\pm7.3\pm1.7~\rm{KeV}, \\
\Gamma(H \to gg)|^{\rm{\overline{MOM}q}, 2l}_{\rm{PMC}}&=&396.1{^{+49.1}_{-26.9}}\pm10.5\pm2.0~\rm{KeV},\\
\Gamma(H \to gg)|^{\rm{\overline{MOM}q}, 3l}_{\rm{PMC}}&=&332.5{^{+25.4}_{-8.7}}\pm7.4\pm1.7~\rm{KeV}.
\end{eqnarray*}
The central values are for all input parameters to be their central values, and the errors are caused by the same way of choosing the input parameters as those of the asymmetric schemes.

\begin{figure}[htb]
\centering
\includegraphics[width=0.235\textwidth]{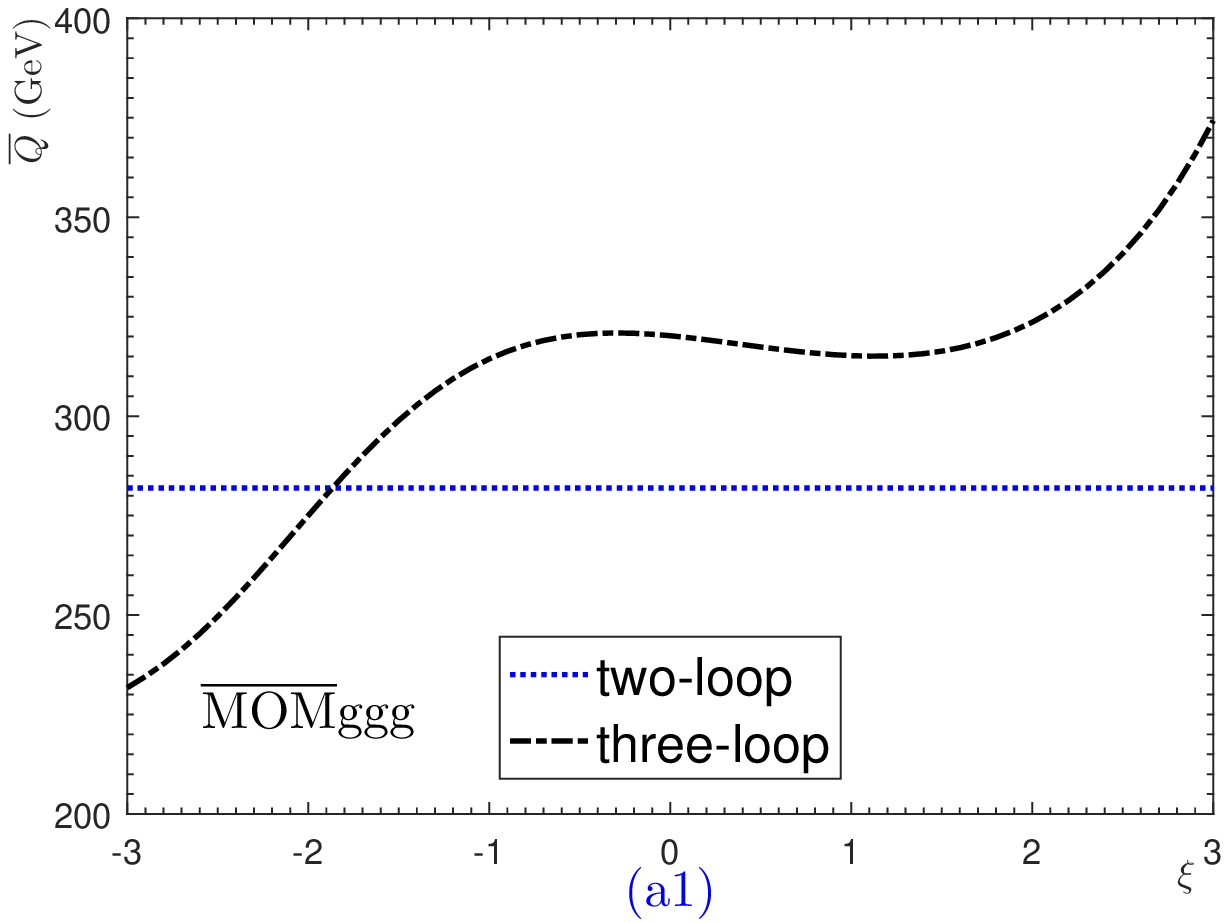}
\includegraphics[width=0.235\textwidth]{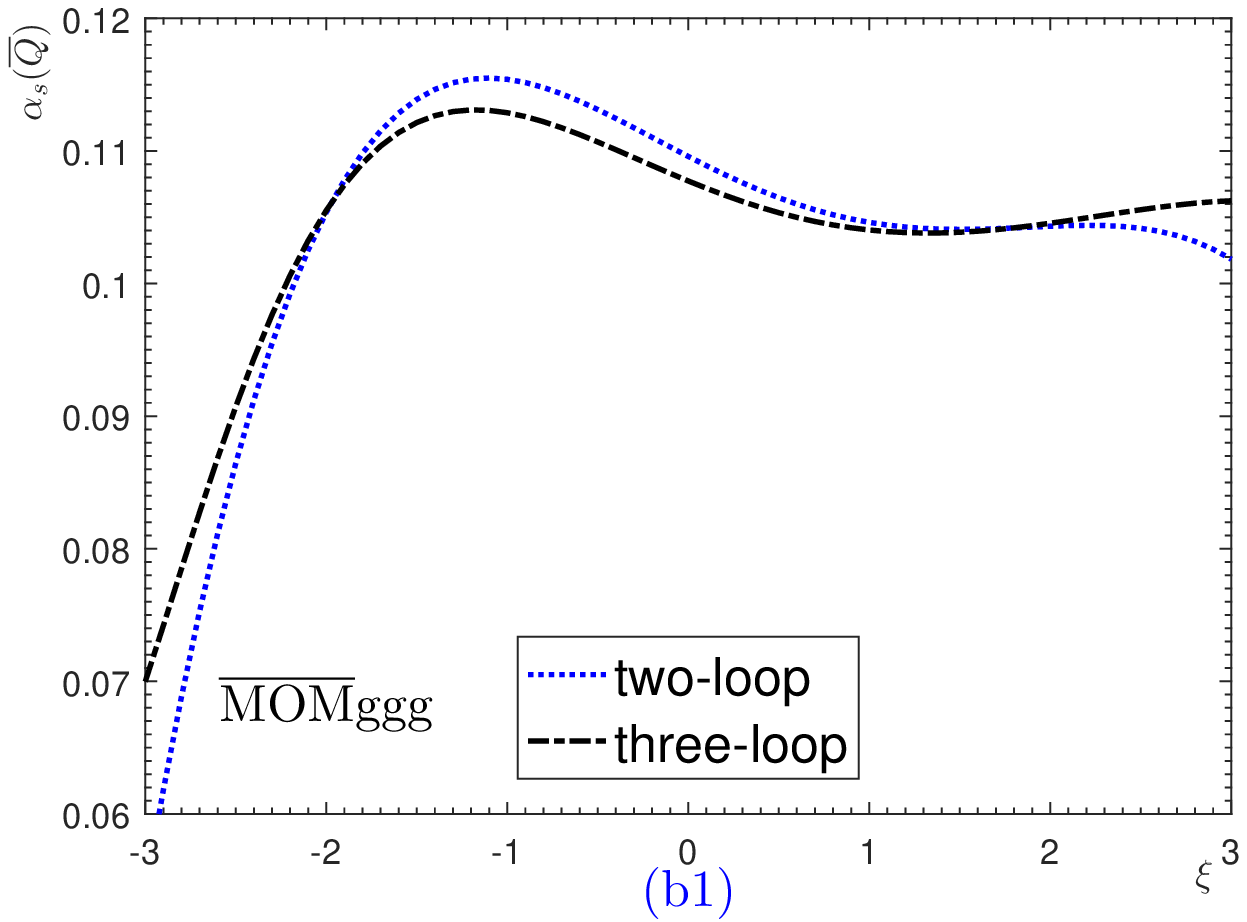}
\includegraphics[width=0.235\textwidth]{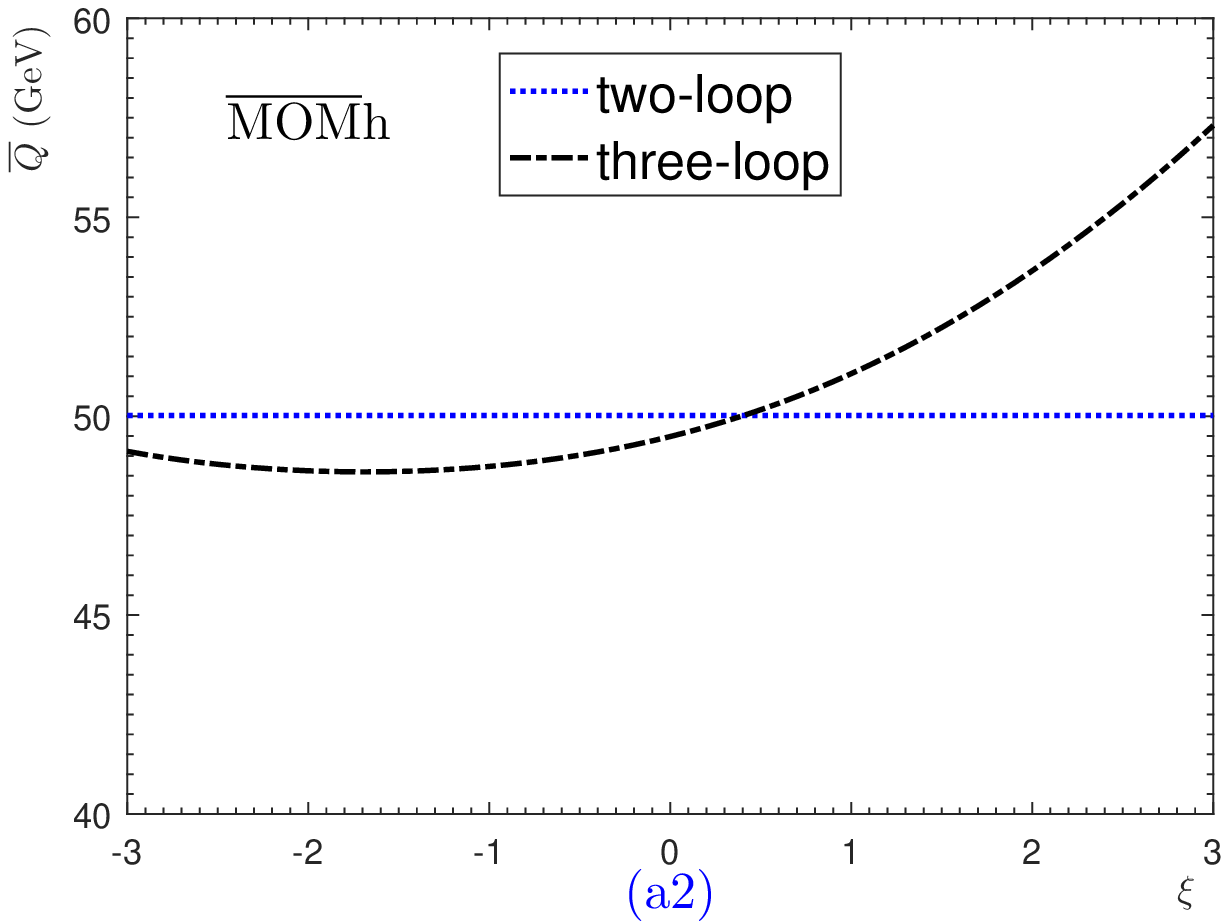}
\includegraphics[width=0.235\textwidth]{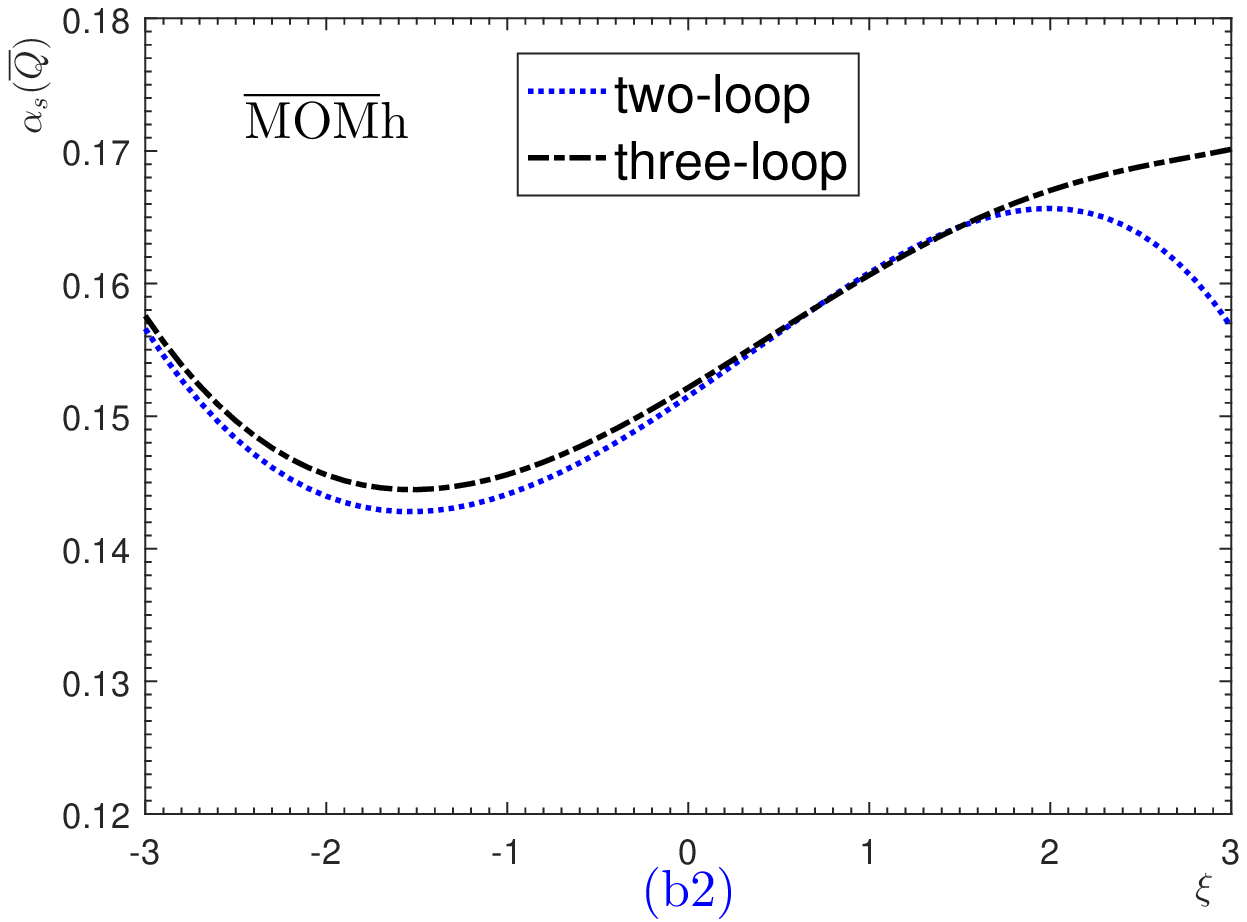}
\includegraphics[width=0.235\textwidth]{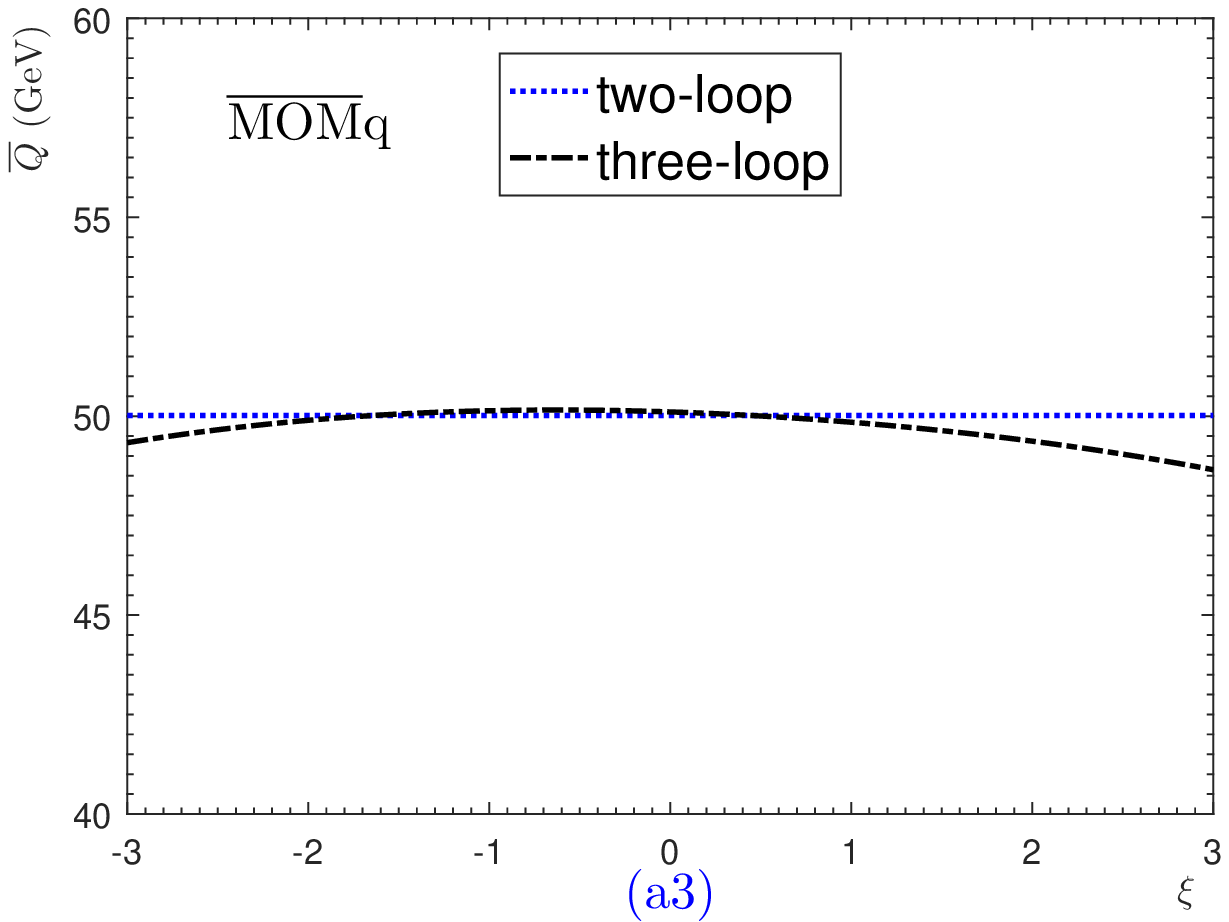}
\includegraphics[width=0.235\textwidth]{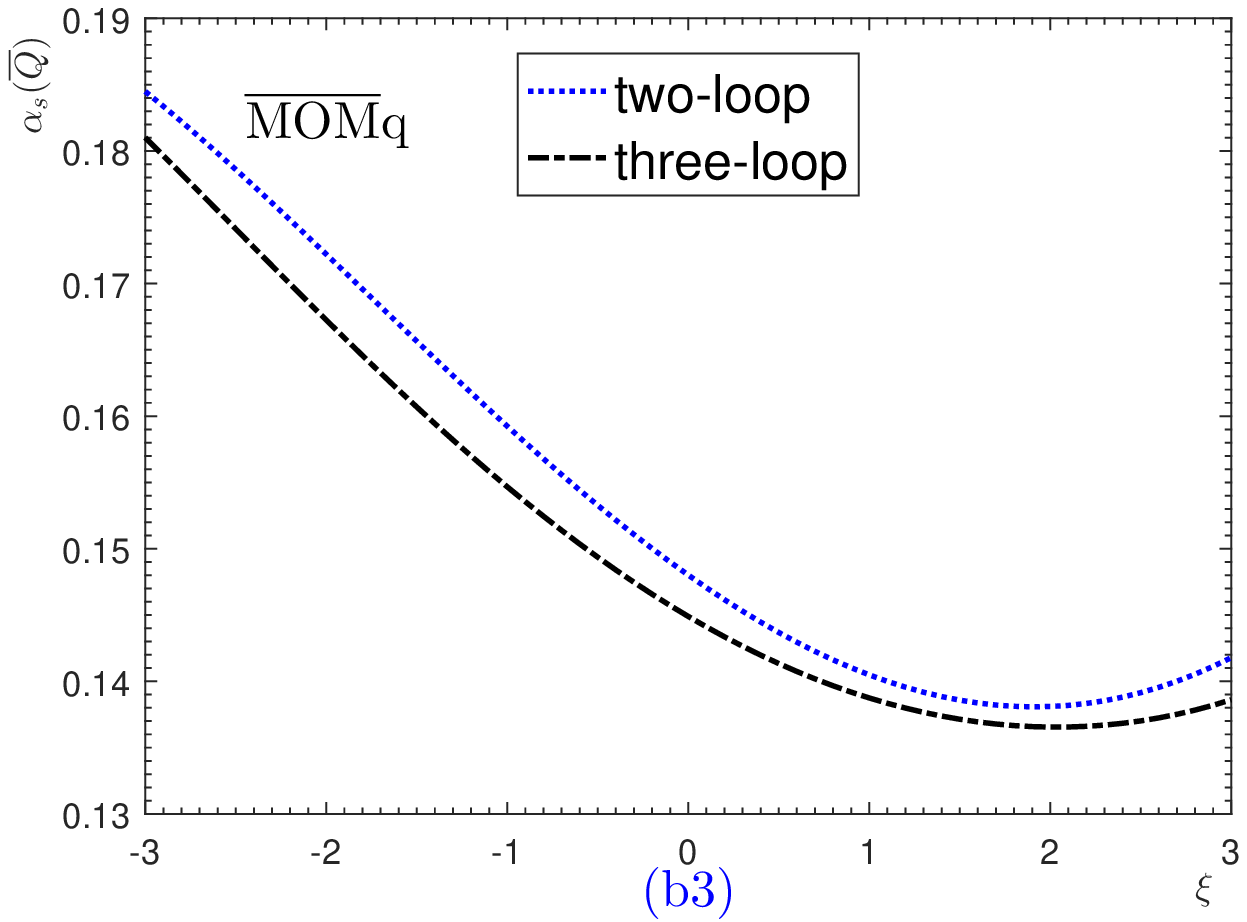}
\caption{The PMC effective scales $\bar{Q}$ (Left column) and their corresponding coupling constants $\alpha_s(\bar{Q})$ (Right column) versus the gauge parameter ($\xi$) of $\Gamma (H\to gg)$. Three symmetric MOM schemes, e.g. $\rm{\overline{MOM}ggg}$, $\rm{\overline{MOM}h}$ and $\rm{\overline{MOM}q}$ schemes, are adopted. The dotted and the dash-dot lines are results up to two-loop and three-loop QCD corrections, respectively.}
\label{single-scale-coupling-Sym-MOM}
\end{figure}

\begin{figure}[htb]
\centering
\includegraphics[width=0.235\textwidth]{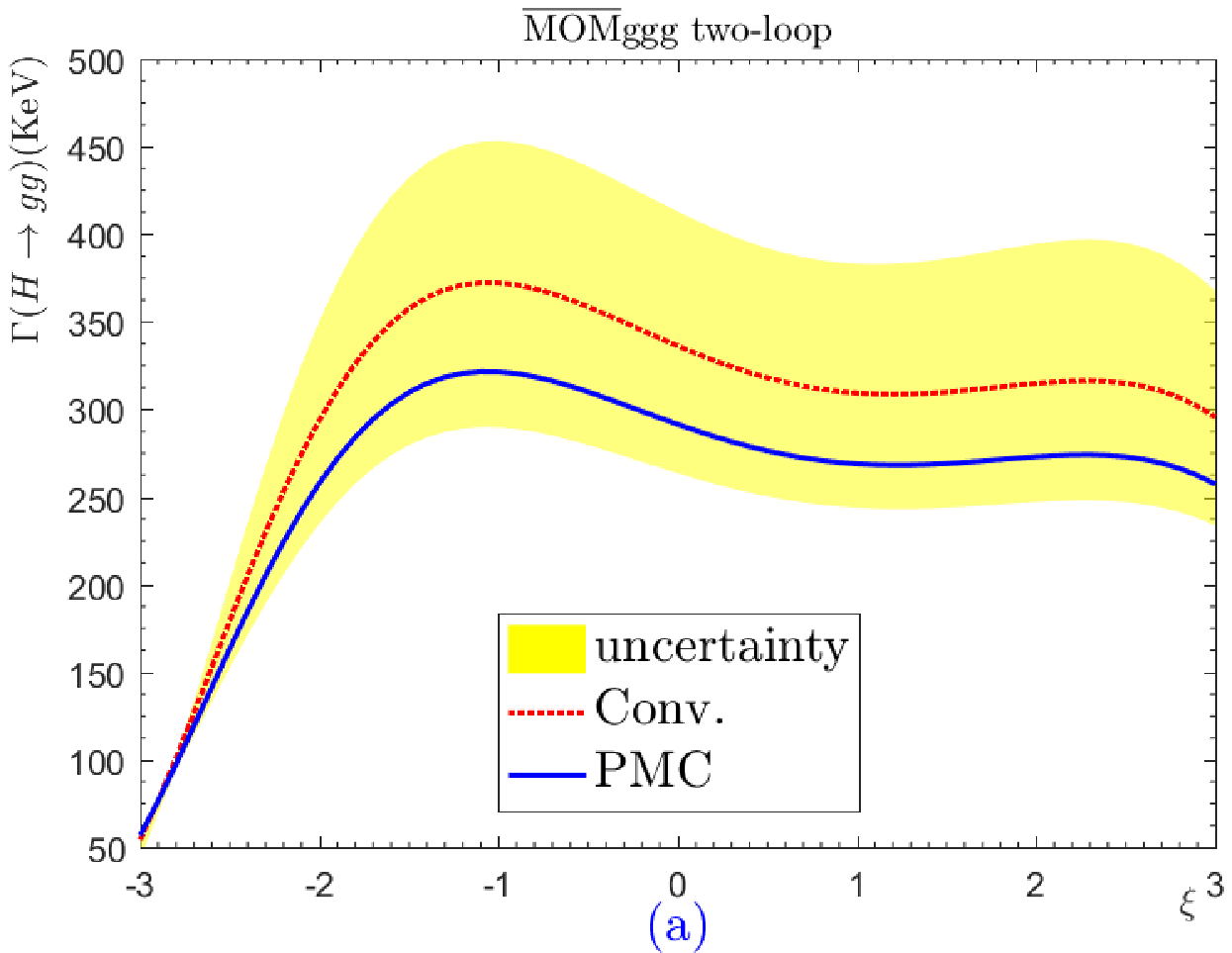}
\includegraphics[width=0.235\textwidth]{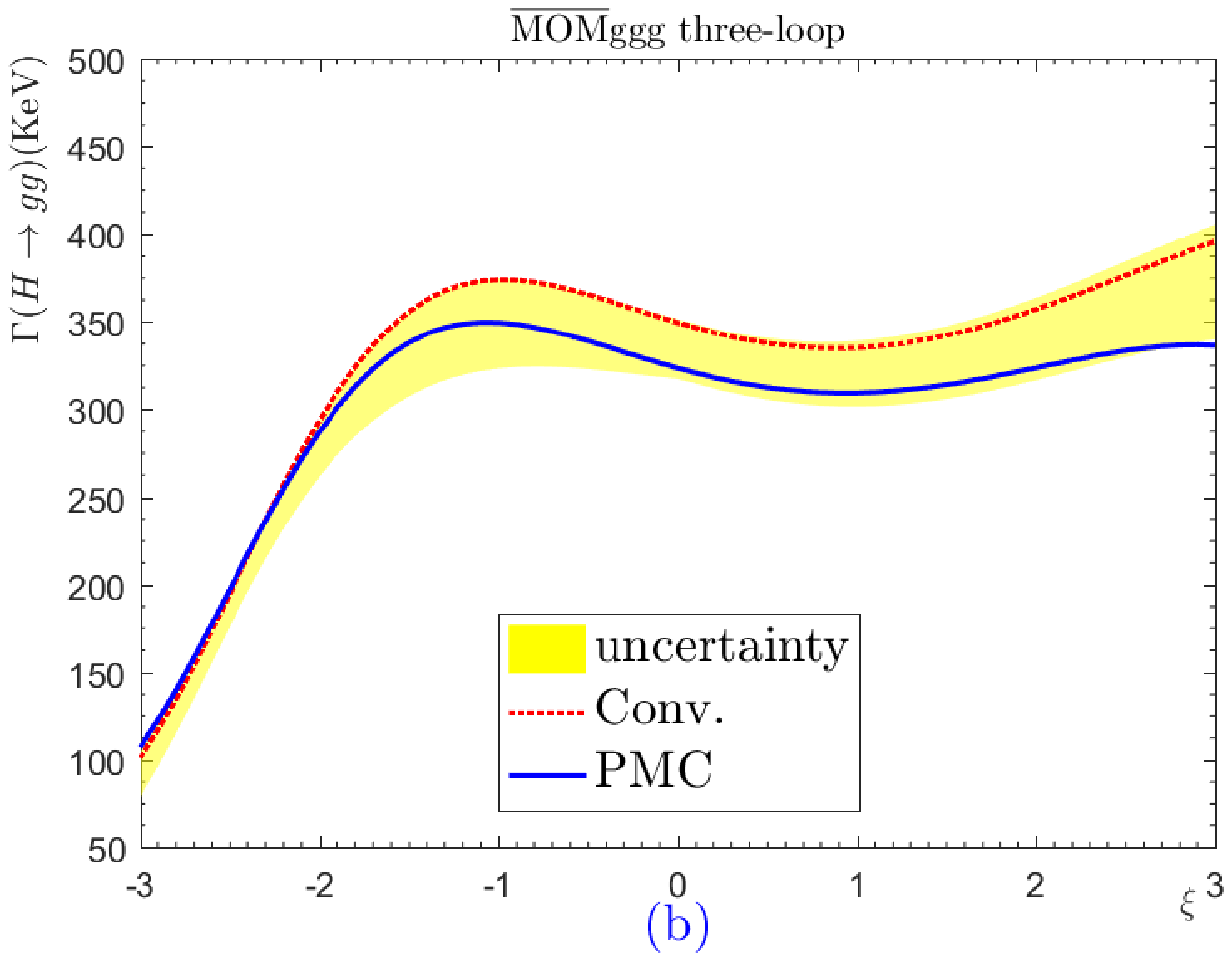}
\includegraphics[width=0.235\textwidth]{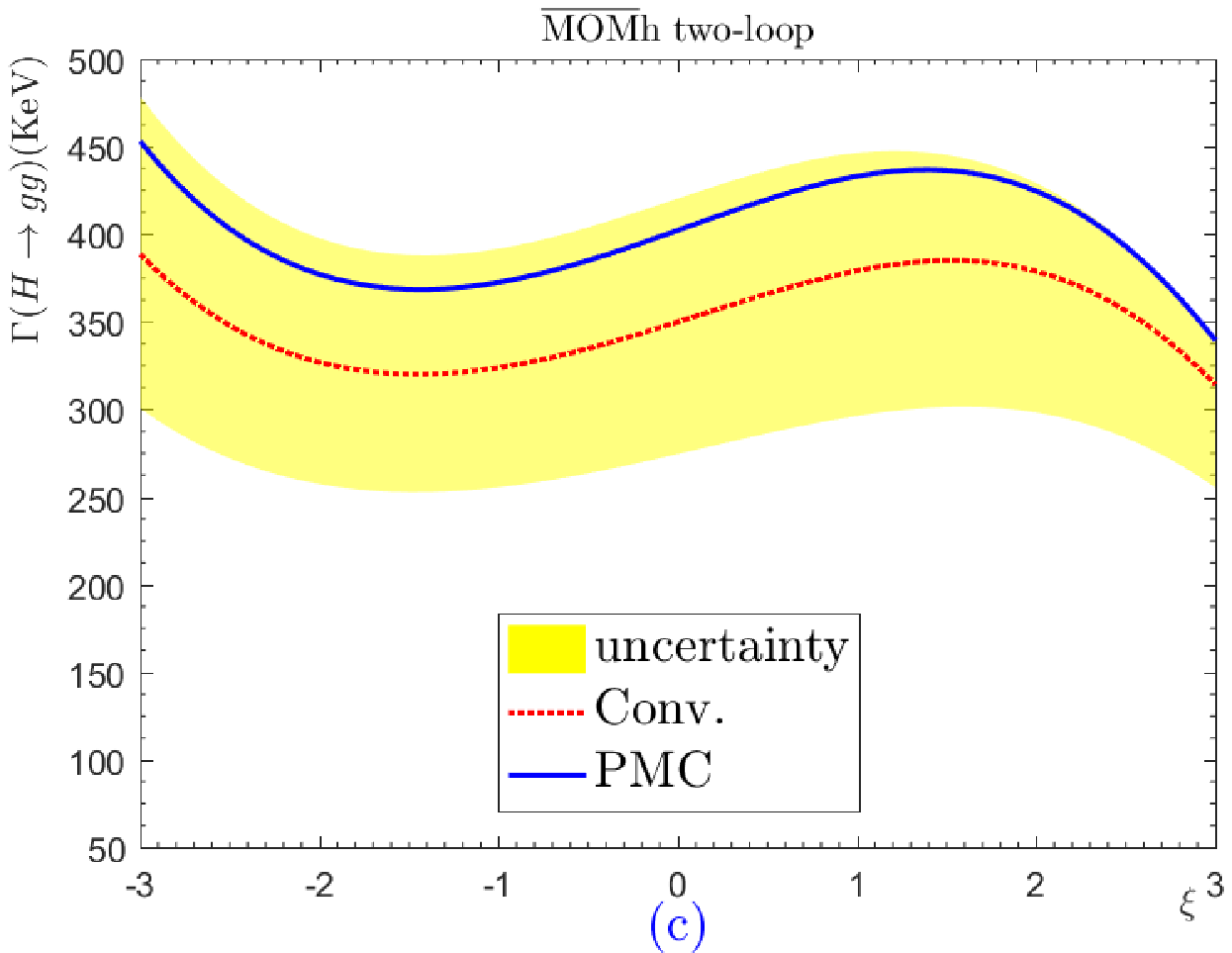}
\includegraphics[width=0.235\textwidth]{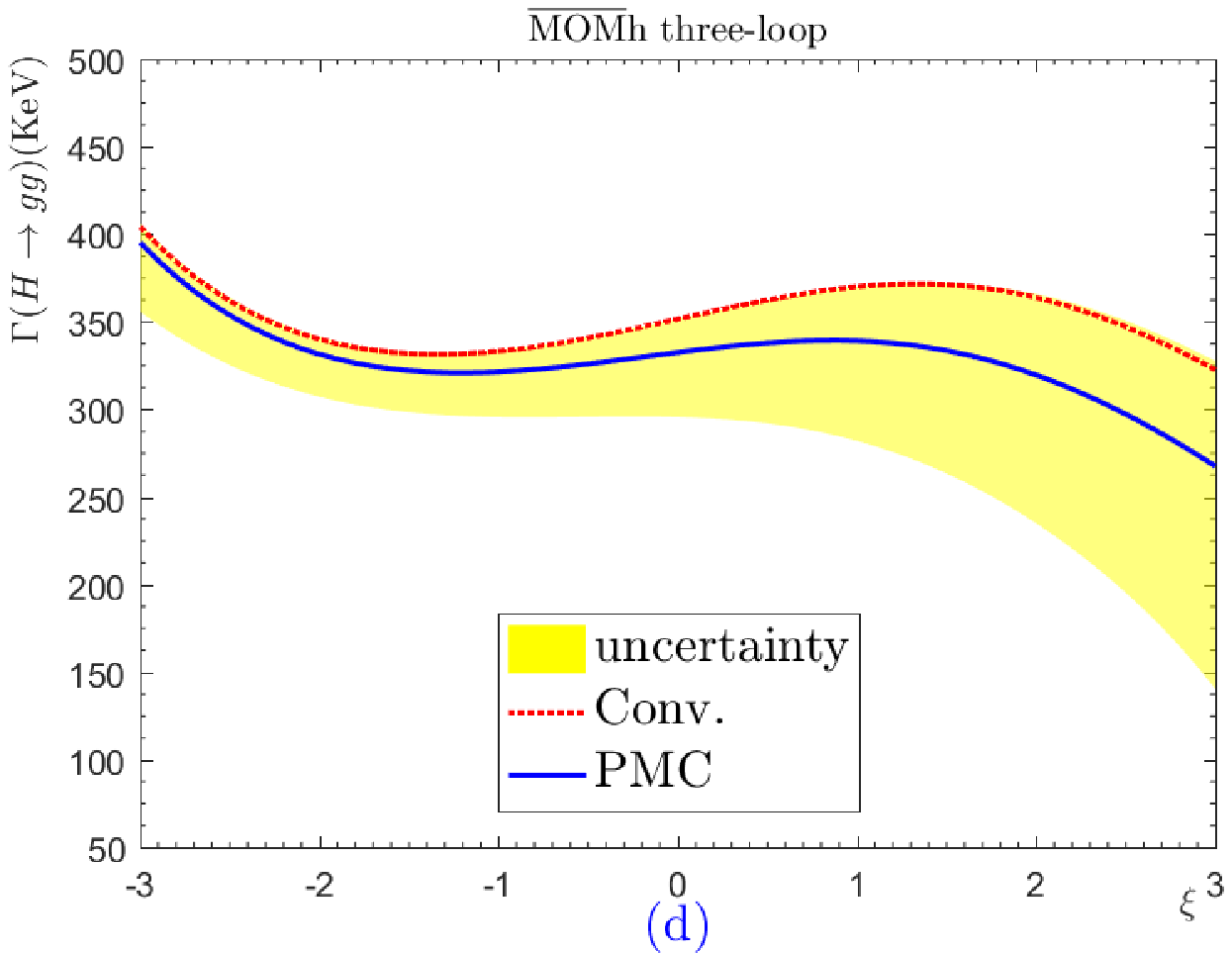}
\includegraphics[width=0.235\textwidth]{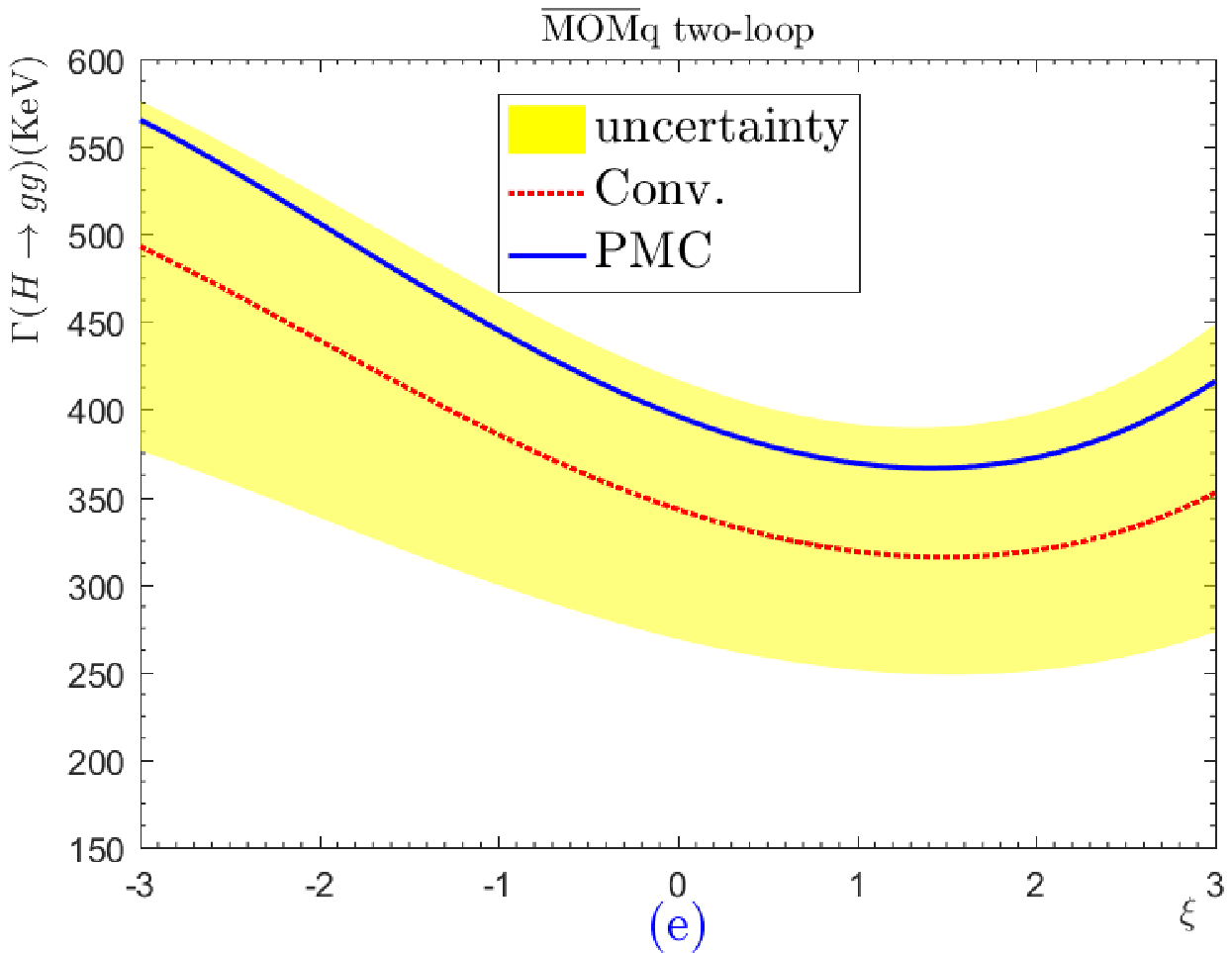}
\includegraphics[width=0.235\textwidth]{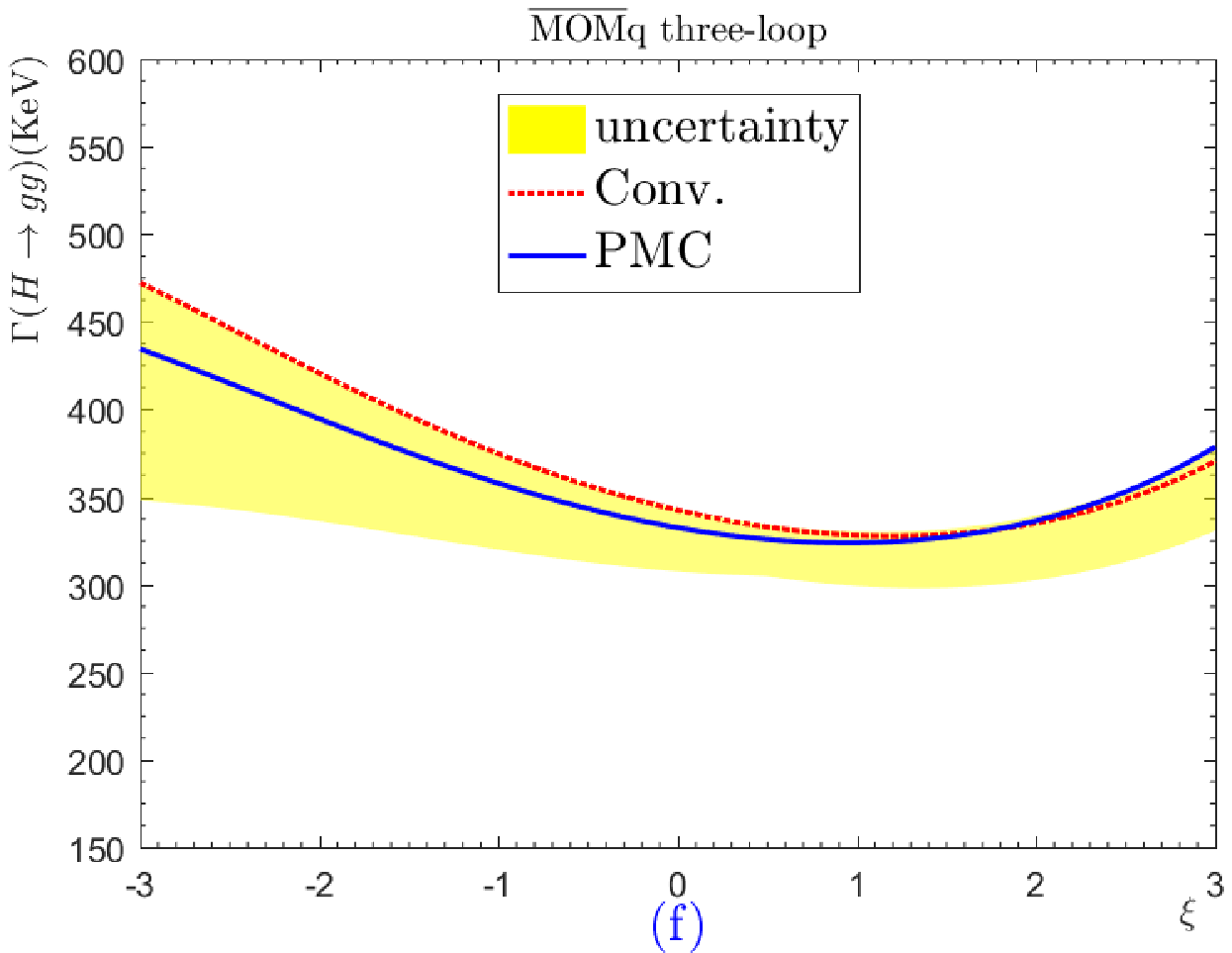}
\caption{Total decay width $\Gamma  (H\to gg)$ versus the gauge parameter ($\xi$) up to two-loop and three-loop levels under three symmetric MOM schemes. The dotted line is for conventional scale setting approach with $\mu = M_H$ and the shaded band shows its renormalization scale uncertainty by varying $\mu \in [M_H/4,4M_H]$. The solid line is the PMC prediction, which is independent to the choice of renormalization scale.}
\label{two-threeloopggg}
\end{figure}

\begin{table}[htb]
\begin{center}
\begin{tabular}{  c c c  c  c c c c c c }
\hline
& ~$ \xi^{\rm MOM}$~          & ~$ -3 $~        & ~$ -2 $~    & ~$ -1 $~        & ~$0$~      & ~$1$~    & ~$2$~ & ~$3$    \\
\hline
& $\Gamma|^{\rm{\overline{MOM}ggg}}_{\rm Conv.}$  & 101.1  &293.9  &374.1  &349.5 &335.3 &357.2  &395.9 \\
\hline
& $\Gamma|^{\rm{\overline{MOM}h}}_{\rm Conv.}$  & 404.3  &340.1 &333.3  &351.6 &370.0 &363.9  &322.6 \\
\hline
& $\Gamma|^{\rm{\overline{MOM}q}}_{\rm Conv.}$  & 472.3  &420.7  &374.7  &342.5 &328.2 &335.2  &370.6 \\
\hline
& $\Gamma|^{\rm{\overline{MOM}ggg}}_{\rm PMC}$  & 107.0  &287.3  &349.6  &323.6 &309.4 &323.8  &336.5 \\
\hline
& $\Gamma|^{\rm{\overline{MOM}h}}_{\rm PMC}$  & 395.2  &331.5 &321.5  &332.5 &339.2 &319.6  &267.5 \\
\hline
& $\Gamma|^{\rm{\overline{MOM}q}}_{\rm PMC}$  & 434.5  &394.7  &357.8  &332.5 &323.9 &336.5  &378.7 \\
\hline
\end{tabular}
\caption{Gauge dependence of the total decay width (in unit: KeV) of $H \to gg$ up to three-loop level under the $\rm{\overline{MOM}ggg}$, $\rm{\overline{MOM}h}$ and $\rm{\overline{MOM}q}$ schemes before and after applying the PMC. Other input parameters are set to be their central values.}
\label{tabletildeconv@pmc}
\end{center}
\end{table}

We present the PMC effective scales $\bar{Q}$ and their corresponding coupling constants $\alpha_s(\bar{Q})$ of $\Gamma (H\to gg)$ under three symmetric MOM schemes in Fig.(\ref{single-scale-coupling-Sym-MOM}). The total decay width $\Gamma  (H\to gg)$ versus the gauge parameter for the cases of three symmetric MOM schemes are presented in Fig.(\ref{two-threeloopggg}). Similar to the case of asymmetric MOM scheme, the gauge dependence cannot be suppressed by eliminating the scale dependence. More explicit, we present the total decay width $\Gamma(H \to gg)$ at several typical gauge parameters in Table.~\ref{tabletildeconv@pmc}.

\section{\label{summary}Summary}

In the paper, we have made a detailed discussion on the gauge dependence of the total decay width $\Gamma(H \to gg)$ up to five-loop level under various MOM schemes. Our main results are:
\begin{itemize}
\item The gauge dependence and the renormalization scale dependence under the MOM schemes are two different things. Figs.(\ref{two-fiveloop}-\ref{two-fiveloopMOMgg}) show that under conventional scale setting approach, the total scale dependence of $\Gamma(H \to gg)$ can be greatly suppressed when more loop terms have been known, due to the correlation of scale dependence among different orders; while the gauge dependence of the pQCD approximant behaves differently for different choices of scales and MOM schemes. After applying the PMC, the conventional renormalization scale dependence can be eliminated, but the gauge dependence of the MOM scheme is still there, which cannot be suppressed by including more loop terms. The gauge dependence may become even larger at higher orders. Thus the gauge dependence is the intrinsic property of MOM schemes. Even though MOM scheme has this weak point, it still has some advantages in dealing with pQCD predictions as explained in the Introduction. Among all the MOM schemes, the gauge dependence of $\Gamma(H\to gg)$ under the $\rm{MOMgg}$ scheme is the smallest and is less than $\pm1\%$. In this sense the MOMgg scheme could be treated as the best type of MOM scheme.

\item By applying the PMC, the scale independent effective momentum flow ($\bar{Q}$) of the process can be fixed by using the RGE, and as shown by Figs.(\ref{single-scale-coupling}, \ref{single-scale-coupling-Sym-MOM}), it differs for various MOM schemes, which ensures the scheme independence of pQCD predictions~\cite{Brodsky:1994eh}. For examples, if setting $\xi^{\rm MOM}\in[-3, 3]$, the effective scale $\bar{Q}$ is $\sim [31, 45]$ GeV for mMOM scheme, $\sim[25, 47]$ GeV for MOMh, MOMq and MOMg schemes, $\sim [109, 122]$ GeV for MOMgg scheme; if setting $\xi^{\rm MOM}$ within a smaller allowable region $[-1, 1]$, the effective scale $\bar{Q}$ changes to $\sim [43, 45]$ GeV for mMOM scheme, $\sim[40, 47]$ GeV for MOMh, MOMq and MOMg scheme, and $\sim [119, 122]$ GeV for MOMgg scheme. The small differences of $\bar{Q}$ for $\xi^{\rm MOM}\in[-1, 1]$ shall be further compensated by the differences of asymptotic scale, leading to a weaker gauge dependence of $\Gamma(H\to gg)$ for $\xi^{\rm MOM} \in [-1,1]$.

\item After applying the PMC, due to the elimination of renormalization scale ambiguity, a more accurate pQCD prediction for $\Gamma(H\to gg)$ can be achieved. It is found that the uncertainties caused by $\Delta\alpha_s(M_Z)$ and $\Delta M_H$ are small, which are around $4\%$ and $1\%$, respectively. By taking $\xi^{\rm MOM}\in[-1,1]$, the total decay width $\Gamma(H \to gg)|^{\rm mMOM}_{\rm PMC}$ shall be changed by about $10\%$, $3\%$, $2\%$ and $5\%$ for $n=2,3,4,5$, respectively; the total decay widthes $\Gamma(H \to gg)|^{\rm MOMh, MOMq, MOMg}_{\rm PMC}$ behave closely, which shall be changed by about $33\%$, $12\%$, $11\%$ and $19\%$ for $n=2,3,4,5$, respectively; and the total decay width $\Gamma(H \to gg)|^{\rm MOMgg}_{\rm PMC}$ is almost independent to the choice of gauge parameter, which shall be changed by only $\sim 1\%$ for $n=3,4,5$. By adding all the mentioned errors in quadrature, we obtain five-loop predictions of $\Gamma(H\to gg)$ under five asymmetric MOM schemes,
    \begin{eqnarray}
    \Gamma(H \to gg)|^{\rm{mMOM}}_{\rm{PMC}} &=& 332.8^{+13.8}_{-8.4}~\rm{KeV}, \\
    \Gamma(H \to gg)|^{\rm{MOMh}}_{\rm{PMC}}  &=& 332.8^{+28.5}_{-35.6}~\rm{KeV}, \\
    \Gamma(H \to gg)|^{\rm{MOMq}}_{\rm{PMC}}  &=& 332.9^{+28.4}_{-35.5}~\rm{KeV},  \\
    \Gamma(H \to gg)|^{\rm{MOMg}}_{\rm{PMC}}  &=& 332.7^{+28.4}_{-35.4}~\rm{KeV}, \\
    \Gamma(H \to gg)|^{\rm{MOMgg}}_{\rm{PMC}} &=&337.9^{+7.6}_{-7.7}~\rm{KeV}.
    \end{eqnarray}
    The MOMgg decay width has the smallest net error due to the small gauge dependence. It is found that the Higgs decay width $\Gamma  (H\to gg)$ varies weakly on the choice of the MOM schemes, being consistent with the renormalization group invariance. Such small differences (less than $\sim 1\%$) among different schemes could be attributed to the unknown higher-order terms, e.g. the unknown  N$^4$LL and higher-order terms in the PMC scale $\bar{Q}$'s perturbative series~\cite{Wu:2019mky}. For example,  with the help of Eq.(\ref{eq:PMCsscaleaQ}), if treating $\pm |\lambda_{3} a^3(Q)|$ as an estimation of the contribution of the unknown N$^4$LL-term of $\bar{Q}$, the change of the total decay width $\Delta\Gamma(H \to gg)$ shall well explain the gaps of the total decay widths of different schemes, e.g. $\Delta\Gamma(H \to gg)|_{\rm{PMC}}\simeq \pm2.8$ KeV for all the MOM schemes.

\item The pQCD convergence of the conventional series varies greatly under different choices of the renormalization scale, due to the mismatching of the perturbative coefficient with the $\alpha_s$-value at the same order; Thus it is improper to use the conventional series to predict the unknown terms. On the other hand, after applying the PMC, the scale-independent coupling $\alpha_s(\bar{Q})$ is determined, and together with the scale-invariant conformal coefficients, one can achieve the intrinsic perturbative nature of the pQCD series and give more reliable prediction of unknown terms. Using the known five-loop prediction of $\Gamma(H \to gg)$ as an explicit example, if choosing $\pm |r_{5,0}a^6_s(\bar{Q})|$ as a conservative estimation of the unknown six-loop prediction, we obtain the ${\cal O}(\alpha_s^6)$-order contribution is $\pm 3.1$ KeV for the mMOM, MOMh, MOMq and MOMg schemes, and $\pm 9.3$ KeV for MOMgg scheme.

\end{itemize}

\noindent{\bf Acknowledgement}: This work is partly supported by the Project Supported by Graduate Research and Innovation Foundation of Chongqing, China (Grant No. CYB19065), the National Natural Science Foundation of China under Grant No.11625520, No.11947406 and No.11905056, the China Postdoctoral Science Foundation under Grant No. 2019M663432, and by the Chongqing Special Postdoctoral Science Foundation under Grant No. XmT2019055.

\appendix

\section{\label{zggq}The relation of the renormalization constants under the mMOM and $\overline{\rm MS}$ schemes}

Eq.(\ref{S-T-identity}) shows that for any scheme $R$, we have
\begin{eqnarray}
Z_a^R=\frac{1}{Z_3^R}\Bigg(\frac{\tilde{Z}_1^R}{\tilde{Z}_3^R}\Bigg)^2,
\end{eqnarray}
and then we obtain
\begin{eqnarray}
\frac{Z_a^{\rm{\overline{MS}}}}{Z_a^{\rm{mMOM}}}&=&\frac{Z_3^{\rm{mMOM}}}{Z_3^{\rm{\overline{MS}}}}\Bigg(\frac{\tilde{Z}_1^{\rm{\overline{MS}}}}{\tilde{Z}_1^{\rm{mMOM}}}\Bigg)^2
\Bigg(\frac{\tilde{Z}_3^{\rm{mMOM}}}{\tilde{Z}_3^{\rm{\overline{MS}}}}\Bigg)^2 \nonumber\\
&=&\frac{Z_3^{\rm{mMOM}}}{Z_3^{\rm{\overline{MS}}}}\Bigg(\frac{\tilde{Z}_3^{\rm{mMOM}}}{\tilde{Z}_3^{\rm{\overline{MS}}}}\Bigg)^2.
\label{zamszamm}
\end{eqnarray}
On the other hand, Eq.(\ref{z3}) leads to
\begin{eqnarray}
\frac{Z_3^{\rm{mMOM}}}{Z_3^{\rm{\overline{MS}}}}=\frac{1+\Pi_A^{\rm{mMOM}}}{1+\Pi_A^{\rm{\overline{MS}}}},
\end{eqnarray}
and Eq.(\ref{tildez3}) leads to
\begin{eqnarray}
\frac{\tilde{Z}_3^{\rm{mMOM}}}{\tilde{Z}_3^{\rm{\overline{MS}}}}=\frac{1+\tilde{\Pi}_c^{\rm{mMOM}}}{1+\tilde{\Pi}_c^{\rm{\overline{MS}}}}.
\end{eqnarray}
At the substraction point $q^2=-\mu^2$, Eqs.(\ref{pia-mm-muequq}, \ref{tildepic-mm-muequq}) lead to
\begin{eqnarray}
\frac{Z_3^{\rm{mMOM}}}{Z_3^{\rm{\overline{MS}}}}&=&\frac{1}{1+\Pi_A^{\rm{\overline{MS}}}},\nonumber \\
\Bigg(\frac{\tilde{Z}_3^{\rm{mMOM}}}{\tilde{Z}_3^{\rm{\overline{MS}}}}\Bigg)^2&=&\frac{1}{(1+\tilde{\Pi}_c^{\rm{\overline{MS}}})^2}.
\label{z3mmz3ms}
\end{eqnarray}
Substituting Eqs.(\ref{zamszamm}, \ref{z3mmz3ms}) into Eq.(\ref{axiMM@MS}), we get the required Eqs.(\ref{aMOM}, \ref{xiMOM}).

\section{Perturbative transformations of the strong couplings and gauge parameters among the MOM schemes and the $\overline{\rm MS}$ scheme}
\label{MSandmMOM}

In this Appendix, we give the perturbative transformations of the strong couplings and gauge parameters among a specific MOM scheme and the $\overline{\rm MS}$ scheme. For convenience, we use the short notations $(\overline{a}, \overline{\xi})$ and $(a, \xi)$ to represent $(a^{\rm{\overline{MS}}}, \xi^{\rm{\overline{MS}}})$ and $(a^{\rm{MOM}}, \xi^{\rm{MOM}})$, respectively. The transformations have firstly been considered in Ref.\cite{Celmaster:1979km} and then been improved in Ref.\cite{Garkusha:2018mua}. Here for self-consistence and for our present needs, we give a more detailed derivation and one-order higher transformations than those of Ref.\cite{Garkusha:2018mua}.

Generally, one can expand the strong couplings and gauge parameters under the MOM scheme over the ones under the $\overline{\rm MS}$ scheme. Up to the present known order and to suit the needs of our present discussions, we have
\begin{eqnarray}
\label{Apendix1}
a&=&\overline{a}+\sum _{i=1}^4 \phi_{i}(\overline{\xi})\overline{a}^{i+1}+\mathcal{O}(\overline{a}^6), \\
\label{Apendix2}
\xi &=&\overline{\xi}\bigg(1+\sum _{n=1}^3\psi_{n}(\overline{\xi})\overline{a}^{n}+\mathcal{O}(\overline{a}^4)\bigg)~.
\end{eqnarray}
And inversely, we have
\begin{eqnarray}
\label{Apendix3}
\overline{a}&=&a+\sum _{n=1}^4 b_{n}(\xi)a^{n+1}+\mathcal{O}(a^6), \\
\label{Apendix4}
\overline{\xi}&=&\xi \bigg(1+\sum _{n=1}^3\chi_{n}(\xi )a^{n}+\mathcal{O}(a^4)\bigg).
\end{eqnarray}
Those two expansions (\ref{Apendix3}) and (\ref{Apendix4}) are important to transform the known $\overline{\rm MS}$ perturbative series of a physical observable to the one under a certain MOM scheme.

Our task is to derive the coefficients $b_i(\xi)$ and $\chi_i(\xi)$ from the known ones $\phi_i(\overline{\xi})$ and $\psi_i(\overline{\xi})$. For the purpose, we first do the following Taylor expansions:
\begin{eqnarray}
\label{Apendix5}
\phi_{i}(\overline{\xi}) &=& \sum_{n=0}^\infty \frac{d^{n}\phi_{i}(\xi)}{n! d\xi^{n}}(\overline{\xi}-\xi)^{n},i=1,2,3,4\ldots, \\
\label{Apendix6}
\psi_{j}(\overline{\xi}) &=& \sum_{n=0}^\infty \frac{d^{n}\psi_{j}(\xi)}{n! d\xi^{n}}(\overline{\xi}-\xi)^{n},j=1,2,3\ldots,
\end{eqnarray}
where
\begin{eqnarray}
\label{deltaxi}
\overline{\xi}-\xi=\xi\bigg(\chi_1(\xi)a+\chi_2(\xi)a^2+\chi_3(\xi)a^3+\mathcal{O}(a^4)\bigg).
\end{eqnarray}
Here $\phi_{i}(\xi)$ and $\psi_{j}(\xi)$ can be derived form Eqs.(\ref{xiMOM}, \ref{aMOMh}, \ref{aMOM}, \ref{aMOMq}, \ref{aMOMg}, \ref{aMOMgg}) with the known results given in Refs.\cite{Chetyrkin:2000dq, Ruijl:2017eht}.

Substituting Eq.(\ref{Apendix3}) into Eq.(\ref{Apendix1}) and using the formula (\ref{Apendix5}) in combination with Eq.(\ref{deltaxi}), then an expression of $b_i(\xi)$ over $\phi_i(\xi)$ and $b_i(\xi)$ can be obtained. Similarly, substituting Eq.(\ref{Apendix4}) into Eq.(\ref{Apendix2}) and using the formula Eq.(\ref{Apendix6}) in combination with Eq.(\ref{deltaxi}), one can obtain the expression of $\chi_i(\xi)$ over $\psi_i(\xi)$ and $\chi_i(\xi)$. It is straightforward to obtain the following set of transformation eqnarrays:
\begin{widetext}
\begin{eqnarray}
\label{v1}
b_1(\xi)&=&-\phi_1(\xi)~, ~~~ \chi_1(\xi)=-\psi_1(\xi)~, \\
\label{v2}
b_2(\xi)&=&-\phi_2(\xi)+2\phi^2_1(\xi)+\xi\psi_1(\xi)\frac{d\phi_1(\xi)}{d\xi}~, \\
\label{v3}
\chi_2(\xi)&=&-\psi_2(\xi)+\psi^2_1(\xi)+\psi_1(\xi)\phi_1(\xi)+\xi\psi_1(\xi)\frac{d\psi_1(\xi)}{d\xi}~, \label{v4} \\
b_3(\xi)&=&-\phi_3(\xi)+5\phi_2(\xi)\phi_1(\xi)-5\phi^3_1(\xi)+\xi\psi_1(\xi)\frac{d\phi_2(\xi)}{d\xi}-\frac{1}{2}\xi^2\psi^2_1(\xi)\frac{d^2\phi_1(\xi)}{d\xi^2} \nonumber\\
&&+\xi\frac{d\phi_1(\xi)}{d\xi}\bigg(\psi_2(\xi)-\psi^2_1(\xi)-5\psi_1(\xi)\phi_1(\xi)-\xi\psi_1(\xi)\frac{d\psi_1(\xi)}{d\xi}\bigg),
\label{v5} \\
\chi_3(\xi)&=&-2\phi_1(\xi)\psi_1(\xi)^2-2 \phi_1(\xi)^2 \psi_1(\xi)+\phi_2(\xi)\psi_1(\xi)+2\phi_1(\xi)\psi_2(\xi)-\psi_1(\xi)^3 +2\psi_2(\xi)\psi_1(\xi)-\psi_3(\xi)  \nonumber\\
&& +\xi \Bigg[-\frac{d\phi_1(\xi)}{d\xi} w_1(\xi)^2+\frac{d\psi_1(\xi)}{d\xi} \left(-2 \phi_1(\xi) \psi_1(\xi)-3 \psi_1(\xi)^2+\psi_2(\xi)\right)+\frac{d\psi_2(\xi)}{d\xi} \psi_1(\xi)\Bigg]  \nonumber\\
&& +\xi^2 \left[-\left(\frac{d\psi_1(\xi)}{d\xi}\right)^2 \psi_1(\xi)-\frac{1}{2}\frac{d^2\psi_1(\xi)}{d\xi^2} \psi_1(\xi)^2\right],
\label{v6} \\
b_4(\xi)&=&14\phi_1(\xi)^4-21 \phi_2(\xi) \phi_1(\xi)^2+6\phi_3(\xi) \phi_1(\xi)+3 \phi_2(\xi)^2-\phi_4(\xi)    \nonumber\\
&+&\xi  \Bigg[\frac{d\phi_1(\xi)}{d\xi} \Bigg(3\psi_1\left(\xi)(2 \phi_1(\xi) \psi_1(\xi)+7 \phi_1(\xi)^2-2 \phi_2(\xi)\right)-6\phi_1(\xi)\psi_2(\xi)+\psi_1(\xi)^3 \nonumber\\
&-&2 \psi_1(\xi)\psi_2(\xi)+\psi_3(\xi)\Bigg)+\frac{d\phi_2(\xi)}{d\xi}\left(-6 \phi_1(\xi) \psi_1(\xi)-\psi_1(\xi)^2+\psi_2(\xi)\right)+\frac{d\phi_3(\xi)}{d\xi} \psi_1(\xi)\Bigg] \nonumber\\
&+&\xi ^2 \Bigg[\frac{d\phi_1(\xi)}{d\xi} \Bigg(\frac{d\psi_1(\xi)}{d\xi} \left(6 \phi_1(\xi) \psi_1(\xi)+3 \psi_1(\xi)^2-\psi_2(\xi)\right)-\frac{d\psi_2(\xi)}{d\xi} \psi_1(\xi)\Bigg)\nonumber\\
&+&\frac{d^2 \phi_1(\xi)}{d\xi^2} \psi_1(\xi)\Bigg(3 \phi_1(\xi) \psi_1(\xi)+\psi_1(\xi)^2-\psi_2(\xi)\Bigg)+3 (\frac{d \phi_1(\xi)}{d\xi})^2 \psi_1(\xi)^2 \nonumber \\
&-&\frac{d \phi_2(\xi)}{d\xi} \frac{d \psi_1(\xi)}{d\xi} \psi_1(\xi)-\frac{1}{2} \frac{d^2 \phi_2(\xi)}{d\xi^2} \psi_1(\xi)^2\Bigg] \nonumber\\
&+&\xi ^3 \Bigg[\frac{d^2 \phi_1(\xi)}{d\xi^2} \frac{d \psi_1(\xi)}{d\xi} \psi_1(\xi)^2+\frac{d \phi_1(\xi)}{d\xi} \Bigg(\frac{1}{2} \frac{d^2 \psi_1(\xi)}{d\xi^2} \psi_1(\xi)^2+(\frac{d \psi_1(\xi)}{d\xi})^2 \psi_1(\xi)\Bigg) \nonumber\\
&+&\frac{1}{6} \frac{d^3 \phi_1(\xi)}{d\xi^3} \psi_1(\xi)^3\Bigg].
\end{eqnarray}
\end{widetext}
The above formulas are adaptable for any MOM schemes, the differences lie in the exact expressions for coefficient functions. It is interesting to find that at the two-loop level, the perturbative predictions under the MOMh, MOMq and MOMg schemes are exactly the same, which can be demonstrated with the help of the formulas (\ref{aMOMh}, \ref{aMOMq}, \ref{aMOMg}). The differences (numerically very small) among those three renormalization schemes start from three-loop level. \\

\section{\label{lambda} The perturbative coefficients of the PMC scale}

The perturbative coefficients $\lambda_{i}$ $(i=0,1,2,3)$ for the expansion of $\ln\bar{Q}^2/Q^2$ over the coupling constant $a(Q)$ up to NNNLL accuracy are
\begin{widetext}
\begin{eqnarray}
\lambda_{0}&=& -\frac{\overline{r}_{2,1}}{\overline{r}_{1,0}},\label{eq:lambda0}\\
\lambda_{1}&=& \frac{(p+1)(\overline{r}_{2,0} \overline{r}_{2,1}- \overline{r}_{1,0} \overline{r}_{3,1})}{p \overline{r}_{1,0}^2}+\frac{(p+1)(\overline{r}_{2,1}^2-\overline{r}_{1,0} \overline{r}_{3,2})}{2 \overline{r}_{1,0}^2}\beta_0,\label{eq:lambda1}
\end{eqnarray}
\begin{eqnarray}
\lambda_{2} &=& \frac{(p+1)^2 \left(\overline{r}_{1,0} \overline{r}_{2,0} \overline{r}_{3,1} - \overline{r}_{2,0}^2 \overline{r}_{2,1} \right)
+ p(p+2) \left(\overline{r}_{1,0} \overline{r}_{2,1} \overline{r}_{3,0}-\overline{r}_{1,0}^2 \overline{r}_{4,1}\right)}{p^2 \overline{r}_{1,0}^3}+\frac{(p+2)\left(\overline{r}_{2,1}^2-\overline{r}_{1,0}\overline{r}_{3,2}\right)}{2\overline{r}_{1,0}^2}\beta_1  \nonumber\\
&&-\frac{(p+1)(2p+1)\overline{r}_{2,0} \overline{r}_{2,1}^2-(p+1)^2 \left(2\overline{r}_{1,0}\overline{r}_{2,1}\overline{r}_{3,1}+\overline{r}_{1,0} \overline{r}_{2,0} \overline{r}_{3,2}\right)+(p+1)(p+2)\overline{r}^2_{1,0} \overline{r}_{4,2}}{2p\overline{r}_{1,0}^3}\beta_0 \nonumber\\
&&+\frac{(p+1)(p+2)\left(\overline{r}_{1,0}\overline{r}_{2,1}\overline{r}_{3,2}-\overline{r}_{1,0}^2 \overline{r}_{4,3}\right)+(p+1)(1+2p)\left(\overline{r}_{1,0} \overline{r}_{2,1} \overline{r}_{3,2}-\overline{r}_{2,1}^3\right)}
{6\overline{r}_{1,0}^3}\beta_0^2,\label{eq:lambda2}  \\
\lambda_{3}&=&\frac{1}{p^3 \overline{r}_{1,0}^4}\Bigg[p \overline{r}_{1,0}^2 \Bigg(\left(p^2+3 p+2\right) \overline{r}_{3,0} \overline{r}_{3,1}+p (p+3) \left(\overline{r}_{2,1}
\overline{r}_{4,0}-\overline{r}_{1,0} \overline{r}_{5,1}\right)\Bigg)+(p+1)^3\left( \overline{r}_{2,0}^3 \overline{r}_{2,1}-\overline{r}_{1,0} \overline{r}_{2,0}^2 \overline{r}_{3,1}\right) \nonumber\\
&&+p \left(p^2+3 p+2\right) \overline{r}_{1,0} \overline{r}_{2,0} \left(\overline{r}_{1,0} \overline{r}_{4,1}-2 \overline{r}_{2,1} \overline{r}_{3,0}\right)\Bigg]+\frac{(p+3) \left(\overline{r}_{2,1}^2-\overline{r}_{1,0} \overline{r}_{3,2}\right)}{2 \overline{r}_{1,0}^2}\beta_2\nonumber\\
&&-\frac{(p+1) \Bigg[\overline{r}_{1,0} \Bigg((p+3) \overline{r}_{1,0} \overline{r}_{4,2}-2 (p+2) \overline{r}_{2,1} \overline{r}_{3,1}\Bigg)+\overline{r}_{2,0} \Bigg((2 p+3) \overline{r}_{2,1}^2-(p+2) \overline{r}_{1,0} \overline{r}_{3,2}\Bigg)\Bigg]}{2 p \overline{r}_{1,0}^3}\beta_1 \nonumber\\
&&-\frac{\left(4 p^2+9 p+3\right) \overline{r}_{2,1}^3-6 \left(p^2+3 p+2\right) \overline{r}_{1,0} \overline{r}_{3,2} \overline{r}_{2,1}+\left(2 p^2+9 p+9\right) \overline{r}_{1,0}^2 \overline{r}_{4,3}}{6 \overline{r}_{1,0}^3}\beta_0 \beta_1\nonumber\\
&&+\frac{1}{2 p^2 \overline{r}_{1,0}^4}\Bigg[(p+1)^2 \overline{r}_{2,0}^2 \Bigg(3 p \overline{r}_{2,1}^2-(p+1) \overline{r}_{1,0} \overline{r}_{3,2}\Bigg)-(p+1)^2 \overline{r}_{1,0} \overline{r}_{2,0} \Bigg(2 (2 p+1) \overline{r}_{2,1} \overline{r}_{3,1}-(p+2) \overline{r}_{1,0} \overline{r}_{4,2}\Bigg)\nonumber\\
&&+\overline{r}_{1,0} \Bigg(-2 p^2 (p+2) \overline{r}_{3,0} \overline{r}_{2,1}^2+2 p \left(p^2+3 p+2\right) \overline{r}_{1,0} \overline{r}_{4,1} \overline{r}_{2,1}+\overline{r}_{1,0} (p+1)^3 \overline{r}_{3,1}^2\nonumber\\
&&+p (p+2) \overline{r}_{1,0} \Bigg((p+1) \overline{r}_{3,0} \overline{r}_{3,2}-(p+3) \overline{r}_{1,0} \overline{r}_{5,2}\Bigg)\Bigg)\Bigg]\beta_0\nonumber\\
&&+\frac{p+1}{6~p~\overline{r}_{1,0}^4}\Bigg[\overline{r}_{2,0} \Bigg(\left(6 p^2+5 p+1\right) \overline{r}_{2,1}^3-3 \left(2 p^2+3 p+1\right) \overline{r}_{1,0} \overline{r}_{3,2} \overline{r}_{2,1}+\left(p^2+3 p+2\right) \overline{r}_{1,0}^2 \overline{r}_{4,3}\Bigg)\nonumber\\
&&+\overline{r}_{1,0} \Bigg(-3 \left(2 p^2+3 p+1\right) \overline{r}_{3,1} \overline{r}_{2,1}^2+3 \left(p^2+3 p+2\right) \overline{r}_{1,0} \overline{r}_{4,2} \overline{r}_{2,1}+3 (p+1)^2 \overline{r}_{1,0} \overline{r}_{3,1} \overline{r}_{3,2}\nonumber\\
&&-\left(p^2+5 p+6\right) \overline{r}_{1,0}^2 \overline{r}_{5,3}\Bigg)\Bigg]\beta_0^2+\frac{p+1}{24 \overline{r}_{1,0}^4}\Bigg[\left(6 p^2+5 p+1\right) \overline{r}_{2,1}^4-6 \left(2 p^2+3 p+1\right) \overline{r}_{1,0} \overline{r}_{3,2} \overline{r}_{2,1}^2\nonumber\\
&&+4 \left(p^2+3 p+2\right) \overline{r}_{1,0}^2 \overline{r}_{4,3} \overline{r}_{2,1}+\overline{r}_{1,0}^2 \Bigg(3 (p+1)^2 \overline{r}_{3,2}^2-\left(p^2+5 p+6\right) \overline{r}_{1,0} \overline{r}_{5,4}\Bigg)\Bigg]\beta_0^3\label{eq:lambda3}.
\end{eqnarray}
\end{widetext}

\section{The $\overline{\rm MS}$-scheme coefficients $r_{i,j}(M_H)$ for $\Gamma(H\to gg)$ up to $\alpha_s^6$-order level}
\label{MShgg}

Using the perturbative coefficients $C_{k}(\mu)$ at the renormalization scale $\mu=M_H$ known in Ref.~\cite{Herzog:2017dtz}, we can obtain the required $\overline{\rm MS}$-scheme coefficients $r_{i,j}(M_H)$, which can be adopted for our present PMC analysis. More explicitly, up to $\alpha_s^6$-order, the coefficients $r_{i,j}(M_H)$ are
\begin{widetext}
\begin{eqnarray}
r_{1,0}(M_H) &=& 16,\\
r_{2,0}(M_H) &=& 288,\\
r_{2,1}(M_H) &=& 56,\\
r_{3,0}(M_H) &=& 3424\ln \frac{M_H^2}{m_t^2}-10560\zeta_{3}-\frac{153160}{9},\\
r_{3,1}(M_H) &=& -\frac{256}{3}\ln \frac{M_H^2}{m_t^2}-160\zeta_{3}+\frac{15424}{9},\\
r_{3,2}(M_H) &=& \frac{2032}{9}-\frac{16{\pi^2}}{3},  \\
r_{4,0}(M_H) &=& \frac{286664}{3}\ln \frac{M_H^2}{m_t^2}+\frac{1196800\zeta_{5}}{3}-\frac{2136260\zeta_{3}}{9}\nonumber\\
&&-\frac{47446802}{81},\\
r_{4,1}(M_H) &=& 856\ln^2 \frac{M_H^2}{m_t^2}+ \frac{39028}{9}\ln \frac{M_H^2}{m_t^2}+\frac{3040\zeta_{5}}{3}\nonumber\\
&&-\frac{712325\zeta_{3}}{18}+856{\pi^2}-\frac{401489}{9}, \\
r_{4,2}(M_H) &=& -\frac{128}{3}\ln^2 \frac{M_H^2}{m_t^2}-\frac{2380}{9}\ln \frac{M_H^2}{m_t^2}-1104\zeta_{3}\nonumber\\
&&-\frac{416{\pi ^2}}{3}+\frac{695671}{81}, \\
r_{4,3}(M_H) &=& -32\zeta_{3}+\frac{28508}{27}-56{\pi^2}, \\
r_{5,0}(M_H) &=& 366368\ln^2 \frac{M_H^2}{m_t^2}-\left(4308480\zeta_{3}+\frac{37289332}{9}\right)\ln \frac{M_H^2}{m_t^2}+\frac{698066944 a_{5}}{945}+\frac{806396224 a_{4}}{567}-\frac{125910400\zeta_{7}}{9} \nonumber\\
&&+\frac{9927911216\zeta_{5}}{945}+8518400\zeta_{3}^2+\frac{121633044056\zeta_{3}}{4725}-\frac{2786175509{\pi^4}}{127575}-\frac{558112{\pi^2}}{3}+\frac{2542274821357}{255150}\nonumber\\
&&-\frac{87258368 \ln^{5} 2}{14175}+\frac{100799528\ln^{4} 2}{1701}+\frac{87258368{\pi^2}\ln^{3} 2}{8505}-\frac{100799528{\pi^2}\ln^{2} 2}{1701}\nonumber\\
&&+\left(\frac{298190944{\pi^4}}{42525}+\frac{109568}{9}{\pi^2}\right)\ln 2,\\
r_{5,1}(M_H) &=& 12916\ln^2 \frac{M_H^2}{m_t^2}+\left(\frac{592412\zeta_{3}}{15}+\frac{46766278}{135}\right)\ln \frac{M_H^2}{m_t^2}-\frac{349033472 a_{5}}{1575}-\frac{404118752 a_{4}}{945}-\frac{69440\zeta_{7}}{9}\nonumber\\
&&+\frac{2655253192\zeta_{5}}{1575}+154880\zeta_{3}^2-\frac{81219195943\zeta _{3}}{47250}+\frac{3225459599{\pi^4}}{425250}+\frac{1364764{\pi ^2}}{45}-\frac{551995547183}{212625}\nonumber\\ &&+\frac{43629184 \ln^{5} 2}{23625}-\frac{50514844 \ln^{4} 2}{2835} - \frac{43629184{\pi ^2} \ln^{3} 2}{14175}+
\frac{50514844{\pi ^2}\ln^{2} 2}{2835}\nonumber\\ &&-\left(\frac{149095472{\pi^4}}{70875}+\frac{8192}{45}{\pi^2}\right)\ln 2, \\
r_{5,2}(M_H) &=& \frac{1712}{5}\ln^3 \frac{M_H^2}{m_t^2}+\frac{25756}{15}\ln^2 \frac{M_H^2}{m_t^2}+\left(\frac{32137\zeta_{3}}{5}-\frac{1712{\pi^2}}{5}+\frac{95279}{45}\right) \ln \frac{M_H^2}{m_t^2}+\frac{2192 a_{4}}{3}\nonumber\\
&&+5144\zeta_{5}+1760\zeta_{3}^2+\left(3520{\pi ^2}-\frac{92524417}{540}\right)\zeta_{3}-\frac{183623{\pi^4}}{900}+\frac{118844{\pi^2}}{15}-\frac{52667603}{405}\nonumber\\
&&+\frac{274\ln^{4} 2}{9}-\frac{274}{9}{\pi^2}\ln^{2} 2,
\end{eqnarray}
\begin{eqnarray}
r_{5,3}(M_H) &=& -\frac{128}{5}\ln^3 \frac{M_H^2}{m_t^2}-\frac{742}{5}\ln^2 \frac{M_H^2}{m_t^2}+\left(\frac{128{\pi^2}}{5}-\frac{17948}{15}\right)\ln \frac{M_H^2}{m_t^2}-480\zeta_{5}+\left(160{\pi^2}-\frac{21104}{3}\right)\zeta_{3}\nonumber\\
&&- \frac{86794{\pi^2}}{45}+\frac{4999417}{108},\\
r_{5,4}(M_H) &=& -448\zeta_{3}+\frac{773024}{135}-\frac{4064{\pi^2}}{9}+\frac{16{\pi^4}}{5}.
\end{eqnarray}
\end{widetext}
Where $\zeta_{n}$ is Riemannian Zeta function, $a_n=\rm{Li_{n}}(\frac{1}{2})={\sum\nolimits_{k = 1}^\infty  {\left( {{2^k}{k^n}} \right)} ^{- 1}}$.

\end{document}